\documentclass[11pt]{article}

\usepackage[T1]{fontenc}
\usepackage{lmodern}
\usepackage[margin=1in]{geometry}
\usepackage{amsmath}
\usepackage{amssymb}
\usepackage[scr=stixtwofancy]{mathalpha}
\DeclareMathAlphabet{\mathrsfs}{U}{rsfs}{m}{n}
\usepackage{mathtools}
\usepackage{booktabs}
\usepackage{float}
\usepackage{etoolbox}
\usepackage{microtype}
\usepackage{comment}
\usepackage{parskip}
\usepackage{tikz}
\usetikzlibrary{arrows.meta,calc,decorations.markings,decorations.pathmorphing,matrix,positioning}
\usepackage{xcolor}
\usepackage[numbers,sort&compress]{natbib}
\usepackage[hidelinks]{hyperref}

\allowdisplaybreaks
\colorlet{GraphRegularColor}{black!76!white}
\colorlet{GraphHighlightColor}{GraphRegularColor}
\colorlet{GraphContractionColor}{GraphRegularColor}

\newcommand{\GraphNodeDiameter}{5pt}
\newcommand{\GraphLineWidth}{1pt}
\tikzset{
  >={Triangle[length=0pt 4,width=0pt 4]},
  vertex/.style={circle,fill=GraphRegularColor,inner sep=0pt,minimum size=\GraphNodeDiameter},
  edge/.style={draw=GraphRegularColor,line width=\GraphLineWidth,line cap=round},
  highlighted vertex/.style={vertex,fill=GraphHighlightColor},
  highlighted edge/.style={edge,draw=GraphHighlightColor},
  contracted vertex/.style={vertex,fill=GraphContractionColor},
  contracted edge/.style={edge,draw=GraphContractionColor,dash pattern=on 1.5pt off 1.5pt},
  omitted edge/.style={edge,decorate,decoration={snake,amplitude=0.7pt,segment length=4pt}},
  graph arrow/.style={->,draw=GraphRegularColor,line width=\GraphLineWidth}
}

\DeclareRobustCommand{\ThreeMelon}{\ensuremath{\ominus}}
\DeclareRobustCommand{\ThreePetal}{%
  \tikz[baseline={([yshift=-0.5ex]current bounding box.center)},x=1em,y=1em]{%
    \foreach \angle in {90,210,330}{%
      \draw[edge,line width=0.4pt,rotate=\angle] (0,0)
        .. controls (0.7,-0.55) and (0.7,0.55) .. (0,0);
    }
    \node[vertex,minimum size=1.2pt] at (0,0) {};
    \pgfresetboundingbox
    \path[use as bounding box] (-0.54,-0.37) rectangle (0.54,0.55);
  }%
}

\patchcmd{\subparagraph}{\bfseries}{\itshape}{}{}

\title{Higher-Loop Vacuum Integrals on $S^D$}
\author{%
  Varun Lochab and Frederik Denef\\[6pt]
  {\small\itshape Department of Physics, Columbia University,}\\
  {\small\itshape 538 West 120th Street, New York, New York 10027, U.S.}\\[6pt]
  {\small
    E-mail: \href{mailto:vv2338@columbia.edu}{vv2338@columbia.edu},
    \href{mailto:fmd7@columbia.edu}{fmd7@columbia.edu}}%
}
\date{}

\begin{document}
\maketitle

\vspace{3cm}

\begin{abstract}
We develop a new framework for evaluating higher-loop integrals for scalar fields on $S^D$. Starting from a Schwinger-parametric form for general higher-loop integrals on the sphere, it combines total-derivative identities with edge-contraction relations into a coupled linear system. To illustrate our method, we solve for all vacuum graphs through two loops. This includes, in particular, the two-loop three-melon, for which we construct the analytic solution for general $D$ and obtain explicit dimensionally regularized results at all integer dimensions.
\end{abstract}

\newpage

\setcounter{tocdepth}{2}
\tableofcontents

\newpage

\section{Introduction}
\label{sec:intro}

Quantum corrections to de~Sitter entropy can be calculated by expanding the Euclidean path integral around the round-sphere saddle \cite{gibbons1977action,anninos2022quantum}. One-loop corrections on the sphere are well understood \cite{christensen1980quantizing,law2021compendium,anninos2022quantum}. At higher loops, interactions lead to integrated products of propagators that are harder to evaluate. Even at two loops, only special cases have been fully worked out \cite{bandaru2024quantum, cacciatori2024loops, castroEtAlToAppear}. We develop a framework for solving higher-loop scalar vacuum integrals on a sphere of arbitrary dimension \(D\). We apply it to all vacuum graphs through two loops and obtain complete results for general spectral parameters.

Our starting point is a theory of massive real scalar fields \(\boldsymbol\phi=(\phi_1,\phi_2,\ldots)\) on the unit sphere \(S^D\) with an arbitrary interaction potential \(\mathcal V(\boldsymbol\phi)\). Its Euclidean action is given by
\begin{equation}
	S[\boldsymbol\phi] = \int_{S^D}d\Omega_D\, \left[ \frac12\sum_n \left( \nabla_\mu\phi_n\nabla^\mu\phi_n +M_n^2\phi_n^2 \right) +\mathcal V(\boldsymbol\phi) \right].
	\label{eq:intro-scalar-action}
\end{equation}
A vacuum graph \(g\) of this theory with \(V\) vertices and \(E\) edges defines the integral
\begin{equation}
	I_g(D;\boldsymbol{\Delta}^{\pm})=\int_{(S^D)^V}\prod_{v=1}^{V}d\Omega_D(\widehat X_v)\prod_{i=1}^{E}G_{\Delta_i^\pm}(\widehat X_{\mathsf a(i)},\widehat X_{\mathsf b(i)}), \qquad \widehat X_v\in S^D.
	\label{eq:intro-physical-graph}
\end{equation}
Here $\mathsf a(i)$ and $\mathsf b(i)$ are the vertices joined by edge $i$, and \(\boldsymbol{\Delta}^{\pm}=(\Delta_1^\pm,\ldots,\Delta_E^\pm)\). The sphere propagator \(G_{\Delta_i^\pm}\) has Klein--Gordon mass squared \(M_i^2=\Delta_i^+\Delta_i^-\), where the weights \(\Delta_i^\pm\) satisfy \(\Delta_i^++\Delta_i^-=D-1\).

Using an ambient-space representation of the sphere propagators and vertex measures in \(\mathbb R^{D+1}\), we recast \eqref{eq:intro-physical-graph} in Schwinger parameters (see Appendix~\ref{app:graph-integrals} for the derivation),
\begin{equation}
	I_g\propto\int_0^\infty\frac{dz}{z}\, z^{(D+1)/2}\prod_{i=1}^E \frac{d\lambda_i}{\lambda_i}\frac{d\mu_i}{\mu_i}\, \lambda_i^{\Delta_i^+}\mu_i^{\Delta_i^-}e^{-zA_g(\boldsymbol{\lambda},\boldsymbol{\mu})}, \qquad \boldsymbol{\beta}:=\left( \frac{D+1}{2}, \boldsymbol{\Delta}^{+}, \boldsymbol{\Delta}^{-} \right).
	\label{eq:intro-Schwinger-integral}
\end{equation}
The exponent vector \(\boldsymbol{\beta}\) contains the dimension \(D\) and the spectral parameters, while the graph polynomial \(A_g(\boldsymbol{\lambda},\boldsymbol{\mu})\) encodes the graph's connectivity. We act on the integrand in \eqref{eq:intro-Schwinger-integral} with \(z\partial_z\), \(\lambda_i\partial_{\lambda_i}\), and \(\mu_i\partial_{\mu_i}\), generating polynomial insertions. Their expansion yields total-derivative identities relating \(I_g\) across a lattice generated by integer shifts of \(\boldsymbol{\beta}\). We then derive insertions that relate the lattice of \(g\) to those of the graphs obtained by contracting its edges. Together, the total-derivative identities and edge-contraction relations form a coupled linear system (Section~\ref{sec:graph-lattices}) that we solve through two loops (Sections~\ref{sec:trees}--\ref{sec:two-loops}). Finally, we discuss how successive contractions of edges and forests---disjoint unions of trees---extend this framework to arbitrary loop order (Section~\ref{sec:conclusion}).

Below, we give a few explicit dimensionally regularized results for the two-loop three-melon integral. Since the three-melon has three edges joining two vertices~(\(\ThreeMelon\)), we label the graph \(\texttt{V\{3\}V}\) and denote its volume-normalized integral by \(\bar I_{\texttt{V\{3\}V}}\). We parameterize its spectral weights and masses by
\begin{equation}
	\Delta_i^\pm=\frac{D-1}{2}\pm\rho_i, \qquad M_i^2=\frac{(D-1)^2}{4}-\rho_i^2.
	\label{eq:intro-spectral-parameters}
\end{equation}
For positive mass squared, \(\rho_i\) is real when \(0<M_i^2\le (D-1)^2/4\) and purely imaginary when \(M_i^2>(D-1)^2/4\). In de Sitter space, these mass ranges correspond to the scalar complementary and principal series of \(\mathrm{SO}(1,D)\), respectively, meeting at \(\rho_i=0\) \cite{sun2025note}. 

We specialize the results here to equal spectral parameters. The general case is given in Section~\ref{sec:three-melon}.

In odd dimensions, the dimensionally regularized results contain a shifted digamma combination symmetric under \(\rho_i\mapsto-\rho_i\). When \(\rho_1=\rho_2=\rho_3=\rho\), the digamma combination is given by
\begin{equation}
	\Psi_\alpha^S := \left(1-\frac{3}{4\sin^2(\pi\rho)}\right) \sum_{\sigma=\pm1}\psi\!\left(\alpha+\sigma\frac{3\rho}{2}\right) +\frac{3}{4\sin^2(\pi\rho)} \sum_{\sigma=\pm1}\psi\!\left(\alpha+\sigma\frac{\rho}{2}\right) -\log\!\left(\frac{\pi}{e^{\gamma_{\mathrm E}-1}}\right).
\end{equation}
The superscript \(S\) labels spherical quantities to distinguish them from their flat-space counterparts obtained later in Section~\ref{sec:three-melon-flat-space-contraction}. We find
\begin{align}
	\bar I_{\texttt{V\{3\}V}}(3-\epsilon;\rho) &= \frac{1}{32\pi^2} \left[\frac1\epsilon-\Psi_{1/2}^S\right] +O(\epsilon), \\
	\bar I_{\texttt{V\{3\}V}}(5-\epsilon;\rho) &= \frac{3\rho^4}{3072\pi^4} \left[ \frac1\epsilon-\Psi_1^S +2(\rho^2-2)\left(\frac{\cot(\pi\rho)}{\rho}\right)^2+\frac56 \right] +O(\epsilon).
\end{align}
The dimensional recurrence in Section~\ref{sec:three-melon-recurrences} transports these two expansions to all higher odd dimensions in steps of four.

For \(D=4-\epsilon\), the polygamma combinations symmetric under \(\rho_i\mapsto-\rho_i\) are
\begin{equation}
	\chi_{i,0}^{S^4} := \sum_{\sigma=\pm1}\psi\!\left(-\frac12+\sigma\rho_i\right) -\log(4\pi)+\gamma_{\mathrm E}-\frac32, \qquad \chi_{i,1}^{S^4} := \sum_{\sigma=\pm1}\psi^{(1)}\!\left(-\frac12+\sigma\rho_i\right) +\frac{\pi^2}{6}.
\end{equation}
For equal spectral parameters, \(\rho_1=\rho_2=\rho_3=\rho\), set \(y^S:=\tfrac14-\rho^2\) and \(\chi_r^{S^4}:=\chi_{i,r}^{S^4}\). We find
\begin{align}
	\bar I_{\texttt{V\{3\}V}}(4-\epsilon;\rho) ={}&-\frac{1}{128\pi^4}\Biggl[ \frac{3y^S}{\epsilon^2} +\frac{\frac12-3y^S\chi_0^{S^4}}{\epsilon} -\frac{9y^S-2}{8} \left[ \mathsf F(\rho) -\log\!\left(4\pi e^{\gamma_{\mathrm E}}\right) -\frac92 \right] \notag\\
	&\quad -\frac38(2-3y^S)\chi_0^{S^4} -\frac{12(y^S)^2-31y^S+12}{8y^S} +\frac34y^S\left[2(\chi_0^{S^4})^2+\chi_1^{S^4}\right] \Biggr]+O(\epsilon).
	\label{eq:intro-three-melon-D4}
\end{align}
This result is transported from the \(D=0-\epsilon\) expansion. At leading order about \(D=0\), the vertices of a vacuum graph with \(E\) edges formally collapse to a single vertex, turning its edges into \(E\) petals of a flower. The three-melon gives a three-petal flower~(\ThreePetal). Here \(\mathsf F(\rho)\) follows from the correction to the flower limit and is most conveniently given by a convergent one-dimensional integral
\begin{equation}
	\mathsf F(\rho) := \int_0^1\frac{d\mathrm w}{1-\mathrm w} \left[ 1-\mathrm w^2 \left( \frac{\pi(1-4\rho^2)}{4\cos(\pi\rho)} {}_2F_1\!\left(\frac12+\rho,\frac12-\rho;2;\mathrm w\right) \right)^3 \right].
	\label{eq:intro-three-melon-F}
\end{equation}
The integral \eqref{eq:intro-three-melon-F} generalizes the flat-space dilogarithmic term \cite{cacciatori2023banana,ford1992standard} to the sphere and recovers it in the flat-space limit, as shown in Section~\ref{sec:three-melon-flat-space-contraction}. For \(D=6-\epsilon\), the dimensional recurrence requires only the convergent \(D=2\) three-melon integral. We omit the resulting regularized expression here for brevity. Iterating the recurrence gives the results at higher even dimensions.

\textit{Note added:} After this paper was completed, \cite{Bassett:2026igj} appeared where a subset of recurrences contained in Table \ref{tab:coupled-system-rows} below (Section \ref{sec:coupled-system}) are derived using hypergeometric identities for propagators. The explicit expression for \(\mathcal{I}_3^+\) stated in \cite{Bassett:2026igj} corresponds to and agrees with the first two terms in \eqref{eq:intro-three-melon-D4} for equal \(\rho\) and \eqref{eq:three-melon-D4-result} for the general case. \newpage

\section{Graph Lattice Framework}
\label{sec:graph-lattices}

In this section, we construct the Schwinger-parameter representation of a scalar vacuum integral directly from its graph data. The derivation is covered in Appendices~\ref{app:embedding-propagator} and~\ref{app:graph-integrals}. %, and it uses the ambient-space representation of the sphere propagator obtained in Appendix~\ref{app:embedding-propagator}.

Suppose the vacuum graph \(g\) has \(V\) vertices and \(E\) edges. Choose an orientation for each edge, and denote the initial and final vertices of edge \(i\) by \(\mathsf a(i)\) and \(\mathsf b(i)\). Here \(\mathsf a\) and \(\mathsf b\) are endpoint maps from edge labels \(i\) to vertex labels \(v\). 

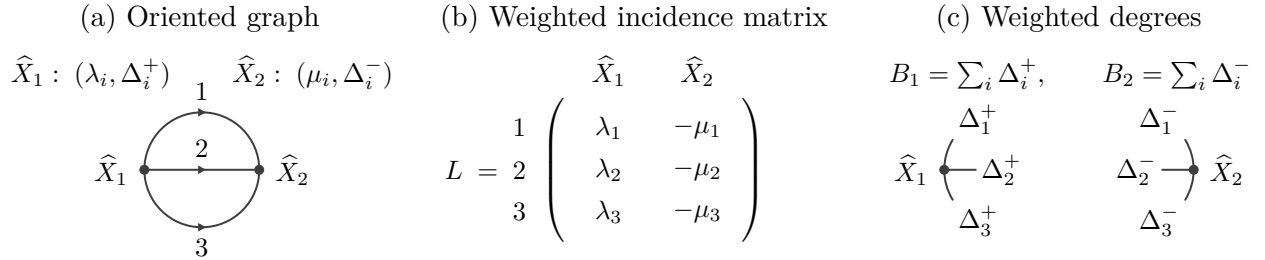
\begin{figure}[ht]
  \centering
  \renewcommand{\GraphLineWidth}{0.75pt}
  \renewcommand{\GraphNodeDiameter}{3.75pt}
  \setlength{\abovecaptionskip}{0pt}
  \begin{tikzpicture}[
    every delimiter/.style={text=black},
    every left delimiter/.style={xshift=3.1pt},
    every right delimiter/.style={xshift=-3.1pt},
    every node/.style={font=\normalfont\small}
  ]
    \node[font=\normalfont\normalsize] at (3.2875,4.6875) {(a) Oriented graph};
    \node[font=\normalfont\normalsize] at (9.0375,4.6875) {(b) Weighted incidence matrix};
    \node[font=\normalfont\normalsize] at (14.7875,4.6875) {(c) Weighted degrees};

    % The manuscript's three-melon geometry.
    \begin{scope}[
      shift={(3.2875,2.6875)},
      xshift=0.81pt, % Balance the visible outer edges of X_1 and X_2.
      mid triangle/.style={
        postaction={decorate},
        decoration={markings,mark=at position 0.5 with {
          \pgftransformxshift{1.6275pt}
          \arrow{>}
        }}
      },
      every node/.style={font=\normalfont\small}
    ]
      \coordinate (x) at (-0.76,0);
      \coordinate (y) at (0.76,0);
      \draw[edge,mid triangle] (x)
        arc[start angle=180,end angle=0,radius=0.76];
      \draw[highlighted edge,mid triangle] (x)--(y);
      \draw[edge,mid triangle] (x)
        arc[start angle=180,end angle=360,radius=0.76];
      \node[highlighted vertex] at (x) {};
      \node[highlighted vertex] at (y) {};
      \node[text=black,left=2.25pt] at (x) {$\widehat X_1$};
      \node[text=black,right=2.25pt] at (y) {$\widehat X_2$};
      \node[text=black] at (0,1.035) {$1$};
      \node[text=black] at (0,0.275) {$2$};
      \node[text=black] at (0,-1.035) {$3$};
      % Keep the left label's right edge and the right label's left edge fixed.
      \node (left-assignment) at (-1.05,1.1125) {\phantom{$(\lambda_i,\Delta_i^+)$}};
      \node (right-assignment) at (1.05,1.1125) {\phantom{$(\mu_i,\Delta_i^-)$}};
    \end{scope}

    % The full 3 by 2 matrix.
    % Offset the matrix right to balance the row labels on its left.
    \matrix (L) [
      matrix of math nodes,
      left delimiter=(,
      right delimiter=),
      text=black,
      nodes={minimum width=1.025cm,minimum height=0.525cm,anchor=center},
      column sep=0.15cm,
      row sep=0cm
    ] at (9.2875,2.6875) {
      |[text=black]|\lambda_1 & |[text=black]|-\mu_1 \\
      |[text=black]|\lambda_2 & |[text=black]|-\mu_2 \\
      |[text=black]|\lambda_3 & |[text=black]|-\mu_3 \\
    };
    \node[text=black] (matrix-column-one) at ($(L-1-1.north)+(0,0.475)$) {$\widehat X_1$};
    \node[text=black] at ($(L-1-2.north)+(0,0.475)$) {$\widehat X_2$};
    % Align (a) and (c) to the unchanged column-label baseline in (b).
    \coordinate (graph-data-label-row) at (matrix-column-one.base);
    \node[text=black,anchor=base east]
      at (left-assignment.east |- graph-data-label-row)
      {$\widehat X_1:\;(\lambda_i,\Delta_i^+)$};
    \node[text=black,anchor=base west]
      at (right-assignment.west |- graph-data-label-row)
      {$\widehat X_2:\;(\mu_i,\Delta_i^-)$};
    \node[text=black,anchor=east]
      at ($(L-1-1.west)+(-0.4375,0)$) {$1$};
    % Keep the row number fixed with twice the usual spacing on both sides of =.
    \node[text=black,anchor=east]
      at ($(L-2-1.west)+(-0.4375,0)$) {$\mathllap{L\;\;{=}\;\;}2$};
    \node[text=black,anchor=east]
      at ($(L-3-1.west)+(-0.4375,0)$) {$3$};

    % Both vertices, with mirrored arcs and matching label clearances.
    \begin{scope}[shift={(14.7875,2.6875)},local bounding box=degree-data]
    \coordinate (vx) at (-1.65,0);
    \coordinate (vy) at (1.65,0);
    % Match (a)'s radius while preserving the arc span and equal edge lengths.
    \pgfmathsetmacro{\edgecutlength}{0.625*0.76/1.1375}
    \pgfmathsetmacro{\edgecutangle}{\edgecutlength/0.76*180/pi}
    \coordinate (p2) at ($(vx)+(\edgecutlength,0)$);
    \draw[edge] (vx)
      arc[start angle=180,end angle={180-\edgecutangle},radius=0.76]
      coordinate (p1);
    \draw[highlighted edge] (vx)--(p2);
    \draw[edge] (vx)
      arc[start angle=180,end angle={180+\edgecutangle},radius=0.76]
      coordinate (p3);
    \node[highlighted vertex] at (vx) {};
    \node[text=black,left=2.25pt] at (vx) {$\widehat X_1$};
    \node[text=black,anchor=south west,inner sep=2pt] at (p1) {$\Delta_1^+$};
    \node[text=black,anchor=west,inner sep=2pt] at (p2) {$\Delta_2^+$};
    \node[text=black,anchor=north west,inner sep=2pt] at (p3) {$\Delta_3^+$};
    \coordinate (q2) at ($(vy)+(-\edgecutlength,0)$);
    \draw[edge] (vy)
      arc[start angle=0,end angle=\edgecutangle,radius=0.76]
      coordinate (q1);
    \draw[highlighted edge] (vy)--(q2);
    \draw[edge] (vy)
      arc[start angle=0,end angle=-\edgecutangle,radius=0.76]
      coordinate (q3);
    \node[highlighted vertex] at (vy) {};
    \node[text=black,right=2.25pt] at (vy) {$\widehat X_2$};
    \node[text=black,anchor=south east,inner sep=2pt] at (q1) {$\Delta_1^-$};
    \node[text=black,anchor=east,inner sep=2pt] at (q2) {$\Delta_2^-$};
    \node[text=black,anchor=north east,inner sep=2pt] at (q3) {$\Delta_3^-$};
    \end{scope}
    % Center the expression's visible ink on the complete labeled vertex diagram.
    \node[text=black,anchor=base,xshift=0.5pt] at (degree-data.center |- graph-data-label-row)
      {$B_1=\sum_i\Delta_i^+,\qquad B_2=\sum_i\Delta_i^-$};
  \end{tikzpicture}
  \caption{Three-melon graph data. Panel (a) shows the chosen edge orientations (\(\widehat X_{\mathsf a(i)}=\widehat X_1\) and \(\widehat X_{\mathsf b(i)}=\widehat X_2\) for \(i=1,2,3\)) and the corresponding endpoint assignments; (b) shows the weighted incidence matrix \(L\), which determines the graph polynomial; and (c) shows the weighted degrees \(B_1\) and \(B_2\), which determine the vertex weights in the normalization.}
  \label{fig:graph-data}
\end{figure}

Associate the pairs \((\lambda_i,\Delta_i^+)\) and \((\mu_i,\Delta_i^-)\) with the endpoints \(\mathsf a(i)\) and \(\mathsf b(i)\), respectively (for the three-melon example, see Figure~\ref{fig:graph-data}(a)). The connectivity of the graph is encoded in an \(E\times V\) weighted incidence matrix \(L\) (see Figure~\ref{fig:graph-data}(b)) defined by
\begin{equation}
	L_{iv} = \lambda_i\delta_{v,\mathsf a(i)} - \mu_i\delta_{v,\mathsf b(i)} ,
\end{equation}
where \(i\) labels edges and \(v\) labels vertices. Row \(i\) records the endpoints of edge \(i\) through the contributions \(\lambda_i\) at \(\mathsf a(i)\) and \(-\mu_i\) at \(\mathsf b(i)\). Any remaining entries are zero.

The weighted incidence matrix defines a polynomial associated with the graph \(g\) through
\begin{equation}
	A_g(\boldsymbol{\lambda},\boldsymbol{\mu}) = \det\!\left(I_V+L^TL\right),
	\label{eq:lattice-incidence-polynomial}
\end{equation}
where \(I_V\) is the \(V\times V\) identity matrix. Reversing an edge exchanges \(\lambda_i\) and \(\mu_i\) and changes the sign of row \(i\) of \(L\), leaving the graph polynomial unchanged.

These graph data determine the vacuum graph integral in Schwinger-parameter form
\begin{equation}
	I_g(D;\boldsymbol{\Delta}^{\pm}) = N_g(D;\boldsymbol{\Delta}^{\pm}) J_g(D;\boldsymbol{\Delta}^{\pm}),
	\label{eq:lattice-I-NJ}
\end{equation}
\begin{equation}
	\text{where}\quad J_g(D;\boldsymbol{\Delta}^{\pm}) = \int_0^\infty\frac{dz}{z}\, z^{(D+1)/2} \prod_{i=1}^{E} \frac{d\lambda_i}{\lambda_i} \frac{d\mu_i}{\mu_i}\, \lambda_i^{\Delta_i^+} \mu_i^{\Delta_i^-} e^{-zA_g(\boldsymbol{\lambda},\boldsymbol{\mu})},
	\label{eq:lattice-Euler-integral}
\end{equation}
\begin{equation}
	\text{and}\quad\hspace{7.62pt} N_g(D;\boldsymbol{\Delta}^{\pm}) = \frac{ 2^{V-E}\pi^{\frac{D+1}{2}(V-E)} }{ \Gamma\!\left(\frac{D+1}{2}\right) \prod_{v=1}^{V}\Gamma(w_v) } \prod_{i=1}^{E} \frac{ 4^{\frac{\Delta_i^++\Delta_i^--D+1}{2}} }{ \Gamma\!\left(\frac{D+1-\Delta_i^+-\Delta_i^-}{2}\right) }.\hspace{33.72pt}
	\label{eq:lattice-normalization}
\end{equation}
The vertex weights \(w_v\) are defined in terms of the weighted degrees \(B_v\) (see Figure~\ref{fig:graph-data}(c)) by
\begin{equation}
	B_v = \sum_{\mathsf a(i)=v}\Delta_i^+ + \sum_{\mathsf b(i)=v}\Delta_i^- \qquad\implies\qquad w_v = \frac{D+1-B_v}{2}.
	\label{eq:lattice-vertex-weight}
\end{equation}

\paragraph{Monomial insertions and the graph lattice.} The total-derivative identities and edge-contraction relations both generate monomial insertions in the integral \(J_g\) \eqref{eq:lattice-Euler-integral}. A general insertion has the form
\begin{equation}
	\mathcal M = z^{p_z}\prod_{i=1}^{E}\lambda_i^{p_i}\mu_i^{q_i}, \qquad p_z,p_i,q_i\in\mathbb Z.
	\label{eq:lattice-monomial-insertion}
\end{equation}
Let \(\boldsymbol{\beta}:=((D+1)/2,\boldsymbol{\Delta}^{+},\boldsymbol{\Delta}^{-})\) denote the exponent vector of \(J_g\) \eqref{eq:lattice-Euler-integral}; multiplication by \(\mathcal M\) shifts the parameters in \(\boldsymbol{\beta}\) as
\begin{equation}
	D\longmapsto D+2p_z, \qquad \boldsymbol{\Delta}^{+}\longmapsto\boldsymbol{\Delta}^{+}+\boldsymbol p, \qquad\text{and}\qquad \boldsymbol{\Delta}^{-}\longmapsto\boldsymbol{\Delta}^{-}+\boldsymbol q.
	\label{eq:lattice-exponent-shifts}
\end{equation}
The propagators in a scalar vacuum graph carry Klein--Gordon mass squared \(M_i^2=\Delta_i^+\Delta_i^-\), with weights satisfying \(\Delta_i^++\Delta_i^-=D-1\). The shifted parameters violate this relation unless \(p_i+q_i=2p_z\) for every edge. General monomial insertions therefore require a larger family of graph integrals, with edge kernels that need not be physical propagators.

To describe this larger family, we let the weights \(\Delta_i^\pm\) vary independently. We then parametrize the dimension and weights as
\begin{equation}
	D=2t+1, \qquad \Delta_i^\pm=u_i\pm\rho_i, \qquad \eta_i=t+1-u_i, \qquad \text{with}\quad \boldsymbol{x}=(t,\boldsymbol{u},\boldsymbol{\rho}).
	\label{eq:lattice-line-parameters}
\end{equation}
Shifts of \(D\) by two then become unit shifts of \(t\). The parameters in \eqref{eq:lattice-line-parameters} satisfy \(\Delta_i^++\Delta_i^-+2\eta_i=D+1\), with the mismatch \(p_i+q_i-2p_z\) under arbitrary shifts absorbed by \(\eta_i\mapsto\eta_i+p_z-\tfrac12(p_i+q_i)\).

With \(\eta_i\), \(B_v\), and \(w_v\) understood as functions of \(\boldsymbol{x}\) through \eqref{eq:lattice-vertex-weight} and \eqref{eq:lattice-line-parameters}, we write \eqref{eq:lattice-I-NJ}--\eqref{eq:lattice-normalization} as
\begin{equation}
	I_g(\boldsymbol{x}) = N_g(\boldsymbol{x})J_g(\boldsymbol{x}),
	\label{eq:lattice-I-NJ-x}
\end{equation}
\begin{equation}
	\text{where}\quad J_g(\boldsymbol{x}) = \int_0^\infty\frac{dz}{z}\, z^{t+1} \prod_{i=1}^{E} \frac{d\lambda_i}{\lambda_i} \frac{d\mu_i}{\mu_i}\, \lambda_i^{u_i+\rho_i}\mu_i^{u_i-\rho_i} e^{-zA_g},
	\label{eq:lattice-Euler-integral-x}
\end{equation}
\begin{equation}
	\text{and}\quad N_g(\boldsymbol{x}) = \frac{ 2^{E+V-2\sum_i\eta_i(\boldsymbol{x})} \pi^{(t+1)(V-E)} }{ \Gamma(t+1) \prod_{i=1}^{E}\Gamma\!\bigl(\eta_i(\boldsymbol{x})\bigr) \prod_{v=1}^{V}\Gamma\!\bigl(w_v(\boldsymbol{x})\bigr) }.
	\label{eq:lattice-normalization-x}
\end{equation}

For a fixed graph \(g\), the parameter values \(\boldsymbol{x}\) related by monomial shifts form its \emph{graph lattice}. Each node \(\boldsymbol{x}\) represents the integral \(I_g(\boldsymbol{x})\) defined by \eqref{eq:lattice-I-NJ-x}--\eqref{eq:lattice-normalization-x}. %Moving between nodes changes the parameters while leaving the graph unchanged.

For the physical graph, \(\eta_i=1\), hence \(u_i=t\) for every edge. Denoting the common value of \(t\) and all \(u_i\) by \(s\), the physical point is
\begin{equation}
	\bar{\boldsymbol{x}} := (s,s\boldsymbol{1}_E,\boldsymbol{\rho}),
	\label{eq:lattice-physical-point}
\end{equation}
where \(\boldsymbol{1}_E\) is the \(E\)-component vector of ones. At this point, \(D=2s+1\), \(\Delta_i^\pm=s\pm\rho_i\), and \(M_i^2=s^2-\rho_i^2\). We distinguish the physical vacuum integral by writing \(I_g(s;\boldsymbol{\rho}):=I_g(\bar{\boldsymbol{x}})\).

Starting from a physical point \(\bar{\boldsymbol{x}}\), the insertion \(\mathcal M\) shifts the parameters according to
\begin{equation}
	\bar{\boldsymbol{x}} \longmapsto \bar{\boldsymbol{x}}+\boldsymbol{\delta}(\mathcal M), \qquad \boldsymbol{\delta}(\mathcal M) = \left( p_z, \frac{\boldsymbol{p}+\boldsymbol{q}}{2}, \frac{\boldsymbol{p}-\boldsymbol{q}}{2} \right).
	\label{eq:lattice-shift-map}
\end{equation}
Writing \(J_g[\mathcal M;\bar{\boldsymbol{x}}]\) for the inserted integral, we have \(J_g[\mathcal M;\bar{\boldsymbol{x}}]=J_g(\bar{\boldsymbol{x}}+\boldsymbol{\delta}(\mathcal M))\). 

To express the insertion in terms of \(I_g\) at the shifted lattice point, define the normalization ratio \(\mathcal N_g(\bar{\boldsymbol{x}};\boldsymbol{\delta}):=N_g(\bar{\boldsymbol{x}})/N_g(\bar{\boldsymbol{x}}+\boldsymbol{\delta})\). Then \(I_g=N_gJ_g\) gives
\begin{equation}
	N_g(\bar{\boldsymbol{x}})J_g[\mathcal M;\bar{\boldsymbol{x}}] = \mathcal N_g(\bar{\boldsymbol{x}};\boldsymbol{\delta}(\mathcal M)) I_g\!\left( \bar{\boldsymbol{x}}+\boldsymbol{\delta}(\mathcal M) \right).
	\label{eq:lattice-shifted-I}
\end{equation}
Monomial insertions therefore act as lattice shifts, with \(\mathcal N_g\) accounting for the change in normalization. For a polynomial insertion, we apply \eqref{eq:lattice-shifted-I} term by term to obtain a finite linear combination of graph integrals at shifted lattice points.

\paragraph{Physical interpretation of the graph lattice.} Each edge of a physical scalar vacuum graph carries a sphere propagator with weights satisfying \(\Delta^++\Delta^-=D-1\). Appendix~\ref{app:embedding-propagator} derives the physical scalar propagator on \(S^D\) as a radial Mellin transform of a massless propagator in the ambient space \(\mathbb R^{D+1}\), whose momentum dependence is \(1/P^2\). For \(\widehat X,\widehat Y\in S^D\), it is given by
\begin{equation}
	G_{\Delta^\pm}(\widehat X,\widehat Y) = \frac{1}{\pi^{D+1}\operatorname{Vol}(\mathbb R_+)} \int_0^\infty\frac{d\lambda}{\lambda} \int_0^\infty\frac{d\mu}{\mu}\, \lambda^{\Delta^+}\mu^{\Delta^-} \int_{\mathbb R^{D+1}}d^{D+1}P\, \frac{e^{-2\mathrm{i}P\cdot(\lambda\widehat X-\mu\widehat Y)}}{4P^2}.
	\label{eq:lattice-interpretation-kernel}
\end{equation}
This representation has the scale redundancy
\begin{equation}
	(\lambda,\mu,P)\mapsto(\mathscr q\lambda,\mathscr q\mu,\mathscr q^{-1}P).
	\label{eq:lattice-interpretation-scale-redundancy}
\end{equation}
Under this rescaling, the phase and Mellin measures are unchanged, while the radial weights, momentum measure, and \(1/P^2\) contribute \(\mathscr q^{\Delta^++\Delta^-}\mathscr q^{-(D+1)}\mathscr q^2=1\). The factor \(\operatorname{Vol}(\mathbb R_+)=\int_0^\infty d\mathscr q/\mathscr q\) divides out the common radial scale.

We briefly verify that this kernel satisfies the sphere Green-function equation
\begin{equation}
	\bigl(-\nabla_{S^D}^2+M^2\bigr)G_{\Delta^\pm}(\widehat X,\widehat Y)=\delta_{S^D}(\widehat X,\widehat Y),
	\label{eq:lattice-interpretation-green}
\end{equation}
with \(M^2=\Delta^+\Delta^-\). For \(X=\mathscr r\widehat X\) in \eqref{eq:lattice-interpretation-kernel}, we assign \(\Delta^+\) to the first endpoint and \(\Delta^-\) to the second. Homogeneity gives \(G_{\Delta^\pm}(X,\widehat Y)=\mathscr r^{-\Delta^+}G_{\Delta^\pm}(\widehat X,\widehat Y)\), so the radial decomposition of the ambient Laplacian yields
\begin{equation}
	-\partial_X^2G_{\Delta^\pm}(X,\widehat Y) = \mathscr r^{-\Delta^+-2} \left[-\nabla_{S^D}^2+\Delta^+\Delta^-\right] G_{\Delta^\pm}(\widehat X,\widehat Y).
	\label{eq:lattice-interpretation-radial-laplacian}
\end{equation}
Applying the same operator to \eqref{eq:lattice-interpretation-kernel}, with \(X\) in place of \(\widehat X\), cancels \(4P^2\); writing the resulting delta function in polar coordinates, we obtain
\begin{equation}
	\begin{aligned}
		-\partial_X^2G_{\Delta^\pm}(X,\widehat Y) &= \frac{1}{\operatorname{Vol}(\mathbb R_+)} \int_0^\infty\frac{d\lambda}{\lambda} \int_0^\infty\frac{d\mu}{\mu}\, \lambda^{\Delta^++2}\mu^{\Delta^-} \delta^{(D+1)}(\lambda\mathscr r\widehat X-\mu\widehat Y) \\
		&= \frac{\delta_{S^D}(\widehat X,\widehat Y)} {\operatorname{Vol}(\mathbb R_+)} \int_0^\infty\frac{d\lambda}{\lambda} \int_0^\infty\frac{d\mu}{\mu}\, \lambda^{\Delta^++2}\mu^{\Delta^--D} \delta(\lambda\mathscr r-\mu) = \mathscr r^{-\Delta^+-2}\delta_{S^D}(\widehat X,\widehat Y).
	\end{aligned}
	\label{eq:lattice-interpretation-contact}
\end{equation}
Comparing the two expressions establishes \eqref{eq:lattice-interpretation-green}.

Replacing \((4P^2)^{-1}\) by \((4P^2)^{-\eta}\) defines the generalized edge kernel (see Appendix~\ref{app:embedding-propagator})
\begin{equation}
	G_{\eta,\Delta^+}(\widehat X,\widehat Y)=\frac{1}{4^\eta\pi^{D+1}\operatorname{Vol}(\mathbb R_+)}\int_0^\infty\frac{d\lambda}{\lambda}\int_0^\infty\frac{d\mu}{\mu}\,\lambda^{\Delta^+}\mu^{\Delta^-}\int_{\mathbb R^{D+1}}d^{D+1}P\,\frac{e^{-2\mathrm{i}P\cdot(\lambda\widehat X-\mu\widehat Y)}}{(P^2)^\eta}.
	\label{eq:lattice-interpretation-general-kernel}
\end{equation}
Scale invariance requires \(\Delta^++\Delta^-+2\eta=D+1\). This identifies the parameter \(\eta\) in the lattice parametrization \eqref{eq:lattice-line-parameters} with that of the generalized edge kernel on \(S^D\). An analogous calculation gives the descent equation
\begin{equation}
	\left[-\nabla_{S^D}^2+\Delta^+(D-1-\Delta^+)\right] G_{\eta,\Delta^+}(\widehat X,\widehat Y) = G_{\eta-1,\Delta^++2}(\widehat X,\widehat Y).
	\label{eq:lattice-interpretation-descent}
\end{equation}
The physical propagator occurs at \(\eta=1\), while \(\eta=0\) gives the sphere delta function.

At each lattice point \(\boldsymbol{x}\) in \eqref{eq:lattice-line-parameters}, \(I_g(\boldsymbol{x})\) is the integral of the graph \(g\) with corresponding dimension and generalized edge kernels. Monomial insertions shift \(\boldsymbol{x}\), moving the integral across the nodes of the same lattice. Total-derivative identities give intra-lattice relations among these integrals; and when a shifted kernel becomes a sphere delta function, the corresponding edge contracts, giving an inter-lattice relation to the lattice of the contracted graph.

%Each node \(I_g(\boldsymbol{x})\) of the graph lattice therefore represents an integral of the same graph \(g\) with a particular choice of generalized edge kernels. Monomial insertions move between nodes by shifting the kernel parameters. Total-derivative identities relate nodes within this lattice, giving intra-lattice relations. When an insertion replaces an edge kernel by a sphere delta function, it contracts that edge, connecting this lattice to that of the contracted graph and giving inter-lattice relations. At any loop order, successive contractions organize these lattices into a hierarchy, with solved graphs supplying the sources for graphs that contract to them.

\subsection{Total-derivative identities}

Assuming vanishing boundary terms, integration by parts in \eqref{eq:lattice-Euler-integral-x} with \(z\partial_z\), \(\lambda_i\partial_{\lambda_i}\), and \(\mu_i\partial_{\mu_i}\) gives
\begin{align}
	J_g[zA_g;\boldsymbol{x}] &= (t+1)J_g(\boldsymbol{x}),
	\label{eq:bulk-z-equation} \\
	J_g\left[ z\lambda_i\frac{\partial A_g}{\partial\lambda_i}; \boldsymbol{x} \right] &= (u_i+\rho_i)J_g(\boldsymbol{x}),
	\label{eq:bulk-lambda-equation} \\
	J_g\left[ z\mu_i\frac{\partial A_g}{\partial\mu_i}; \boldsymbol{x} \right] &= (u_i-\rho_i)J_g(\boldsymbol{x}),
	\label{eq:bulk-Euler-equations}
\end{align}
where the last two identities hold for every edge \(i\).

To express these identities as relations among lattice nodes, write the monomial expansion of \(A_g\) and the shift in \(\boldsymbol{x}\) associated with each term in \(zA_g\) as
\begin{equation}
	A_g = \sum_{m\in\mathcal A_g} c_m \boldsymbol{\lambda}^{\boldsymbol{p}_m} \boldsymbol{\mu}^{\boldsymbol{q}_m}, \qquad \boldsymbol{\delta}_m = \boldsymbol{\delta}\!\left( z\,\boldsymbol{\lambda}^{\boldsymbol{p}_m} \boldsymbol{\mu}^{\boldsymbol{q}_m} \right) = \left( 1, \frac{\boldsymbol{p}_m+\boldsymbol{q}_m}{2}, \frac{\boldsymbol{p}_m-\boldsymbol{q}_m}{2} \right).
	\label{eq:bulk-polynomial-expansion}
\end{equation}
Here \(\mathcal A_g\) indexes the monomials of \(A_g\), and the multi-index notation means
\begin{equation}
	\boldsymbol{\lambda}^{\boldsymbol{p}_m} = \prod_{i=1}^{E}\lambda_i^{p_{mi}}, \qquad \boldsymbol{\mu}^{\boldsymbol{q}_m} = \prod_{i=1}^{E}\mu_i^{q_{mi}}.
	\label{eq:bulk-multi-index-notation}
\end{equation}

The insertion rule \eqref{eq:lattice-shifted-I} applies at any base point \(\boldsymbol{x}\), with \(\mathcal N_g(\boldsymbol{x};\boldsymbol{\delta})=N_g(\boldsymbol{x})/N_g(\boldsymbol{x}+\boldsymbol{\delta})\). Substituting \eqref{eq:bulk-polynomial-expansion} into \eqref{eq:bulk-z-equation}--\eqref{eq:bulk-Euler-equations} and restoring the normalization gives
\begin{equation}
	\begin{aligned}
		\sum_{m\in\mathcal A_g} c_m \mathcal N_g(\boldsymbol{x};\boldsymbol{\delta}_m) I_g(\boldsymbol{x}+\boldsymbol{\delta}_m) &= (t+1)I_g(\boldsymbol{x}), \\
		\sum_{m\in\mathcal A_g} p_{mi}c_m \mathcal N_g(\boldsymbol{x};\boldsymbol{\delta}_m) I_g(\boldsymbol{x}+\boldsymbol{\delta}_m) &= (u_i+\rho_i)I_g(\boldsymbol{x}), \\
		\sum_{m\in\mathcal A_g} q_{mi}c_m \mathcal N_g(\boldsymbol{x};\boldsymbol{\delta}_m) I_g(\boldsymbol{x}+\boldsymbol{\delta}_m) &= (u_i-\rho_i)I_g(\boldsymbol{x}).
	\end{aligned}
	\label{eq:bulk-I-equations}
\end{equation}
At each base point \(\boldsymbol{x}\), these give \(2E+1\) linear relations within the lattice of \(g\).

\subsection{Edge-contraction relations}

Let \(i:a\to b\) be a physical edge, \(\eta_i=1\), with distinct endpoints; the remaining edges may be general. Replacing its propagator by the sphere delta function identifies \(a\) and \(b\), giving the contracted graph \(g/i\).\footnote{Edge contraction \(g/i\) removes edge \(i\) and identifies its endpoints. Edge deletion \(g\setminus i\) removes only the edge, leaving all vertices intact. In the graph integral, deletion is implemented by the zero-mode cut discussed in Appendix~\ref{app:spectral-cuts}. Other spectral cuts are also discussed there.} Its integral is \(I_{g/i}(\pi_i\boldsymbol{x})\), where \(\pi_i\) removes \(u_i,\rho_i\) and leaves the other parameters unchanged.

Appendix~\ref{app:contact-contractions} constructs an insertion that makes this replacement and evaluates it in two ways: as a delta-function edge and as a polynomial insertion in the Schwinger-parameter integral. The second evaluation introduces the contact polynomial
\begin{equation}
	Q_i = \det\!\left( I_V+L_{\,\widehat i}^{T}L_{\,\widehat i} \right),
	\label{eq:contact-Q-definition}
\end{equation}
where \(L_{\,\widehat i}\) is obtained from \(L\) by removing the row for edge \(i\) while retaining all \(V\) vertex columns.

Equating the two evaluations gives
\begin{equation}
	4 \frac{ \Gamma(w_a)\Gamma(w_b) }{ \Gamma(w_a-\xi)\Gamma(w_b+\xi-1) } N_g(\boldsymbol{x}) J_g\!\left[ z\lambda_i^{2\xi}\mu_i^{2(1-\xi)}Q_i; \boldsymbol{x} \right] = I_{g/i}(\pi_i\boldsymbol{x}).
	\label{eq:contact-fixed-equations}
\end{equation}
The identity holds for general \(\xi\). For the lattice relations, we use \(\xi=1,0,\tfrac12\), giving the polynomial insertions \(z\lambda_i^2Q_i\), \(z\mu_i^2Q_i\), and \(z\lambda_i\mu_iQ_i\), respectively.

To express \eqref{eq:contact-fixed-equations} in terms of parent-graph nodes, expand
\begin{equation}
	Q_i = \sum_{m\in\mathcal Q_i} d_{im}\, \boldsymbol{\lambda}^{\boldsymbol p_{im}} \boldsymbol{\mu}^{\boldsymbol q_{im}}.
	\label{eq:contact-Q-expansion}
\end{equation}
Here \(\mathcal Q_i\) indexes the monomials of \(Q_i\), with exponent vectors \(\boldsymbol p_{im}\) and \(\boldsymbol q_{im}\). Including the prefactor \(z\lambda_i^{2\xi}\mu_i^{2(1-\xi)}\), each monomial produces the shift
\begin{equation}
	\boldsymbol{\delta}_{im}(\xi) = \boldsymbol{\delta}\!\left( z\lambda_i^{2\xi}\mu_i^{2(1-\xi)} \boldsymbol{\lambda}^{\boldsymbol p_{im}} \boldsymbol{\mu}^{\boldsymbol q_{im}} \right) = \left( 1,\, \boldsymbol e_i+ \frac{\boldsymbol p_{im}+\boldsymbol q_{im}}{2},\, (2\xi-1)\boldsymbol e_i+ \frac{\boldsymbol p_{im}-\boldsymbol q_{im}}{2} \right),
	\label{eq:contact-shift}
\end{equation}
where \(\boldsymbol e_i\) is the \(i\)-th unit vector. Applying \eqref{eq:lattice-shifted-I} term by term gives the edge-contraction relation
\begin{equation}
	4 \frac{ \Gamma(w_a)\Gamma(w_b) }{ \Gamma(w_a-\xi)\Gamma(w_b+\xi-1) } \sum_{m\in\mathcal Q_i} d_{im}\, \mathcal N_g\!\left( \boldsymbol{x}; \boldsymbol{\delta}_{im}(\xi) \right) I_g\!\left( \boldsymbol{x}+\boldsymbol{\delta}_{im}(\xi) \right) = I_{g/i}(\pi_i\boldsymbol{x}).
	\label{eq:contact-I-equations}
\end{equation}
Each term on the left is an integral of the original graph; the full sum gives the contracted graph integral. Together, \eqref{eq:bulk-I-equations} and \eqref{eq:contact-I-equations} couple the lattice of \(g\) to those of its one-edge contractions.

The construction also extends to simultaneous contraction of a forest, a possibly disconnected union of trees, but single-edge contractions suffice through two loops.

\subsection{Coupled lattice equations}
\label{sec:coupled-system}

The total-derivative identities \eqref{eq:bulk-I-equations} and edge-contraction relations \eqref{eq:contact-I-equations} involve both physical graph integrals and auxiliary integrals with general edge kernels. We combine these equations to eliminate the auxiliary integrals and obtain relations among physical graph integrals. Table~\ref{tab:coupled-system-rows} collects the equations entering the system.

\begin{table}[ht]
\centering
\small
\renewcommand{\arraystretch}{1.35}
\begin{tabular*}{\textwidth}{
  @{\extracolsep{\fill}}
  p{0.24\textwidth}
  p{0.70\textwidth}
  @{}
}
\toprule
Row & Relation\\
\midrule
\(z\partial_z\)
&
\(\displaystyle
  \sum_{m\in\mathcal A_g} c_m
  \mathcal N_g(\boldsymbol{x};\boldsymbol{\delta}_m)
  I_g(\boldsymbol{x}+\boldsymbol{\delta}_m)
  =(t+1)I_g(\boldsymbol{x})
\)\\
\(\lambda_i\partial_{\lambda_i}\)
&
\(\displaystyle
  \sum_{m\in\mathcal A_g} p_{mi}c_m
  \mathcal N_g(\boldsymbol{x};\boldsymbol{\delta}_m)
  I_g(\boldsymbol{x}+\boldsymbol{\delta}_m)
  =(u_i+\rho_i)I_g(\boldsymbol{x})
\)\\
\(\mu_i\partial_{\mu_i}\)
&
\(\displaystyle
  \sum_{m\in\mathcal A_g} q_{mi}c_m
  \mathcal N_g(\boldsymbol{x};\boldsymbol{\delta}_m)
  I_g(\boldsymbol{x}+\boldsymbol{\delta}_m)
  =(u_i-\rho_i)I_g(\boldsymbol{x})
\)\\
\mbox{Edge contraction \((i,\xi)\)}
&
\(\displaystyle
  4
  \frac{
    \Gamma(w_a)\Gamma(w_b)
  }{
    \Gamma(w_a-\xi)\Gamma(w_b+\xi-1)
  }
  \sum_{m\in\mathcal Q_i} d_{im}
  \mathcal N_g(\boldsymbol{x};\boldsymbol{\delta}_{im}(\xi))
  I_g(\boldsymbol{x}+\boldsymbol{\delta}_{im}(\xi))
  =I_{g/i}(\pi_i\boldsymbol{x})
\)\\
\bottomrule
\end{tabular*}
\caption{Total-derivative identities and edge-contraction relations. The edge-contraction equation applies when edge \(i\) is physical, \(u_i=t\), and has distinct endpoints.}
\label{tab:coupled-system-rows}
\end{table}
Choose a finite collection of these equations at selected base points. The parent-graph nodes appearing in these equations form a finite stencil. Let \(\boldsymbol I_g\) collect their integral values, and let \(\boldsymbol I_{\partial g}\) collect the contracted-graph integrals on the right-hand sides. Moving all parent-graph terms to the left gives
\begin{equation}
	\mathsf E_g\boldsymbol{I}_g=0, \qquad \mathsf C_g\boldsymbol{I}_g = \mathsf B_g\boldsymbol{I}_{\partial g}.
	\label{eq:coupled-graph-system}
\end{equation}
Here \(\mathsf E_g\) and \(\mathsf C_g\) contain the coefficients of the total-derivative and edge-contraction equations, respectively, while \(\mathsf B_g\) records which contracted integral appears in each edge-contraction equation. 

Equivalently, moving the source terms to the left gives
\begin{equation}
	\begin{pmatrix}
		\mathsf E_g & 0\\
		\mathsf C_g & -\mathsf B_g
	\end{pmatrix} \begin{pmatrix}
		\boldsymbol I_g\\
		\boldsymbol I_{\partial g}
	\end{pmatrix} =0.
	\label{eq:coupled-graph-system-matrix}
\end{equation}
We call the physical parent-graph integrals and contracted-graph integrals to be retained in a relation the targets, and the remaining parent-graph integrals auxiliary. We place the auxiliary columns first and perform Gaussian elimination. Rows in which all auxiliary coefficients vanish give relations among physical parent-graph integrals and the contracted sources. If the desired relation does not appear, we add equations at shifted base points and enlarge the stencil to include every new integral they introduce. We repeat this step until elimination yields the desired relation. The resulting relations hold as meromorphic identities in the parameters.

For trees and one-loop cycles, equations based at the physical point \(\bar{\boldsymbol x}\) suffice, even though these equations contain nonphysical shifted integrals. In both cases, the auxiliary integrals can be eliminated by hand.

The two-loop three-melon \(\texttt{V\{3\}V}\) requires additional total-derivative identities at shifted base points. We choose physical targets related by dimensional or spectral shifts, then eliminate the auxiliary integrals to find recurrences among them. We use these recurrences, together with initial data and analytic conditions, to construct its meromorphic solution and perform dimensional regularization.

\section{Trees}
\label{sec:trees}

Let \(\mathcal T_E\) be a connected tree with \(E\) edges and \(E+1\) integrated vertices. We derive a leaf recursion by combining a total-derivative identity with an edge-contraction relation at the physical point \(\bar{\boldsymbol{x}}\) \eqref{eq:lattice-physical-point}. Writing \(L_E\) for the weighted incidence matrix, the graph polynomial is \(A_{\mathcal T_E}=\det(I_{E+1}+L_E^TL_E)\).
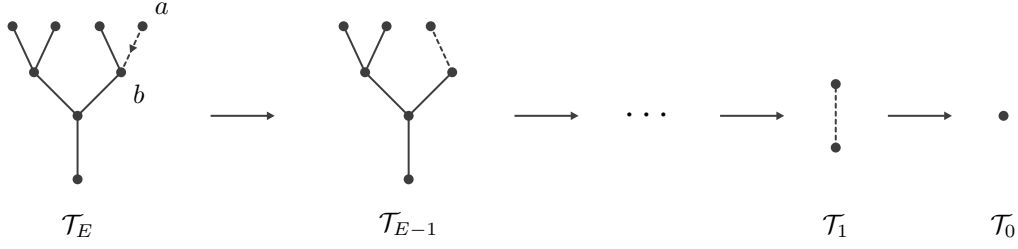
\begin{figure}[ht]
\centering
\renewcommand{\GraphLineWidth}{0.75pt}
\renewcommand{\GraphNodeDiameter}{3.75pt}
\begin{tikzpicture}[
  removed vertex/.style={contracted vertex},
  removed edge/.style={contracted edge},
  flow/.style={graph arrow},
  annotation/.style={graph arrow},
  every node/.style={font=\small}
]
  % Edge lengths in cm; the bottom edge is 0.8 times its original 1.05 cm.
  \def\TreeEdgeLength{0.84}
  \pgfmathsetmacro{\TreeBranchLength}{0.97*\TreeEdgeLength}
  \pgfmathsetmacro{\TreeTwigLength}{0.8*\TreeEdgeLength}

  \begin{scope}[shift={(2.0,2.2)}]
    \coordinate (c)   at (0,0);
    \coordinate (ul)  at (135:\TreeBranchLength);
    \coordinate (ur)  at (45:\TreeBranchLength);
    \coordinate (ull) at ($(ul)+(115:\TreeTwigLength)$);
    \coordinate (ulr) at ($(ul)+(65:\TreeTwigLength)$);
    \coordinate (url) at ($(ur)+(115:\TreeTwigLength)$);
    \coordinate (urr) at ($(ur)+(65:\TreeTwigLength)$);
    \coordinate (d)   at (0,-\TreeEdgeLength);

    \draw[edge] (c)--(ul) (c)--(ur) (c)--(d);
    \draw[edge] (ul)--(ull) (ul)--(ulr);
    \draw[edge] (ur)--(url);
    \draw[removed edge,
      postaction={decorate},
      decoration={markings,mark=at position 0.5 with {
        \pgftransformxshift{1.6275pt}
        \arrow{>}
      }}
    ] (urr)--(ur);

    \foreach \p in {c,ul,ull,ulr,url,d}
      \node[vertex] at (\p) {};
    \foreach \p in {ur,urr}
      \node[removed vertex] at (\p) {};

    \node[text=black,above right=1pt] at (urr) {$a$};
    \node[text=black,below right=1pt] at (ur) {$b$};

    \node[text=black] at (0,-1.45) {$\mathcal T_E$};
  \end{scope}

  \draw[flow] (3.76,2.2) -- (4.61,2.2);

  \begin{scope}[shift={(6.38,2.2)}]
    \coordinate (c)   at (0,0);
    \coordinate (ul)  at (135:\TreeBranchLength);
    \coordinate (ur)  at (45:\TreeBranchLength);
    \coordinate (ull) at ($(ul)+(115:\TreeTwigLength)$);
    \coordinate (ulr) at ($(ul)+(65:\TreeTwigLength)$);
    \coordinate (url) at ($(ur)+(115:\TreeTwigLength)$);
    \coordinate (d)   at (0,-\TreeEdgeLength);

    \draw[edge] (c)--(ul) (c)--(ur) (c)--(d);
    \draw[edge] (ul)--(ull) (ul)--(ulr);
    \draw[removed edge] (ur)--(url);

    \foreach \p in {c,ul,ull,ulr,d}
      \node[vertex] at (\p) {};
    \foreach \p in {ur,url}
      \node[removed vertex] at (\p) {};

    \node[text=black] at (0,-1.45) {$\mathcal T_{E-1}$};
  \end{scope}

  \draw[flow] (7.78,2.2) -- (8.63,2.2);
  \node[text=black,font=\Large] at (9.57,2.2) {$\cdots$};
  \draw[flow] (10.50,2.2) -- (11.35,2.2);

  \coordinate (single-bottom) at (12.03,{2.2-0.5*\TreeEdgeLength});
  \coordinate (single-top) at (12.03,{2.2+0.5*\TreeEdgeLength});
  \draw[removed edge] (single-bottom) -- (single-top);
  \node[removed vertex] at (single-bottom) {};
  \node[removed vertex] at (single-top) {};
  \node[text=black] at (12.03,0.75) {$\mathcal T_1$};

  \draw[flow] (12.72,2.2) -- (13.57,2.2);
  \node[vertex] at (14.25,2.2) {};
  \node[text=black] at (14.25,0.75) {$\mathcal T_0$};
\end{tikzpicture}
\caption{Successive leaf contractions reduce \(\mathcal T_E\) to the single vertex \(\mathcal T_0\). At each step, the dashed edge is contracted. In the first graph, it joins the leaf vertex \(a\) to its neighbor \(b\).}
\label{fig:tree-leaf-peeling}
\end{figure}

Choose a leaf vertex \(a\), as in Figure~\ref{fig:tree-leaf-peeling}, and orient its unique incident edge \(i\) from \(a\) to \(b\). With \(\boldsymbol e_v\in\mathbb R^{E+1}\) the standard basis column vectors, the corresponding row of \(L_E\) is
\begin{equation}
	\boldsymbol\ell_i = \lambda_i\boldsymbol e_a^T-\mu_i\boldsymbol e_b^T.
\end{equation}
Remove row \(i\) from \(L_E\) to form \(L_{\,\widehat i}\). Since \(a\) is a leaf, its column in \(L_{\,\widehat i}\) is zero. The remaining columns form the incidence matrix \(L_{E-1}\) of the contracted tree \(\mathcal T_{E-1}:=\mathcal T_E/i\). 

Ordering the vertex columns of \(L_{\,\widehat i}\) with \(a\) first gives
\begin{equation}
	I_{E+1}+L_{\,\widehat i}^TL_{\,\widehat i} = \begin{pmatrix}
		1&0\\
		0&I_E+L_{E-1}^TL_{E-1}
	\end{pmatrix},
	\label{eq:tree-line-deleted-block}
\end{equation}
so the contact polynomial \eqref{eq:contact-Q-definition} reduces to
\begin{equation}
	Q_i = \det(I_{E+1}+L_{\,\widehat i}^TL_{\,\widehat i}) = \det(I_E+L_{E-1}^TL_{E-1}) = A_{\mathcal T_{E-1}}.
	\label{eq:tree-Q-is-smaller-polynomial}
\end{equation}
Restoring row \(i\) gives
\begin{equation}
	A_{\mathcal T_E} = \det\left( I_{E+1}+L_{\,\widehat i}^TL_{\,\widehat i} +\boldsymbol\ell_i^T\boldsymbol\ell_i \right).
	\label{eq:tree-restored-polynomial}
\end{equation}
We apply the matrix determinant lemma,
\begin{equation}
	\det(B+v^Tv) = \det(B) \left( 1+vB^{-1}v^T \right),
\end{equation}
to \eqref{eq:tree-restored-polynomial} with \(B=I_{E+1}+L_{\,\widehat i}^{T}L_{\,\widehat i}\) and \(v=\boldsymbol{\ell}_i\). Using the block form \eqref{eq:tree-line-deleted-block}, we obtain
\begin{equation}
	A_{\mathcal T_E} = Q_i(1+\lambda_i^2) + \mu_i^2R_i,
	\label{eq:tree-leaf-polynomial}
\end{equation}
where \(R_i\) depends only on the Euler variables of the remaining edges. 

Differentiating \eqref{eq:tree-leaf-polynomial} with respect to \(\lambda_i\), we get
\begin{equation}
	\lambda_i \frac{\partial A_{\mathcal T_E}}{\partial\lambda_i} = 2\lambda_i^2Q_i.
	\label{eq:tree-leaf-derivative}
\end{equation}
At \(\bar{\boldsymbol{x}}\), where \(\Delta_i^\pm=s\pm\rho_i\), substituting \eqref{eq:tree-leaf-derivative} into the total-derivative identity in \(\lambda_i\) \eqref{eq:bulk-lambda-equation} and restoring the normalization gives
\begin{equation}
	2N_{\mathcal T_E}(\bar{\boldsymbol{x}}) J_{\mathcal T_E}[z\lambda_i^2Q_i;\bar{\boldsymbol{x}}] = \Delta_i^+I_{\mathcal T_E}(\bar{\boldsymbol{x}}).
	\label{eq:tree-normalized-bulk-row}
\end{equation}
The same inserted integral appears in the edge-contraction relation \eqref{eq:contact-fixed-equations} with \(\xi=1\). At the leaf, \(B_a=\Delta_i^+\), so the vertex-weight formula \eqref{eq:lattice-vertex-weight} gives \(2(w_a-1)=D-1-\Delta_i^+=\Delta_i^-\). The contraction relation thus becomes
\begin{equation}
	2\Delta_i^- N_{\mathcal T_E}(\bar{\boldsymbol{x}}) J_{\mathcal T_E}[z\lambda_i^2Q_i;\bar{\boldsymbol{x}}] = I_{\mathcal T_{E-1}}(\pi_i\bar{\boldsymbol{x}}).
	\label{eq:tree-leaf-contact-row}
\end{equation}
Eliminating the common inserted integral between \eqref{eq:tree-normalized-bulk-row} and \eqref{eq:tree-leaf-contact-row} yields the leaf recursion
\begin{equation}
	I_{\mathcal T_E}(\bar{\boldsymbol{x}}) = \frac{1}{\Delta_i^+\Delta_i^-} I_{\mathcal T_{E-1}}(\pi_i\bar{\boldsymbol{x}}) = \frac{1}{s^2-\rho_i^2} I_{\mathcal T_{E-1}}(\pi_i\bar{\boldsymbol{x}}).
	\label{eq:tree-leaf-peeling-rule}
\end{equation}
If edge \(i\) is oriented toward the leaf, the total-derivative identity in \(\mu_i\) \eqref{eq:bulk-Euler-equations} and the \(\xi=0\) specialization of \eqref{eq:contact-fixed-equations} give the same recursion.

Repeated application of \eqref{eq:tree-leaf-peeling-rule} contracts every edge, leaving the one-vertex tree \(\mathcal T_0\), as shown in Figure~\ref{fig:tree-leaf-peeling}. The one-vertex tree has \(A_{\mathcal T_0}=1\). Using \eqref{eq:lattice-I-NJ-x}--\eqref{eq:lattice-normalization-x}, we write its integral as
\begin{equation}
	I_{\mathcal T_0}(s) = \frac{2\pi^{s+1}}{\Gamma(s+1)^2} \int_0^\infty \frac{dz}{z}\, z^{s+1}e^{-z}.
\end{equation}
The \(z\)-integral evaluates to \(\Gamma(s+1)\), leaving
\begin{equation}
	I_{\mathcal T_0}(s) = \frac{2\pi^{s+1}}{\Gamma(s+1)} = \Omega_{2s+1}.
	\label{eq:tree-one-vertex-value}
\end{equation}
The integral for any tree with \(E\) edges is therefore
\begin{equation}
	\boxed{\displaystyle I_{\mathcal T_E}(s,\boldsymbol{\rho}) = \Omega_{2s+1} \prod_{i=1}^{E} \frac{1}{s^2-\rho_i^2}. }
	\label{eq:general-tree-solution}
\end{equation}
The integral depends on the edge parameters but not on the shape of the tree or the order of leaf contractions. This result also follows directly from the spectral form \eqref{eq:propagator-spectral-form}. Integrating a leaf selects the zero mode,
\begin{equation}
	\begin{aligned}
		\int d\Omega_D(\widehat X_a)\,G_{\Delta_i^\pm}(\widehat X_a,\widehat X_b) &=\sum_{\ell=0}^{\infty}\sum_{\boldsymbol{m}} \frac{Y_{\ell\boldsymbol{m}}^{*}(\widehat X_b)}{(s+\ell)^2-\rho_i^2} \int d\Omega_D(\widehat X_a)\,Y_{\ell\boldsymbol{m}}(\widehat X_a) \\
		&=\sum_{\ell=0}^{\infty}\frac{\delta_{\ell0}}{(s+\ell)^2-\rho_i^2} =\frac{1}{s^2-\rho_i^2}.
	\end{aligned}
\end{equation}
Repeating this for all edges leaves one unconstrained vertex, giving \eqref{eq:general-tree-solution}.

The normalized zero-mode cut defined in Appendix~\ref{app:spectral-cuts} deletes an edge while retaining all vertices. Cutting every edge of the tree therefore leaves \(V=E+1\) unconstrained vertices. Applying this prescription to the tree result \eqref{eq:general-tree-solution}, we obtain
\begin{equation}
	\lim_{\boldsymbol{\rho}\to s\boldsymbol{1}_E} \left[\prod_{i=1}^{E}\Omega_{2s+1}(s^2-\rho_i^2)\right] I_{\mathcal T_E}(s;\boldsymbol{\rho}) = \Omega_{2s+1}^{E+1} = \Omega_{2s+1}^{V}.
\end{equation}

Finally, note that the derivation of the leaf recursion \eqref{eq:tree-leaf-peeling-rule} requires only that \(a\) be a leaf, so it also applies when the rest of the graph contains loops. Repeated leaf contractions remove any tree attached at a single vertex, leaving the loop core and a factor \(1/(s^2-\rho_i^2)\) for each removed edge.

\section{One loop}
\label{sec:one-loop}

A connected one-loop graph consists of a single cycle with possible trees attached to its vertices. Let the cycle have \(n\) edges with spectral parameters \(\rho_1,\ldots,\rho_n\), and denote the set of edges outside the cycle by \(\mathsf T(g)\).

For \(n=1\), the cycle is the self-loop petal, denoted by \(\texttt{V}^{\circ}\). For \(n\geq2\), we write \(\texttt{V1}\,\texttt{V}^{\,n-2}\texttt{V1}\). This gives \(\texttt{V1V1}\) for the two-edge cycle, \(\texttt{V1VV1}\) for the triangle, and so on (see Appendix~\ref{app:nomenclature} for nomenclature).

At the physical point \(\bar{\boldsymbol{x}}\), repeated application of the leaf recursion \eqref{eq:tree-leaf-peeling-rule} gives
\begin{equation}
	I_g(s;\boldsymbol{\rho}) = \left( \prod_{e\in\mathsf T(g)} \frac{1}{s^2-\rho_e^2} \right) \begin{cases}
		I_{\texttt{V}^{\circ}}(s;\rho_1), &n=1, \\[2mm]
		I_{\texttt{V1}\,\texttt{V}^{\,n-2}\texttt{V1}} (s;\rho_1,\ldots,\rho_n), &n\geq2.
	\end{cases}
	\label{eq:one-loop-core-reduction}
\end{equation}
We first solve the petal recurrences and then express a general cycle as a sum of petal integrals.

\subsection{Petal}
\label{subsec:one-loop-petal}
The petal consists of one vertex and one self-loop. At the physical point, its vertex weight is \(w=1\), and its incidence matrix, graph polynomial, and normalization are
\begin{equation}
	L_{\texttt{V}^{\circ}} = \begin{pmatrix}
		\lambda-\mu
	\end{pmatrix}, \qquad A_{\texttt{V}^{\circ}} = 1+(\lambda-\mu)^2, \qquad N_{\texttt{V}^{\circ}}(s) = \frac{1}{\Gamma(s+1)}.
	\label{eq:one-loop-self-loop-data}
\end{equation}
Substituting these data into the \(\lambda\) and \(\mu\) total-derivative identities \eqref{eq:bulk-I-equations} gives
\begin{equation}
	\begin{aligned}
		2(s+1)\Big[ I_{\texttt{V}^{\circ}}(s+1;\rho+1) -I_{\texttt{V}^{\circ}}(s+1;\rho) \Big] &=(s+\rho)I_{\texttt{V}^{\circ}}(s;\rho), \\
		2(s+1)\Big[ I_{\texttt{V}^{\circ}}(s+1;\rho-1) -I_{\texttt{V}^{\circ}}(s+1;\rho) \Big] &=(s-\rho)I_{\texttt{V}^{\circ}}(s;\rho).
	\end{aligned}
	\label{eq:one-petal-euler-recurrences}
\end{equation}
Adding the first equation to the second evaluated at \(\rho+1\) gives the spectral recurrence
\begin{equation}
	(\rho+1-s)I_{\texttt{V}^{\circ}}(s;\rho+1) =(s+\rho)I_{\texttt{V}^{\circ}}(s;\rho).
	\label{eq:one-petal-spectral-recurrence}
\end{equation}
Using \eqref{eq:one-petal-spectral-recurrence} at \(s+1\) in the first equation of \eqref{eq:one-petal-euler-recurrences} then gives the dimensional recurrence
\begin{equation}
	2(s+1)(2s+1)I_{\texttt{V}^{\circ}}(s+1;\rho) +(s^2-\rho^2)I_{\texttt{V}^{\circ}}(s;\rho)=0.
	\label{eq:one-petal-dimensional-recurrence}
\end{equation}
The general solution of the homogeneous dimensional recurrence is
\begin{equation}
	I_{\texttt{V}^{\circ}}(s;\rho) =\mathrsfs{G}(s;\rho)\,\mathrm p(s;\rho), \qquad \mathrsfs{G}(s;\rho) = \frac{\Gamma(1-2s)\Gamma(s+\rho)\Gamma(s-\rho)} {\Gamma(1-s)\Gamma(s+1)}.
	\label{eq:one-petal-gamma-factor}
\end{equation}
Here \(\mathrm p(s;\rho)\) is the periodic ambiguity left by the dimensional recurrence. The two recurrences and the symmetry \(\lambda\leftrightarrow\mu\) require it to satisfy
\begin{equation}
	\mathrm p(s+1;\rho)=\mathrm p(s;\rho),\qquad \mathrm p(s;\rho+1)=-\mathrm p(s;\rho),\qquad \mathrm p(s;-\rho)=\mathrm p(s;\rho).
	\label{eq:one-petal-ambiguity-properties}
\end{equation}

The analysis in Appendix~\ref{app:petal-analyticity-and-growth} shows that \(\mathrm p(s;\rho)\) itself is entire in \(\rho\) and satisfies
\begin{equation}
	\left|\mathrm p(s;\sigma+\mathrm i t)\right| = O\!\left(|t|^{1-2s}e^{\pi|t|}\right), \qquad t\to\pm\infty, \qquad 0<s<\tfrac12, \qquad |\sigma|\leq\tfrac12.
	\label{eq:one-petal-growth-bound}
\end{equation}
Together with antiperiodicity, these properties leave only the modes \(e^{\pm\mathrm i\pi\rho}\). Evenness therefore gives
\begin{equation}
	\mathrm p(s;\rho)=c(s)\cos(\pi\rho).
	\label{eq:one-petal-ambiguity}
\end{equation}
The zero-mode cut in Appendix~\ref{app:spectral-cuts} fixes \(c(s)\). Deleting the self-loop leaves a single vertex, whose integral is \(\Omega_{2s+1}\), so the cut formula gives
\begin{equation}
	\lim_{\rho\to s} (s^2-\rho^2)I_{\texttt{V}^{\circ}}(s;\rho)=1.
	\label{eq:one-petal-zero-mode-cut}
\end{equation}
Substituting \eqref{eq:one-petal-gamma-factor} and \eqref{eq:one-petal-ambiguity} into \eqref{eq:one-petal-zero-mode-cut} gives \(c(s)=1\). Substituting \(\mathrm p(s;\rho)=\cos(\pi\rho)\) into \eqref{eq:one-petal-gamma-factor} and continuing meromorphically in \(s\) gives
\begin{equation}
	\boxed{ I_{\texttt{V}^{\circ}}(s;\rho) = \frac{ \Gamma(1-2s) \Gamma(s+\rho) \Gamma(s-\rho) \cos(\pi\rho) }{ \Gamma(1-s) \Gamma(s+1) }. }
	\label{eq:one-petal-solution}
\end{equation}

\subsection{General cycle}

For \(n\geq2\), label the cycle's vertices cyclically by \(\widehat X_1,\ldots,\widehat X_n\), and orient edge \(i\) as
\begin{equation}
	i:\widehat X_i\longrightarrow \widehat X_{i+1}, \qquad \widehat X_{n+1}:=\widehat X_1.
	\label{eq:one-loop-ring-orientation}
\end{equation}
The weighted incidence matrix is
\begin{equation}
	(L_n)_{iv} = \lambda_i\delta_{v,i} - \mu_i\delta_{v,i+1}, \qquad i,v=1,\ldots,n,
	\label{eq:one-loop-ring-incidence}
\end{equation}
where indices are understood cyclically. The cycle polynomial is
\begin{equation}
	A_n := A_{\texttt{V1}\,\texttt{V}^{\,n-2}\texttt{V1}} = \det(I_n+L_n^TL_n) = \det(I_n+L_nL_n^T),
	\label{eq:one-loop-ring-polynomial}
\end{equation}
and the contact polynomial for edge \(i\) is
\begin{equation}
	Q_i := \det\!\left( I_n+ L_{n,\,\widehat i}^{T} L_{n,\,\widehat i} \right),
	\label{eq:one-loop-ring-contact-polynomials}
\end{equation}
where \(L_{n,\,\widehat i}\) is formed by removing row \(i\) from \(L_n\).

\begin{figure}[ht]
\centering
\renewcommand{\GraphLineWidth}{0.75pt}
\renewcommand{\GraphNodeDiameter}{3.75pt}
\begin{tikzpicture}[
  selected vertex/.style={contracted vertex},
  selected edge/.style={
    contracted edge,
    postaction={decorate},
    decoration={markings,mark=at position 0.5 with {
      \pgftransformxshift{1.6275pt}
      \arrow{>}
    }}
  },
  every node/.style={font=\small,text=black}
]
  % A regular m-gon has side length 2 R sin(pi/m). Match the left cycle.
  \pgfmathsetmacro{\cycleedgelength}{2*0.78*1.25*sin(180/8)}
  \pgfmathsetmacro{\cyclerightscale}{\cycleedgelength/(2*1.08*sin(180/7))}
  % The n-edge cycle. The two contacts used in the subtraction are marked.
  \begin{scope}[shift={(1.65,1.72)},scale=0.78]
    \foreach \p [count=\k from 0] in {x1,x2,x3,x4,x5,x6,x7,xn}
      \coordinate (\p) at ({90-\k*360/8}:1.25);

    \draw[selected edge] (xn)--(x1);
    \draw[selected edge] (x1)--(x2);
    \draw[edge] (x2)--(x3)--(x4)--(x5);
    \draw[omitted edge] (x5)--(x6);
    \draw[edge] (x6)--(x7)--(xn);

    \foreach \p in {x3,x4,x5,x6,x7}
      \node[vertex] at (\p) {};
    \foreach \p in {xn,x1,x2}
      \node[selected vertex] at (\p) {};

    \node[text=black] at (-0.54,1.49) {$\rho_n$};
    \node[text=black] at (0.54,1.49) {$\rho_1$};
    \node[text=black] at (1.47,0.48) {$\rho_2$};
    \node[text=black] at (-1.67,0.52) {$\rho_{n-1}$};
  \end{scope}

  \node[font=\Large] at (3.35,1.72) {$=$};
  \node at (4.50,1.72)
    {$\displaystyle\frac{1}{\rho_n^2-\rho_1^2}$};
  \node[font=\huge] at (5.40,1.72) {$[$};

  % Contract edge 1: parameters rho_2,...,rho_n remain.
  \begin{scope}[shift={(6.75,1.72)},scale=\cyclerightscale]
    \foreach \p [count=\k from 0] in {a,b,c,d,e,f,g}
      \coordinate (\p) at ({90-\k*360/7}:1.08);
    \draw[edge] (g)--(a)--(b)--(c)--(d);
    \draw[omitted edge] (d)--(e);
    \draw[edge] (e)--(f)--(g);
    \foreach \p in {a,b,c,d,e,f,g}
      \node[vertex] at (\p) {};
    \node[text=black] at (-0.50,1.35) {$\rho_n$};
    \node[text=black] at (0.50,1.35) {$\rho_2$};
  \end{scope}

  \node[font=\Large] at (8.20,1.72) {$-$};

  % Contract edge n: parameters rho_1,...,rho_{n-1} remain.
  \begin{scope}[shift={(9.65,1.72)},scale=\cyclerightscale]
    \foreach \p [count=\k from 0] in {a,b,c,d,e,f,g}
      \coordinate (\p) at ({90-\k*360/7}:1.08);
    \draw[edge] (g)--(a)--(b)--(c)--(d);
    \draw[omitted edge] (d)--(e);
    \draw[edge] (e)--(f)--(g);
    \foreach \p in {a,b,c,d,e,f,g}
      \node[vertex] at (\p) {};
    % Retain the figure's footprint while lowering the longer label.
    \node at (-0.63,1.47) {\phantom{$\rho_{n-1}$}};
    % Match the label baselines and balance their visible inner edges about a.
    \node[text=black] at (-0.708,1.3415) {$\rho_{n-1}$};
    \node[text=black] at (0.50,1.35) {$\rho_1$};
  \end{scope}

  \node[font=\huge] at (11.02,1.72) {$]$};
\end{tikzpicture}
\caption{One-step reduction of the $n$-edge cycle. Contracting the dashed edges $1$ and $n$ gives the first and second cycles on the right, respectively. Wavy segments indicate omitted parts of the cycle.}
\label{fig:one-loop-cycle-reduction}
\end{figure}
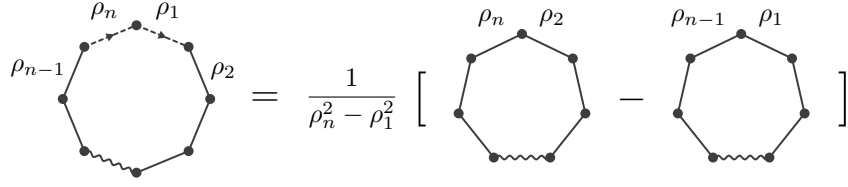

We reduce the cycle by comparing contractions of the two edges meeting at \(\widehat X_1\). Edge \(1\) starts there, while edge \(n\) ends there. Applying the edge-contraction relation \eqref{eq:contact-fixed-equations} with \(\xi=1\) for edge \(1\) and \(\xi=0\) for edge \(n\), then subtracting, gives
\begin{equation}
	\begin{aligned}
		&4(w_1-1) N_{\texttt{V1}\,\texttt{V}^{\,n-2}\texttt{V1}}(\bar{\boldsymbol{x}}) J_{\texttt{V1}\,\texttt{V}^{\,n-2}\texttt{V1}} \left[z(\lambda_1^2Q_1-\mu_n^2Q_n);\bar{\boldsymbol{x}}\right] \\
		&\qquad \qquad \qquad = \begin{cases}
			I_{\texttt{V}^{\circ}}(s;\rho_2) - I_{\texttt{V}^{\circ}}(s;\rho_1), &n=2, \\[2mm]
			I_{\texttt{V1}\,\texttt{V}^{\,n-3}\texttt{V1}} (s;\rho_2,\ldots,\rho_n) - I_{\texttt{V1}\,\texttt{V}^{\,n-3}\texttt{V1}} (s;\rho_1,\ldots,\rho_{n-1}), &n\geq3.
		\end{cases}
	\end{aligned}
	\label{eq:one-loop-ring-contact-difference}
\end{equation}
Now let us consider the difference \(z(\lambda_1^2 Q_1-\mu_n^2 Q_n)\). It can be checked that
\begin{equation}
	z(\lambda_1^2Q_1-\mu_n^2Q_n) = \frac12 \left( z\lambda_1 \frac{\partial A_n}{\partial\lambda_1} - z\mu_n \frac{\partial A_n}{\partial\mu_n} \right).
	\label{eq:one-loop-ring-local-identity}
\end{equation}
At \(\bar{\boldsymbol{x}}\), the total-derivative identities \eqref{eq:bulk-lambda-equation} and \eqref{eq:bulk-Euler-equations} give
\begin{align}
	J_{\texttt{V1}\,\texttt{V}^{\,n-2}\texttt{V1}} \left[ z\lambda_1 \frac{\partial A_n}{\partial\lambda_1}; \bar{\boldsymbol{x}} \right] &= (s+\rho_1) J_{\texttt{V1}\,\texttt{V}^{\,n-2}\texttt{V1}}(\bar{\boldsymbol{x}}), \\
	J_{\texttt{V1}\,\texttt{V}^{\,n-2}\texttt{V1}} \left[ z\mu_n \frac{\partial A_n}{\partial\mu_n}; \bar{\boldsymbol{x}} \right] &= (s-\rho_n) J_{\texttt{V1}\,\texttt{V}^{\,n-2}\texttt{V1}}(\bar{\boldsymbol{x}}).
\end{align}
Subtracting these relations and using \eqref{eq:one-loop-ring-local-identity} gives
\begin{equation}
	J_{\texttt{V1}\,\texttt{V}^{\,n-2}\texttt{V1}} \left[ z(\lambda_1^2Q_1-\mu_n^2Q_n); \bar{\boldsymbol{x}} \right] = \frac{\rho_1+\rho_n}{2} J_{\texttt{V1}\,\texttt{V}^{\,n-2}\texttt{V1}}(\bar{\boldsymbol{x}}).
	\label{eq:one-loop-ring-bulk-reduction}
\end{equation}
The vertex-weight formula \eqref{eq:lattice-vertex-weight} at \(\widehat X_1\) gives \(2(w_1-1)=\rho_n-\rho_1\). Multiplying \eqref{eq:one-loop-ring-bulk-reduction} by the prefactor in \eqref{eq:one-loop-ring-contact-difference} and using \(I_g=N_gJ_g\), we obtain
\begin{equation}
	4(w_1-1) N_{\texttt{V1}\,\texttt{V}^{\,n-2}\texttt{V1}}(\bar{\boldsymbol{x}}) J_{\texttt{V1}\,\texttt{V}^{\,n-2}\texttt{V1}} \!\left[ z(\lambda_1^2Q_1-\mu_n^2Q_n); \bar{\boldsymbol{x}} \right] = (\rho_n^2-\rho_1^2) I_{\texttt{V1}\,\texttt{V}^{\,n-2}\texttt{V1}} (\bar{\boldsymbol{x}}).
	\label{eq:one-loop-ring-normalized-bulk-reduction}
\end{equation}
Combining \eqref{eq:one-loop-ring-contact-difference} with \eqref{eq:one-loop-ring-normalized-bulk-reduction} eliminates the inserted integrals and gives
\begin{equation}
	\begin{aligned}
		&(\rho_n^2-\rho_1^2) I_{\texttt{V1}\,\texttt{V}^{\,n-2}\texttt{V1}} (s;\rho_1,\ldots,\rho_n) \\
		&\qquad \qquad \qquad= \begin{cases}
			I_{\texttt{V}^{\circ}}(s;\rho_2) - I_{\texttt{V}^{\circ}}(s;\rho_1), &n=2, \\[2mm]
			I_{\texttt{V1}\,\texttt{V}^{\,n-3}\texttt{V1}} (s;\rho_2,\ldots,\rho_n) - I_{\texttt{V1}\,\texttt{V}^{\,n-3}\texttt{V1}} (s;\rho_1,\ldots,\rho_{n-1}), &n\geq3.
		\end{cases}
	\end{aligned}
	\label{eq:one-loop-ring-reduction}
\end{equation}
For \(\rho_1^2\ne\rho_n^2\), this gives the one-step reduction shown in Figure~\ref{fig:one-loop-cycle-reduction}.
For pairwise distinct squared spectral parameters, iterating \eqref{eq:one-loop-ring-reduction} down to the petal gives\footnote{The zero-mode cut of edge \(k\) deletes that edge, leaving a tree (see Appendix~\ref{app:spectral-cuts} on spectral cuts). Only the \(i=k\) term contributes to this cut, and \eqref{eq:one-petal-zero-mode-cut} gives
\begin{equation*}
	\lim_{\rho_k\to\pm s}\Omega_{2s+1}(s^2-\rho_k^2)\sum_{i=1}^{n}\frac{I_{\texttt{V}^{\circ}}(s;\rho_i)}{\prod_{j\ne i}(\rho_i^2-\rho_j^2)} = \Omega_{2s+1}\prod_{j\ne k}\frac{1}{s^2-\rho_j^2} = I_{\mathcal T_{n-1}}(\pi_k\bar{\boldsymbol{x}}).
\end{equation*}
This agrees with \eqref{eq:general-tree-solution} for a tree with \(n-1\) edges.}
\begin{equation}
	I_{\texttt{V1}\,\texttt{V}^{\,n-2}\texttt{V1}} (s;\rho_1,\ldots,\rho_n) = \sum_{i=1}^{n} \frac{ I_{\texttt{V}^{\circ}}(s;\rho_i) }{ \displaystyle \prod_{\substack{j=1\\j\neq i}}^{n} (\rho_i^2-\rho_j^2) }, \qquad n\geq2.
	\label{eq:general-ring-explicit}
\end{equation}
Substituting the petal solution \eqref{eq:one-petal-solution} gives
\begin{equation}
	\boxed{\displaystyle I_{\texttt{V1}\,\texttt{V}^{\,n-2}\texttt{V1}} (s;\rho_1,\ldots,\rho_n) = \frac{ \Gamma(1-2s) }{ \Gamma(1-s)\Gamma(s+1) } \sum_{i=1}^{n} \frac{ \Gamma(s+\rho_i) \Gamma(s-\rho_i) \cos(\pi\rho_i) }{ \displaystyle \prod_{\substack{j=1\\j\neq i}}^{n} (\rho_i^2-\rho_j^2) }, \qquad n\geq2. }
	\label{eq:general-ring-solution}
\end{equation}
Coincident squared spectral parameters are handled by taking limits of the full sum in \eqref{eq:general-ring-explicit}. When \(\rho_1=\cdots=\rho_n=\rho\), this gives
\begin{equation}
	I_{\texttt{V1}\,\texttt{V}^{\,n-2}\texttt{V1}} (s;\rho,\ldots,\rho) = \frac{1}{(n-1)!} \left( \frac{1}{2\rho} \frac{\partial}{\partial\rho} \right)^{n-1} I_{\texttt{V}^{\circ}}(s;\rho), \qquad n\geq2.
	\label{eq:general-ring-equal-rho}
\end{equation}
If only some \(\rho_i^2\) coincide, take the corresponding limits while keeping the others distinct.

The polynomial identity \eqref{eq:one-loop-ring-local-identity} uses only the two edges meeting at a vertex with no other incident edges (a degree-two vertex). The same reduction therefore applies to a chain whose internal vertices have degree two, even when its endpoints belong to a larger graph.

\section{Two loops}
\label{sec:two-loops}

For connected two-loop graphs, the reductions in Sections~\ref{sec:trees} and~\ref{sec:one-loop} leave two-petal graphs and the three-melon \(\texttt{V\{3\}V}\). The two-petal integrals factorize into known one-loop and tree integrals, so the three-melon is the only new integral at two loops.

\subsection{Two-petal}

Rotational invariance makes the coincident propagator independent of its position. Since
\begin{equation}
	I_{\texttt{V}^{\circ}}(s;\rho) = \int_{S^{2s+1}} d\Omega_{2s+1}(\widehat X)\, G_\rho(\widehat X,\widehat X),
\end{equation}
we have
\begin{equation}
	G_\rho(\widehat X,\widehat X) = \frac{ I_{\texttt{V}^{\circ}}(s;\rho) }{ \Omega_{2s+1} }.
	\label{eq:one-petal-factor}
\end{equation}
Attaching a petal of parameter \(\rho\) at any vertex therefore multiplies the graph integral by \(I_{\texttt{V}^{\circ}}(s;\rho)/\Omega_{2s+1}\). For two petals sharing a vertex, this gives
\begin{equation}
	I_{\texttt{V}^{\circ\texttt{\{2\}}}}(s;\rho_i,\rho_j) = \frac{ I_{\texttt{V}^{\circ}}(s;\rho_i) I_{\texttt{V}^{\circ}}(s;\rho_j) }{ \Omega_{2s+1} }.
	\label{eq:two-loop-reducible-graphs}
\end{equation}
The same rule evaluates any two-petal decoration, with one factor \eqref{eq:one-petal-factor} for each petal. For example, if the two petals are attached to opposite endpoints of an edge with parameter \(\rho_k\), the same rule and the tree formula \eqref{eq:general-tree-solution} give
\begin{equation}
	I_{\texttt{V}^{\circ}\texttt{V}^{\circ}} (s;\rho_i,\rho_j,\rho_k) = \frac{ I_{\texttt{V}^{\circ}}(s;\rho_i) I_{\texttt{V}^{\circ}}(s;\rho_j) }{ \Omega_{2s+1} } \frac{1}{s^2-\rho_k^2}.
\end{equation}

\subsection{Three-melon}
\label{sec:three-melon}

\begin{figure}[ht]
\centering
\renewcommand{\GraphLineWidth}{0.75pt}
\renewcommand{\GraphNodeDiameter}{3.75pt}
\begin{tikzpicture}[
  removed edge/.style={contracted edge},
  flow/.style={graph arrow},
  every node/.style={font=\small}
]
  % Contract edge 3.
  \begin{scope}[shift={(7.00,-0.25)}]
    \coordinate (x) at (-0.76,0);
    \coordinate (y) at (0.76,0);
    \draw[edge] (x)
      arc[start angle=180,end angle=0,radius=0.76];
    \draw[edge] (x)--(y);
    \draw[removed edge] (x)
      arc[start angle=180,end angle=360,radius=0.76];
    \node[contracted vertex] at (x) {};
    \node[contracted vertex] at (y) {};
    \node[text=black,left=3pt] at (x) {$\widehat X_1$};
    \node[text=black,right=3pt] at (y) {$\widehat X_2$};
    \node[text=black] at (0,0.92) {$\rho_1$};
    \node[text=black] at (0,0.16) {$\rho_2$};
    \node[text=black] at (0,-0.97) {$\rho_3$};
  \end{scope}

  % Contract edge 1.
  \begin{scope}[shift={(7.00,4.65)}]
    \coordinate (x) at (-0.76,0);
    \coordinate (y) at (0.76,0);
    \draw[removed edge] (x)
      arc[start angle=180,end angle=0,radius=0.76];
    \draw[edge] (x)--(y);
    \draw[edge] (x)
      arc[start angle=180,end angle=360,radius=0.76];
    \node[contracted vertex] at (x) {};
    \node[contracted vertex] at (y) {};
    \node[text=black,left=3pt] at (x) {$\widehat X_1$};
    \node[text=black,right=3pt] at (y) {$\widehat X_2$};
    \node[text=black] at (0,0.92) {$\rho_1$};
    \node[text=black] at (0,0.16) {$\rho_2$};
    \node[text=black] at (0,-0.97) {$\rho_3$};
  \end{scope}

  % Contract edge 2.
  \begin{scope}[shift={(7.00,2.20)}]
    \coordinate (x) at (-0.76,0);
    \coordinate (y) at (0.76,0);
    \draw[edge] (x)
      arc[start angle=180,end angle=0,radius=0.76];
    \draw[removed edge] (x)--(y);
    \draw[edge] (x)
      arc[start angle=180,end angle=360,radius=0.76];
    \node[contracted vertex] at (x) {};
    \node[contracted vertex] at (y) {};
    \node[text=black,left=3pt] at (x) {$\widehat X_1$};
    \node[text=black,right=3pt] at (y) {$\widehat X_2$};
    \node[text=black] at (0,0.92) {$\rho_1$};
    \node[text=black] at (0,0.16) {$\rho_2$};
    \node[text=black] at (0,-0.97) {$\rho_3$};
  \end{scope}

  % Centre the arrows in the visible gap between X_2 and the petal labels.
  \draw[flow] (9.353,4.65) -- (10.253,4.65);
  \draw[flow] (9.353,2.20) -- (10.253,2.20);
  \draw[flow] (9.353,-0.25) -- (10.253,-0.25);

  % The three corresponding double self-loop graphs.
  \coordinate (m3) at (12.55,-0.25);
  \draw[edge] (m3)
    .. controls (11.35,0.53) and (11.35,-1.03) .. (m3);
  \draw[edge] (m3)
    .. controls (13.75,0.53) and (13.75,-1.03) .. (m3);
  \node[vertex] at (m3) {};
  \node[text=black,below=5.5pt] at (m3) {$\widehat X_1=\widehat X_2$};
  \node[text=black] at (11.35,-0.25) {$\rho_1$};
  \node[text=black] at (13.75,-0.25) {$\rho_2$};

  \coordinate (m1) at (12.55,4.65);
  \draw[edge] (m1)
    .. controls (11.35,5.43) and (11.35,3.87) .. (m1);
  \draw[edge] (m1)
    .. controls (13.75,5.43) and (13.75,3.87) .. (m1);
  \node[vertex] at (m1) {};
  \node[text=black,below=5.5pt] at (m1) {$\widehat X_1=\widehat X_2$};
  \node[text=black] at (11.35,4.65) {$\rho_2$};
  \node[text=black] at (13.75,4.65) {$\rho_3$};

  \coordinate (m2) at (12.55,2.20);
  \draw[edge] (m2)
    .. controls (11.35,2.92) and (11.35,1.48) .. (m2);
  \draw[edge] (m2)
    .. controls (13.75,2.92) and (13.75,1.48) .. (m2);
  \node[vertex] at (m2) {};
  \node[text=black,below=5.5pt] at (m2) {$\widehat X_1=\widehat X_2$};
  \node[text=black] at (11.35,2.20) {$\rho_3$};
  \node[text=black] at (13.75,2.20) {$\rho_1$};
\end{tikzpicture}
\caption{Three-melon edge contractions. In each row, contracting the dashed edge merges \(\widehat X_1\) and \(\widehat X_2\), leaving the other two edges as petals at the merged vertex.}
\label{fig:three-melon-contraction}
\end{figure}
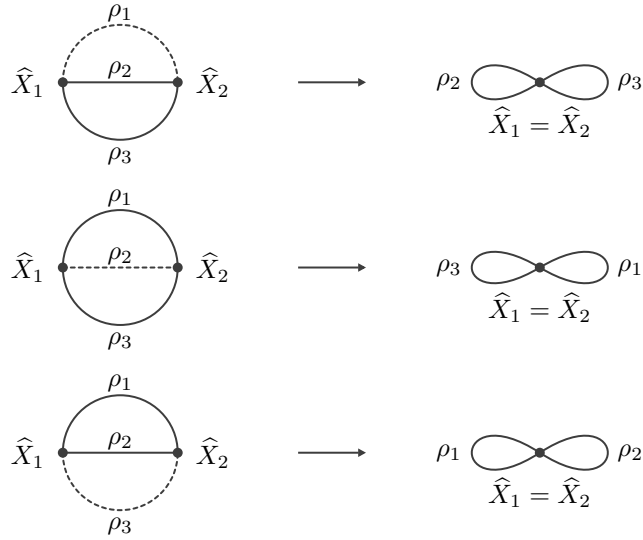

The three-melon \(\texttt{V\{3\}V}\) consists of two vertices \(\widehat X_1\) and \(\widehat X_2\) joined by three parallel edges. Orienting all three edges from \(\widehat X_1\) to \(\widehat X_2\) gives the weighted incidence matrix
\begin{equation}
	L_{\texttt{V\{3\}V}} = \begin{pmatrix}
		\lambda_1&-\mu_1\\
		\lambda_2&-\mu_2\\
		\lambda_3&-\mu_3
	\end{pmatrix}.
	\label{eq:three-melon-incidence}
\end{equation}
The corresponding graph polynomial is
\begin{equation}
	A_{\texttt{V\{3\}V}} = 1+ \sum_{i=1}^{3} \left(\lambda_i^2+\mu_i^2\right) + \sum_{1\leq i<j\leq3} \left( \lambda_i\mu_j-\lambda_j\mu_i \right)^2.
	\label{eq:three-melon-polynomial}
\end{equation}
At the physical point \(\bar{\boldsymbol{x}}\) defined in \eqref{eq:lattice-physical-point}, the integral is \(I_{\texttt{V\{3\}V}}(\bar{\boldsymbol{x}})=N_{\texttt{V\{3\}V}}(\bar{\boldsymbol{x}})J_{\texttt{V\{3\}V}}(\bar{\boldsymbol{x}})\), with the normalization and Euler integral given by \eqref{eq:lattice-I-NJ-x}--\eqref{eq:lattice-normalization-x}. The total-derivative identities \eqref{eq:bulk-z-equation}--\eqref{eq:bulk-Euler-equations} give
\begin{align}
	J_{\texttt{V\{3\}V}} [zA_{\texttt{V\{3\}V}};\bar{\boldsymbol{x}}] &= (s+1)J_{\texttt{V\{3\}V}}(\bar{\boldsymbol{x}}),
	\label{eq:three-melon-scale-row} \\
	J_{\texttt{V\{3\}V}}\left[ z\lambda_i \frac{\partial A_{\texttt{V\{3\}V}}}{\partial\lambda_i}; \bar{\boldsymbol{x}} \right] &= (s+\rho_i)J_{\texttt{V\{3\}V}}(\bar{\boldsymbol{x}}),
	\label{eq:three-melon-lambda-row} \\
	J_{\texttt{V\{3\}V}}\left[ z\mu_i \frac{\partial A_{\texttt{V\{3\}V}}}{\partial\mu_i}; \bar{\boldsymbol{x}} \right] &= (s-\rho_i)J_{\texttt{V\{3\}V}}(\bar{\boldsymbol{x}}), \qquad i=1,2,3.
	\label{eq:three-melon-bulk-rows}
\end{align}
Contracting edge \(i\) identifies \(\widehat X_1\) and \(\widehat X_2\), leaving the other two edges \(j\) and \(k\) as petals at the merged vertex. Figure~\ref{fig:three-melon-contraction} shows the three possible contractions. Throughout Section~\ref{sec:three-melon}, \((i,j,k)\) denotes a cyclic permutation of \((1,2,3)\). The contact polynomial is
\begin{equation}
	Q_i = 1+ \sum_{\ell\neq i} \left(\lambda_\ell^2+\mu_\ell^2\right) + \left( \lambda_j\mu_k-\lambda_k\mu_j \right)^2.
	\label{eq:three-melon-contact-minor}
\end{equation}
The edge-contraction relation \eqref{eq:contact-fixed-equations}, specialized to \(\xi=1\) and \(\xi=0\), gives
\begin{align}
	-2(s+\rho_1+\rho_2+\rho_3) N_{\texttt{V\{3\}V}}(\bar{\boldsymbol{x}}) J_{\texttt{V\{3\}V}}[z\lambda_i^2Q_i;\bar{\boldsymbol{x}}] &= I_{\texttt{V}^{\circ\texttt{\{2\}}}}(s;\rho_j,\rho_k), \nonumber\\
	-2(s-\rho_1-\rho_2-\rho_3) N_{\texttt{V\{3\}V}}(\bar{\boldsymbol{x}}) J_{\texttt{V\{3\}V}}[z\mu_i^2Q_i;\bar{\boldsymbol{x}}] &= I_{\texttt{V}^{\circ\texttt{\{2\}}}}(s;\rho_j,\rho_k).
	\label{eq:three-melon-contact-rows}
\end{align}
There are two equations for each cyclic choice of \((i,j,k)\), with the same two-petal source \eqref{eq:two-loop-reducible-graphs}.

The seven total-derivative identities \eqref{eq:three-melon-scale-row}--\eqref{eq:three-melon-bulk-rows} involve polynomials of degree at most four in \(\lambda_i,\mu_i\). The contraction insertions in \eqref{eq:three-melon-contact-rows}, however, contain the degree-six terms
\begin{equation}
	\lambda_i^2 \left( \lambda_j\mu_k-\lambda_k\mu_j \right)^2, \qquad \mu_i^2 \left( \lambda_j\mu_k-\lambda_k\mu_j \right)^2.
	\label{eq:three-melon-contact-degree-six}
\end{equation}
No linear combination of the seven total-derivative identities at \(\bar{\boldsymbol{x}}\), with coefficients depending only on the graph parameters, can reproduce these insertions.

For trees and one-loop cycles, identities at the physical point suffice to obtain the contraction combinations used in the reductions. For the three-melon, we also need to include identities at shifted base points in the system, as described in Section~\ref{sec:coupled-system}.
% The contraction relations can also be used at shifted physical nodes. Any
% resulting shifts in the two-petal sources can be reduced using the petal
% result \eqref{eq:one-petal-solution}. Here the six equations at
% \(\bar{\boldsymbol{x}}\) suffice, so we keep these sources unshifted.

To build this coupled system, we restrict the base points at which we add total-derivative identities. Writing \(\boldsymbol{\delta}=(\delta_t,\boldsymbol{\delta}_u,\boldsymbol{\delta}_\rho)\), the allowed shifts satisfy
\begin{equation}
	0\leq\delta_t,(\boldsymbol{\delta}_u)_i\leq c_D, \qquad \left|(\boldsymbol{\delta}_\rho)_i\right|\leq c_\rho, \qquad i=1,2,3.
\end{equation}
Here \(c_D\) bounds the shifts of \(t\) and \(\boldsymbol{u}\), while \(c_\rho\) bounds the spectral shifts.

We begin with seven total-derivative identities \eqref{eq:three-melon-scale-row}--\eqref{eq:three-melon-bulk-rows} at each chosen physical target node and the six contraction equations \eqref{eq:three-melon-contact-rows} at \(\bar{\boldsymbol{x}}\). Expanding their polynomial insertions into monomials and applying \eqref{eq:lattice-shifted-I} identifies the shifted integrals. At each new node within the bounds, we add the seven total-derivative identities \eqref{eq:bulk-z-equation}--\eqref{eq:bulk-Euler-equations} and repeat the procedure for any further shifted integrals they introduce. Every integral appearing in an included equation is retained, even if its node lies outside the bounds. %The six contraction equations remain at \(\bar{\boldsymbol{x}}\) throughout, so their three two-petal sources are unshifted. %Every integral appearing in the equations is retained, including those outside the bounds; only the base points at which we add identities are restricted.

After \(h\) iterations, the shifts appearing in the collected equations define the finite stencil \(\mathcal S^{(h)}_{c_D,c_\rho}\). Combining the collected total-derivative identities with the six contraction equations at \(\bar{\boldsymbol{x}}\) gives the coupled system in which we seek dimensional and spectral recurrences.

The corresponding three-melon integrals form the vector
\begin{equation}
	\boldsymbol{I}_{\texttt{V\{3\}V}} = \left( I_{\texttt{V\{3\}V}} \bigl( \bar{\boldsymbol{x}}+\boldsymbol{\delta} \bigr) \right)_{\boldsymbol{\delta}\in \mathcal S^{(h)}_{c_D,c_\rho}}.
\end{equation}
The three known contraction sources form
\begin{equation}
	\boldsymbol{I}_{\texttt{V}^{\circ\texttt{\{2\}}}} = \begin{pmatrix}
		I_{\texttt{V}^{\circ\texttt{\{2\}}}}(s;\rho_2,\rho_3) \\[2pt]
		I_{\texttt{V}^{\circ\texttt{\{2\}}}}(s;\rho_3,\rho_1) \\[2pt]
		I_{\texttt{V}^{\circ\texttt{\{2\}}}}(s;\rho_1,\rho_2)
	\end{pmatrix}.
\end{equation}
The coupled system is
\begin{equation}
	\begin{pmatrix}
		\mathsf E_{\texttt{V\{3\}V}} & 0 \\[2pt]
		\mathsf C_{\texttt{V\{3\}V}} & -\mathsf B_{\texttt{V\{3\}V}}
	\end{pmatrix} \begin{pmatrix}
		\boldsymbol{I}_{\texttt{V\{3\}V}} \\[2pt]
		\boldsymbol{I}_{\texttt{V}^{\circ\texttt{\{2\}}}}
	\end{pmatrix} =0.
\end{equation}

We retain the chosen physical target integrals and the three known contraction sources as targets for the recurrence search. All other three-melon integrals are auxiliary, including those at other physical nodes. We order the auxiliary columns first and perform exact Gaussian elimination in C++ at generic parameters. Rows in which all auxiliary coefficients vanish give relations among the targets. If no such relation appears, we enlarge the system and repeat. 

We obtain the dimensional recurrence at \((c_D,c_\rho,h)=(3,1,3)\) and the spectral recurrence at \((1,2,1)\). We state the resulting \(s\mapsto s+2\) and \(\rho_i\mapsto\rho_i+2\) recurrences below.

\subsubsection{Recurrence relations}
\label{sec:three-melon-recurrences}

It is convenient to work with the three-melon and petal integrals per unit sphere volume,
\begin{equation}
	\bar I_{\texttt{V\{3\}V}}(s;\boldsymbol{\rho}) := \frac{I_{\texttt{V\{3\}V}}(s;\boldsymbol{\rho})}{\Omega_{2s+1}}, \qquad \bar I_{\texttt{V}^{\circ}}(s;\rho) := \frac{I_{\texttt{V}^{\circ}}(s;\rho)}{\Omega_{2s+1}} = \frac{\Gamma(s+\rho)\Gamma(s-\rho)\cos(\pi\rho)} {(4\pi)^{s+\frac12}\Gamma(s+\tfrac12)\cos(\pi s)}.
	\label{eq:three-melon-recurrence-density}
\end{equation}
To write the two-petal sources compactly, define the petal matrix for an ordered triple \((\alpha,\beta,\gamma)\) by
\begin{equation}
	\overline{\mathsf P}(s;\alpha,\beta,\gamma) := \operatorname{diag} \left( \bar I_{\texttt{V}^{\circ}}(s;\alpha), \bar I_{\texttt{V}^{\circ}}(s;\beta), \bar I_{\texttt{V}^{\circ}}(s;\gamma) \right), \qquad \overline{\mathsf P}(s;\boldsymbol{\rho}) := \overline{\mathsf P}(s;\rho_1,\rho_2,\rho_3).
	\label{eq:three-melon-petal-matrix}
\end{equation}
The diagonal entries of its adjugate are the two-petal contraction sources in density form. In the dimensional recurrence, these sources are weighted by the polynomial
\begin{equation}
	\mathsf M(u) = (2s+1) \left[ 8s^2\operatorname{tr}\mathsf Z - \operatorname{tr} \left( \operatorname{adj}\mathsf Z \right) \right]u - 2s \left[ u+2(2s+1)^2 \right] \left[ u^2+ \operatorname{tr} \left( \operatorname{adj}\mathsf Z \right) \right],
	\label{eq:three-melon-channel-polynomial}
\end{equation}
where
\begin{equation}
	\mathsf Z := \operatorname{diag} \left( \mathsf z_1,\mathsf z_2,\mathsf z_3 \right), \qquad \mathsf z_k := s^2+\rho_k^2-\sum_{j\neq k}\rho_j^2.
	\label{eq:three-melon-channel-data}
\end{equation}

Finally, for each cyclic permutation \((i,j,k)\), define the shifted K\"all\'en polynomial\footnote{Whenever \(\rho_i+q,\rho_j,\rho_k\) form a triangle, its area squared is \(-\mathcal K_q^{(i)}/16\).}
\begin{equation}
	\mathcal K_q^{(i)} := \prod_{\sigma_j,\sigma_k=\pm1} \left( \rho_i+q+\sigma_j\rho_j+\sigma_k\rho_k \right), \qquad \mathcal K_q:=\mathcal K_q^{(1)}.
	\label{eq:three-melon-analytic-shifted-kallen}
\end{equation}
The dimensional recurrence is then
\begin{equation}
	\boxed{ (4\pi)^4(2s)_4\, \bar I_{\texttt{V\{3\}V}}(s+2;\boldsymbol{\rho}) - \mathcal K_s\mathcal K_{-s}\, \bar I_{\texttt{V\{3\}V}}(s;\boldsymbol{\rho}) = \frac{1}{2s+1} \operatorname{tr} \left[ \mathsf M(\mathsf Z) \operatorname{adj} \overline{\mathsf P}(s;\boldsymbol{\rho}) \right]. }
	\label{eq:three-melon-s-recurrence}
\end{equation}
Here \((a)_n:=\Gamma(a+n)/\Gamma(a)=a(a+1)\cdots(a+n-1)\) denotes the Pochhammer symbol, or rising factorial, for positive integers \(n\), with \((a)_0=1\).

For each cyclic permutation \((i,j,k)\), define
\begin{equation}
	\mathsf R_i := \operatorname{diag} \left( 2(s-\rho_i-2)(s-\rho_i-1), \mathsf z_k-2(s+\rho_i), \mathsf z_j-2(s+\rho_i) \right).
	\label{eq:three-melon-rho-source-matrix}
\end{equation}
The spectral recurrence is then
\begin{equation}
	\boxed{ \mathcal K_{2-s}^{(i)}\, \bar I_{\texttt{V\{3\}V}} \left( s;\boldsymbol{\rho}+2\boldsymbol{e}_i \right) - \mathcal K_s^{(i)}\, \bar I_{\texttt{V\{3\}V}}(s;\boldsymbol{\rho}) = \frac{ 2(2s-1)(\rho_i+1) }{ (s-\rho_i-2)_2 } \operatorname{tr} \left[ \mathsf R_i \operatorname{adj} \overline{\mathsf P}(s;\rho_i,\rho_j,\rho_k) \right]. }
	\label{eq:three-melon-rho-recurrences}
\end{equation}

These recurrences\footnote{Both recurrences reproduce the two-edge cycle \eqref{eq:general-ring-explicit} under the zero-mode cut of Appendix~\ref{app:spectral-cuts},
\begin{equation*}
	\lim_{\rho_i\to-s}\Omega_{2s+1}(s^2-\rho_i^2)\, \bar I_{\texttt{V\{3\}V}}(s;\boldsymbol{\rho}) = \frac{\bar I_{\texttt{V}^{\circ}}(s;\rho_j)-\bar I_{\texttt{V}^{\circ}}(s;\rho_k)}{\rho_j^2-\rho_k^2}.
\end{equation*}
} propagate the integral between physical nodes on the three-melon lattice. Since \(D=2s+1\), the dimensional recurrence advances \(D\) by four. For equal spectral parameters, the dimensional recurrence follows by setting \(\rho_1=\rho_2=\rho_3=\rho\) in \eqref{eq:three-melon-s-recurrence}. In Section~\ref{sec:three-melon-analytical-solution}, we solve it at generic \(s\) and \(\rho_i\) to construct the meromorphic solution. 

For dimensional regularization, the recurrence \eqref{eq:three-melon-s-recurrence} provides a shortcut: it propagates Laurent series directly. Given the expansion about dimension \(D\), expanding the recurrence coefficients and known two-petal sources in the same regulator determines the expansion about dimension \(D+4\). Seed expansions about \(D=0,1,2,3\) therefore suffice. The first three are obtained by expanding the integral representation, while the \(D=3\) expansion follows from the analytical solution together with the spectral recurrence \eqref{eq:three-melon-rho-recurrences}. Iteration then gives the Laurent expansions about all higher integer dimensions, as developed in Section~\ref{sec:three-melon-dimensional-regularization}.

\subsubsection{Analytical solution}
\label{sec:three-melon-analytical-solution}

We construct the meromorphic solution at generic \(s\) from the dimensional recurrence \eqref{eq:three-melon-s-recurrence}. Dividing by the coefficient of the shifted integral gives
\begin{equation}
	\bar I_{\texttt{V\{3\}V}}(s+2;\boldsymbol{\rho}) = \mathsf A_s(s)\, \bar I_{\texttt{V\{3\}V}}(s;\boldsymbol{\rho}) + \mathsf C_s(s).
	\label{eq:three-melon-analytic-recurrence}
\end{equation}
The homogeneous coefficient and contraction source are
\begin{equation}
	\mathsf A_s(s)=\frac{\mathcal K_s\mathcal K_{-s}}{(4\pi)^4(2s)_4}, \qquad \mathsf C_s(s)= \frac{1}{(4\pi)^4(2s)_4(2s+1)} \operatorname{tr}\! \left[ \mathsf M(\mathsf Z)\operatorname{adj}\overline{\mathsf P}(s;\boldsymbol{\rho}) \right].
	\label{eq:three-melon-analytic-coefficients}
\end{equation}

For \(\boldsymbol{\sigma}\in\{\pm1\}^3\), set \(\rho_{\boldsymbol{\sigma}}:=\sigma_1\rho_1+\sigma_2\rho_2+\sigma_3\rho_3\). A solution of the homogeneous recurrence is
\begin{equation}
	\mathrsfs{H}(s;\boldsymbol{\rho}) = \frac{ \displaystyle \prod_{\boldsymbol{\sigma}\in\{\pm1\}^3} \Gamma\!\left(\frac{s+\rho_{\boldsymbol{\sigma}}}{2}\right) }{ 256\pi^{2s+3}\Gamma(2s) }, \qquad \mathrsfs{H}(s+2;\boldsymbol{\rho}) =\mathsf A_s(s)\mathrsfs{H}(s;\boldsymbol{\rho}).
	\label{eq:three-melon-analytic-propagator}
\end{equation}
Using this homogeneous solution, we can write the general solution of \eqref{eq:three-melon-analytic-recurrence} as
\begin{equation}
	\bar I_{\texttt{V\{3\}V}}(s;\boldsymbol{\rho}) = \underbrace{ \mathrm p(s;\boldsymbol{\rho})\mathrsfs{H}(s;\boldsymbol{\rho}) \vphantom{\displaystyle \sum_{n=0}^{\infty} \frac{\mathsf C_s(s+2n)} {\mathrsfs{H}(s+2n+2;\boldsymbol{\rho})} } }_{\bar I_{\texttt{V\{3\}V}}^{\,\mathrm{hom}}(s;\boldsymbol{\rho})} \underbrace{ -\,\,\mathrsfs{H}(s;\boldsymbol{\rho}) \sum_{n=0}^{\infty} \frac{\mathsf C_s(s+2n)} {\mathrsfs{H}(s+2n+2;\boldsymbol{\rho})} }_{\bar I_{\texttt{V\{3\}V}}^{\,\mathrm{part}}(s;\boldsymbol{\rho})}, \qquad \mathrm p(s+2;\boldsymbol{\rho})=\mathrm p(s;\boldsymbol{\rho}).
	\label{eq:three-melon-difference-tail}
\end{equation}
The infinite sum is understood by analytic continuation. The arbitrary coefficient \(\mathrm p\), with period two in \(s\), carries the homogeneous ambiguity. We first evaluate the particular solution and then return to this coefficient.

The contraction source \eqref{eq:three-melon-analytic-coefficients} separates into three channels,
\begin{equation}
	\mathsf C_s(s) = \sum_{\mathrm{cyc}(k,i,j)} \mathsf C_{s,k\vert ij}(s;\boldsymbol{\rho}), \qquad \mathsf C_{s,k\vert ij}(s;\boldsymbol{\rho}) := \frac{ \mathsf M(\mathsf z_k) \bar I_{\texttt{V}^{\circ}}(s;\rho_i) \bar I_{\texttt{V}^{\circ}}(s;\rho_j) }{ (4\pi)^4(2s)_4(2s+1) },
	\label{eq:three-melon-analytic-source-sum}
\end{equation}
where \(k\vert ij\) denotes the contraction of edge \(k\), leaving edges \(i\) and \(j\) as petals. By linearity,
\begin{equation}
	\bar I_{\texttt{V\{3\}V}}^{\,\mathrm{part}}(s;\boldsymbol{\rho}) = -\mathrsfs{H}(s;\boldsymbol{\rho}) \sum_{\mathrm{cyc}(k,i,j)} \sum_{n=0}^{\infty} \frac{ \mathsf C_{s,k\vert ij}(s+2n;\boldsymbol{\rho}) }{ \mathrsfs{H}(s+2n+2;\boldsymbol{\rho}) }.
	\label{eq:three-melon-analytic-particular-sum}
\end{equation}
For a fixed channel, the summand in \eqref{eq:three-melon-analytic-particular-sum} separates into a shifted polynomial and a Pochhammer factor,
\begin{equation}
	\frac{ \mathsf C_{s,k\vert ij}(s+2n;\boldsymbol{\rho}) }{ \mathrsfs{H}(s+2n+2;\boldsymbol{\rho}) } = \frac{ \mathsf C_{s,k\vert ij}(s;\boldsymbol{\rho}) }{ \mathrsfs{H}(s+2;\boldsymbol{\rho})\mathsf M(\mathsf z_k) } \left.\mathsf M(\mathsf z_k)\right|_{s\to s+2n} 16^n h_{k\vert ij,n}.
	\label{eq:three-melon-tail-summand-factorization}
\end{equation}
The remaining factor \(h_{k\vert ij,n}\) has generating function
\begin{equation}
	\mathrm G_{k\vert ij}(\zeta) := \sum_{n=0}^{\infty}h_{k\vert ij,n}\zeta^n = {}_{11}F_{10}\!\left( \begin{matrix}
		1,\ \dfrac{s}{2},\ \dfrac{s+1}{2},\  \left\{ \dfrac{s+\tau\rho_r}{2}, \dfrac{s+1+\tau\rho_r}{2} \right\}_{\substack{r\in\{i,j\}\\\tau=\pm1}} \\[3pt]
		\dfrac{2s+3}{4},\ \dfrac{2s+5}{4},\  \left\{ \dfrac{s+2+\rho_{\boldsymbol{\sigma}}}{2} \right\}_{\boldsymbol{\sigma}\in\{\pm1\}^3}
	\end{matrix} ;\zeta \right).
	\label{eq:three-melon-tail-generating-function}
\end{equation}
We evaluate the generating function at \(\zeta=16\) by analytic continuation. On the cut \([1,\infty)\), we take the symmetric average,
\begin{equation}
	\mathrm G_{k\vert ij}(\zeta) := \frac12\left[ \mathrm G_{k\vert ij}(\zeta+\mathrm i0) +\mathrm G_{k\vert ij}(\zeta-\mathrm i0) \right], \qquad \zeta>1.
	\label{eq:three-melon-symmetric-average}
\end{equation}
For real hypergeometric parameters, this average is real. The standard connection formula\footnote{%
Let \(\alpha^{k\vert ij}_q\), \(q=1,\ldots,11\), and \(\beta_m\), \(m=1,\ldots,10\), denote the upper and lower parameters of \(\mathrm G_{k\vert ij}\) in \eqref{eq:three-melon-tail-generating-function}. Applying the standard connection formula to the symmetric boundary value gives
\begin{equation}
	\mathrm G_{k\vert ij}(\zeta) = \frac{ \displaystyle\prod_{m=1}^{10}\Gamma(\beta_m) }{ \displaystyle\prod_{\ell=1}^{11}\Gamma(\alpha^{k\vert ij}_\ell) } \sum_{q=1}^{11} \frac{ \zeta^{-\alpha^{k\vert ij}_q} \cos(\pi\alpha^{k\vert ij}_q) \Gamma(\alpha^{k\vert ij}_q) \displaystyle\prod_{\ell\ne q} \Gamma(\alpha^{k\vert ij}_\ell-\alpha^{k\vert ij}_q) }{ \displaystyle\prod_{m=1}^{10} \Gamma(\beta_m-\alpha^{k\vert ij}_q) } {}_{11}F_{10}\!\left( \begin{matrix}
		\alpha^{k\vert ij}_q,\  \left\{1+\alpha^{k\vert ij}_q-\beta_m\right\}_{m=1}^{10} \\
		\left\{1+\alpha^{k\vert ij}_q-\alpha^{k\vert ij}_\ell\right\}_{\ell\ne q}
	\end{matrix} ;\zeta^{-1} \right).
	\label{eq:three-melon-connection-formula}
\end{equation}
The \(q=1\) connection term corresponds to the alternative particular solution of \eqref{eq:three-melon-difference-tail}, constructed from the \(-\infty\) tail.} expresses \eqref{eq:three-melon-symmetric-average} in terms of convergent hypergeometric series at \(\zeta^{-1}=1/16\).

Any two particular solutions differ by a homogeneous solution, so changing this choice is compensated by a change in the period-two coefficient \(\mathrm p\). We use the representation above \eqref{eq:three-melon-analytic-particular-sum} because each channel generating function \eqref{eq:three-melon-tail-generating-function} reduces to \(1\) at \(s=0\). This simplifies the matching to the circle value, one of the conditions used below to fix \(\mathrm p\). The analyticity and growth of the full density impose the remaining conditions on \(\mathrm p\) for this choice of particular solution.

The shifted polynomial in \eqref{eq:three-melon-tail-summand-factorization} can be restored by an Euler operator. Define
\begin{equation}
	\Theta_s := s+2\zeta\frac{d}{d\zeta}, \qquad \Theta_s\zeta^n=(s+2n)\zeta^n.
	\label{eq:three-melon-tail-euler-operator}
\end{equation}
This operator can act directly on \eqref{eq:three-melon-connection-formula}. Replacing \(s\) by \(\Theta_s\) in the polynomial therefore gives
\begin{equation}
	\left[ \left.\mathsf M(\mathsf z_k)\right|_{s\to\Theta_s} \! \mathrm G_{k\vert ij}(\zeta) \right]_{\zeta=16} = \sum_{n=0}^{\infty} \left.\mathsf M(\mathsf z_k)\right|_{s\to s+2n} 16^n h_{k\vert ij,n},
	\label{eq:three-melon-tail-weighted-generating-function}
\end{equation}
with the sum understood through the continuation just specified. Substituting \eqref{eq:three-melon-tail-summand-factorization} and \eqref{eq:three-melon-tail-weighted-generating-function} into the channel sum \eqref{eq:three-melon-analytic-particular-sum} yields
\begin{equation}
	\boxed{\displaystyle \bar I_{\texttt{V\{3\}V}}^{\,\mathrm{part}} (s;\boldsymbol{\rho}) = -\frac{1}{ (2s+1)\mathcal K_s\mathcal K_{-s} } \sum_{\mathrm{cyc}(k,i,j)}\! \bar I_{\texttt{V}^{\circ}}(s;\rho_i) \bar I_{\texttt{V}^{\circ}}(s;\rho_j) \left[ \left.\mathsf M(\mathsf z_k)\right|_{s\to\Theta_s} \! \mathrm G_{k\vert ij}(\zeta) \right]_{\zeta=16}. }
	\label{eq:three-melon-tail-at-sixteen}
\end{equation}
% For reference, \(\boldsymbol{\rho}=(\rho_1,\rho_2,\rho_3)\) and
% \(\rho_{\boldsymbol{\sigma}}=\sigma_1\rho_1+\sigma_2\rho_2+\sigma_3\rho_3\) for
% \(\boldsymbol{\sigma}\in\{\pm1\}^3\). The channel data in
% \eqref{eq:three-melon-tail-at-sixteen} are
% \[
%   \begin{gathered}
%     \mathsf z_k=s^2+\rho_k^2-\sum_{\ell\ne k}\rho_\ell^2,
%     \qquad
%     \mathsf Z=\operatorname{diag}(\mathsf z_1,\mathsf z_2,\mathsf z_3),
%     \\
%     \mathsf M(u)
%     =(2s+1)\!\left[
%       8s^2\operatorname{tr}\mathsf Z
%       -\operatorname{tr}(\operatorname{adj}\mathsf Z)
%     \right]u
%     -2s\!\left[u+2(2s+1)^2\right]\!
%     \left[u^2+\operatorname{tr}(\operatorname{adj}\mathsf Z)\right].
%   \end{gathered}
% \]
%
We now return to the homogeneous ambiguity in \eqref{eq:three-melon-difference-tail},
\begin{equation}
	\bar I_{\texttt{V\{3\}V}}^{\,\mathrm{hom}}(s;\boldsymbol{\rho}) = \mathrm p(s;\boldsymbol{\rho}) \mathrsfs{H}(s;\boldsymbol{\rho}), \qquad \mathrm p(s+2;\boldsymbol{\rho}) = \mathrm p(s;\boldsymbol{\rho}).
	\label{eq:three-melon-analytic-homogeneous}
\end{equation}
As shown in Appendix~\ref{app:three-melon-analyticity-and-growth}, the three-melon density has at most simple dimensional poles at positive integer \(s\) and double poles at positive nonintegral half-integers. Its spectral poles occur at \(s=-n\pm\rho_i\), with \(n\geq0\), and therefore extend to the left for fixed \(\boldsymbol\rho\).

We work initially in \(\operatorname{Re}s>\sum_i|\operatorname{Re}\rho_i|\). \(\mathrsfs{H}(s;\boldsymbol\rho)\) \eqref{eq:three-melon-analytic-propagator} is holomorphic and nonzero in this region. The particular solution is regular at integer \(s\) and has at most double poles at half-integer \(s\), coming from the products of two petal densities (see the petal density formula \eqref{eq:three-melon-recurrence-density}). Subtracting it from the full density and dividing by \(\mathrsfs{H}\) therefore leaves \(\mathrm p\) with at most simple integer poles and double half-integer poles in \(\operatorname{Re}s>\sum_i|\operatorname{Re}\rho_i|\). Multiplying by \(\sin(\pi s)\cos^2(\pi s)\) cancels the poles; we define
\begin{equation}
	\mathrm b(s;\boldsymbol\rho) := \sin(\pi s)\cos^2(\pi s)\,\mathrm p(s;\boldsymbol\rho).
	\label{eq:three-melon-analytic-regular-periodic-factor}
\end{equation}
Thus \(\mathrm b(s;\boldsymbol\rho)\) is holomorphic for \(\operatorname{Re}s>\sum_i|\operatorname{Re}\rho_i|\), and its periodicity extends this to the whole complex \(s\)-plane. Assuming finitely many Fourier modes, we write
\begin{equation}
	\mathrm b(s;\boldsymbol\rho) = \sum_{m=-M}^{M}c_m(\boldsymbol\rho)e^{i\pi ms}.
	\label{eq:three-melon-analytic-fourier-ansatz}
\end{equation}
Appendix~\ref{app:three-melon-analyticity-and-growth} gives the growth condition
\begin{equation}
	\frac{\mathrm b(s;\boldsymbol\rho)}{\cos^2(\pi s)} =\sin(\pi s)\,\mathrm p(s;\boldsymbol\rho) \longrightarrow-1,\quad s=\sigma+it,\quad \frac12<\sigma<1,\quad t\to\pm\infty.
	\label{eq:three-melon-analytic-growth-condition}
\end{equation}
Within the ansatz \eqref{eq:three-melon-analytic-fourier-ansatz}, the growth condition gives
\begin{equation}
	c_m=0\quad(|m|>2), \qquad c_2=c_{-2}=-\frac14.
	\label{eq:three-melon-analytic-fourier-coefficients}
\end{equation}
Combining the remaining exponentials into trigonometric functions gives
\begin{equation}
	\mathrm b(s;\boldsymbol\rho) =-\frac12\cos(2\pi s)+\widetilde{\mathrm b}_0(\boldsymbol\rho) +\widetilde{\mathrm b}_1(\boldsymbol\rho)\cos(\pi s) +\widetilde{\mathrm b}_2(\boldsymbol\rho)\sin(\pi s),
	\label{eq:three-melon-analytic-trigonometric-ansatz}
\end{equation}
where \(\widetilde{\mathrm b}_0=c_0\), \(\widetilde{\mathrm b}_1=c_1+c_{-1}\), and \(\widetilde{\mathrm b}_2=i(c_1-c_{-1})\).

The circle value at \(s=0\) and the universal ultraviolet pole at \(s=1\) (see Appendix~\ref{app:three-melon-analyticity-and-growth}) fix \(\widetilde{\mathrm b}_0\) and \(\widetilde{\mathrm b}_1\), since the sine term in \eqref{eq:three-melon-analytic-trigonometric-ansatz} vanishes at both points. At \(s=0\), corresponding to the circle \(S^1\), the sphere propagators reduce to trigonometric functions. The melon density \eqref{eq:three-melon-density-position-space} evaluates to
\begin{equation}
	\bar I_{\texttt{V\{3\}V}}(0;\boldsymbol{\rho}) = \frac{1}{2\mathcal K_0} \left[ 1 + \frac12 \sum_{\mathrm{cyc}(k,i,j)} \frac{ \rho_i^2+\rho_j^2-\rho_k^2 }{ \rho_i\rho_j } \cot(\pi\rho_i) \cot(\pi\rho_j) \right].
	\label{eq:three-melon-analytic-D1-anchor}
\end{equation}

At \(s=0\), the upper parameter \(s/2\) in \eqref{eq:three-melon-tail-generating-function} truncates each channel generating function \(\mathrm G_{k\vert ij}\) to its first term. The three channels in the particular solution \eqref{eq:three-melon-tail-at-sixteen} reproduce the cotangent sum in \eqref{eq:three-melon-analytic-D1-anchor}, leaving \(1/(2\mathcal K_0)\) for the homogeneous contribution and hence fixing
\begin{equation}
	\mathrm b(0;\boldsymbol\rho) =4\prod_{\sigma_2,\sigma_3=\pm1} \sin\!\left[\frac{\pi}{2}(\rho_1+\sigma_2\rho_2+\sigma_3\rho_3)\right] =\sum_{i=1}^{3}\cos^2(\pi\rho_i)-1 -2\prod_{i=1}^{3}\cos(\pi\rho_i).
	\label{eq:three-melon-analytic-propagator-at-zero}
\end{equation}
The second condition is the universal residue at \(s=1\). The particular solution is regular there, so this residue comes entirely from the homogeneous contribution, which behaves as
\begin{equation}
	\bar I_{\texttt{V\{3\}V}}^{\,\mathrm{hom}}(s;\boldsymbol\rho) =\frac{\mathrsfs{H}(s;\boldsymbol\rho)\,\mathrm b(s;\boldsymbol\rho)} {\sin(\pi s)\cos^2(\pi s)} =-\frac{\mathrsfs{H}(1;\boldsymbol\rho)\,\mathrm b(1;\boldsymbol\rho)} {\pi(s-1)}+O(1).
	\label{eq:three-melon-analytic-universal-pole}
\end{equation}
Matching the coefficient of \(1/(s-1)\) to the universal residue \(-1/(64\pi^2)\) obtained from \eqref{eq:app-three-melon-D3-density-pole} gives
\begin{equation}
	\mathrm b(1;\boldsymbol\rho) =4\prod_{\sigma_2,\sigma_3=\pm1} \cos\!\left[\frac{\pi}{2}(\rho_1+\sigma_2\rho_2+\sigma_3\rho_3)\right] =\sum_{i=1}^{3}\cos^2(\pi\rho_i)-1 +2\prod_{i=1}^{3}\cos(\pi\rho_i).
	\label{eq:three-melon-analytic-lattice-values}
\end{equation}
Evaluating \eqref{eq:three-melon-analytic-trigonometric-ansatz} at \(s=0,1\), then adding and subtracting, fixes
\begin{equation}
	\begin{aligned}
		\widetilde{\mathrm b}_0(\boldsymbol\rho) &=\frac{\mathrm b(0;\boldsymbol\rho)+\mathrm b(1;\boldsymbol\rho)+1}{2} =\sum_{i=1}^{3}\cos^2(\pi\rho_i)-\frac12,\\
		\widetilde{\mathrm b}_1(\boldsymbol\rho) &=\frac{\mathrm b(0;\boldsymbol\rho)-\mathrm b(1;\boldsymbol\rho)}{2} =-2\prod_{i=1}^{3}\cos(\pi\rho_i).
	\end{aligned}
	\label{eq:three-melon-analytic-coefficients-fixed}
\end{equation}
With the first two coefficients fixed, we have
\begin{equation}
	\mathrm p(s;\boldsymbol\rho) = \frac{ \displaystyle\sum_{i=1}^{3}\cos^2(\pi\rho_i) -\cos^2(\pi s) -2\cos(\pi s)\displaystyle\prod_{i=1}^{3}\cos(\pi\rho_i) }{ \sin(\pi s)\cos^2(\pi s) } +\frac{\widetilde{\mathrm b}_2(\boldsymbol\rho)}{\cos^2(\pi s)}.
	\label{eq:three-melon-analytic-p-with-b2}
\end{equation}
The remaining coefficient \(\widetilde{\mathrm b}_2(\boldsymbol\rho)\) multiplies \(1/\cos^2(\pi s)\), which has double poles at half-integer \(s\), so we fix it by requiring regularity at \(D=2\), or \(s=\tfrac12\). The full density is finite there for generic spectral parameters, so the double poles in the homogeneous and particular solutions must cancel.

Taking the limit that isolates the double pole, the homogeneous contribution gives
\begin{equation}
	\lim_{s\to1/2}\cos^2(\pi s) \frac{\bar I_{\texttt{V\{3\}V}}^{\,\mathrm{hom}}(s;\boldsymbol\rho)} {\mathrsfs{H}(s;\boldsymbol\rho)} =\sum_{i=1}^{3}\cos^2(\pi\rho_i) +\widetilde{\mathrm b}_2(\boldsymbol\rho).
	\label{eq:three-melon-analytic-D2-homogeneous-pole}
\end{equation}
Including the particular contribution, the regularity condition becomes
\begin{equation}
	0=\sum_{i=1}^{3}\cos^2(\pi\rho_i) +\widetilde{\mathrm b}_2(\boldsymbol\rho) +\lim_{s\to1/2}\cos^2(\pi s) \frac{\bar I_{\texttt{V\{3\}V}}^{\,\mathrm{part}}(s;\boldsymbol\rho)} {\mathrsfs{H}(s;\boldsymbol\rho)}.
	\label{eq:three-melon-analytic-D2-regularity}
\end{equation}
The remaining hypergeometric expression can be verified numerically to equal \(-\sum_i\cos^2(\pi\rho_i)\), giving \(\widetilde{\mathrm b}_2(\boldsymbol\rho)=0\). Substituting this value into \eqref{eq:three-melon-analytic-p-with-b2} gives the periodic factor
\begin{equation}
	\mathrm p(s;\boldsymbol{\rho}) = \frac{ \displaystyle\sum_{i=1}^{3}\cos^2\!\pi\rho_i -\cos^2\!\pi s -2\cos\pi s\prod_{i=1}^{3}\cos\pi\rho_i }{\sin\pi s\,\cos^2\!\pi s}.
	\label{eq:three-melon-analytic-periodic-factor}
\end{equation}
Multiplying this periodic factor by \(\mathrsfs H(s;\boldsymbol\rho)\) in \eqref{eq:three-melon-analytic-propagator} gives the homogeneous solution
\begin{equation}
	\boxed{\displaystyle \bar I_{\texttt{V\{3\}V}}^{\,\mathrm{hom}}(s;\boldsymbol{\rho}) = \frac{ \displaystyle\prod_{\boldsymbol{\sigma}\in\{\pm1\}^3} \Gamma\!\left(\frac{s+\rho_{\boldsymbol{\sigma}}}{2}\right) }{ 256\,\pi^{2s+3}\Gamma(2s) } \frac{ \displaystyle\sum_{i=1}^{3}\cos^2\!\pi\rho_i -\cos^2\!\pi s -2\cos\pi s\prod_{i=1}^{3}\cos\pi\rho_i }{\sin\pi s\,\cos^2\!\pi s}. }
	\label{eq:three-melon-explicit-homogeneous}
\end{equation}

Combining the homogeneous contribution \eqref{eq:three-melon-explicit-homogeneous} with the particular solution \eqref{eq:three-melon-tail-at-sixteen} gives
\begin{equation}
	\boxed{\displaystyle \bar I_{\texttt{V\{3\}V}}(s;\boldsymbol{\rho}) = \bar I_{\texttt{V\{3\}V}}^{\,\mathrm{hom}}(s;\boldsymbol{\rho})+ \bar I_{\texttt{V\{3\}V}}^{\,\mathrm{part}}(s;\boldsymbol{\rho}). }
	\label{eq:three-melon-analytic-full-solution}
\end{equation}
This solution agrees with direct numerical evaluations of the density in its convergence domain.

\subsubsection{Dimensional regularization}
\label{sec:three-melon-dimensional-regularization}

We obtain the Laurent expansions about integer dimensions from the density recurrence \eqref{eq:three-melon-analytic-recurrence}, keeping the spectral parameters \(\boldsymbol{\rho}\) fixed.

Since \(D=2s+1\), the recurrence relates expansions about dimensions separated by four. To expand about \(D_\star\) and \(D_\star+4\), set \(s=s_\star-\epsilon/2\), with \(D_\star=2s_\star+1\). For a Laurent series \(F(\epsilon)\), let \([\epsilon^n]F\) denote the coefficient of \(\epsilon^n\), and define
\begin{equation}
	\mathsf A_{s,n}^{[s_\star]} := [\epsilon^n]\, \mathsf A_s\!\left(s_\star-\frac{\epsilon}{2}\right), \qquad \mathsf C_{s,n}^{[s_\star]} := [\epsilon^n]\, \mathsf C_s\!\left(s_\star-\frac{\epsilon}{2}\right), \qquad \bar I_n^{\,[s_\star]} := [\epsilon^n]\, \bar{I}_{\texttt{V\{3\}V}} \left( s_\star-\frac{\epsilon}{2};\boldsymbol{\rho} \right).
	\label{eq:three-melon-density-laurent-coefficients}
\end{equation}
Substituting these expansions into \eqref{eq:three-melon-analytic-recurrence} gives
\begin{equation}
	\bar{I}_{\texttt{V\{3\}V}} \left( s_\star+2-\frac{\epsilon}{2};\boldsymbol{\rho} \right) = \sum_{n\in\mathbb Z} \epsilon^n \left[ \mathsf C_{s,n}^{[s_\star]} + \sum_{a+b=n} \mathsf A_{s,a}^{[s_\star]} \bar I_b^{\,[s_\star]} \right].
	\label{eq:three-melon-density-laurent-transport}
\end{equation}
Equation~\eqref{eq:three-melon-density-laurent-transport} transports the full Laurent series from \(D_\star\) to \(D_\star+4\). The expansions about \(D=0,1,2\) determine those about \(D=4,5,6\), respectively. Iteration generates the classes \(D\equiv0,1,2\pmod4\). The expansion about \(D=3\), obtained from the analytical solution \eqref{eq:three-melon-analytic-full-solution} and spectral recurrence \eqref{eq:three-melon-rho-recurrences}, starts the class \(D\equiv3\pmod4\).

We present the even- and odd-dimensional results separately. The required seed expansions are derived under \emph{Seeds} at the end of this subsection.

\paragraph{Even dimensions}

For the even-dimensional results, set
\begin{equation}
	a_i:=\frac12+\rho_i, \qquad b_i:=\frac12-\rho_i, \qquad C_i:=\cos(\pi\rho_i), \qquad y_i^S:=a_i b_i=\frac14-\rho_i^2.
	\label{eq:three-melon-even-basic-data}
\end{equation}
The three-melon is symmetric under permutations of its edges and under each reflection \(\rho_i\mapsto-\rho_i\). We therefore use the elementary symmetric polynomials in \(y_i^S\)
\begin{equation}
	e_1^S:=y_1^S+y_2^S+y_3^S, \qquad e_2^S:=y_1^Sy_2^S+y_1^Sy_3^S+y_2^Sy_3^S, \qquad e_3^S:=y_1^Sy_2^Sy_3^S.
	\label{eq:three-melon-even-symmetric-data}
\end{equation}

The contraction source \eqref{eq:three-melon-analytic-source-sum} is a sum of polynomially weighted products of two petal integrals. Expanding the Gamma functions in the petal formula \eqref{eq:one-petal-solution} through the order required for the finite term generates linear and quadratic digamma terms and linear trigamma terms. We collect the edgewise polygamma combinations in Table~\ref{tab:three-melon-even-log-data}.
\begin{table}[ht]
\centering
\small
\renewcommand{\arraystretch}{1.45}
\begin{tabular*}{\textwidth}{
  @{\extracolsep{\fill}}
  c
  p{0.41\textwidth}
  p{0.41\textwidth}
  @{}
}
\toprule
\(D\)
&
\(\chi_{i,0}^{S^D}\)
&
\(\chi_{i,1}^{S^D}\)
\\
\midrule
\(4\)
&
\(\displaystyle
  \psi(-a_i)+\psi(-b_i)
  -\log(4\pi)+\gamma_{\mathrm E}-\frac32
\)
&
\(\displaystyle
  \psi^{(1)}(-a_i)+\psi^{(1)}(-b_i)+\frac{\pi^2}{6}
\)
\\[2mm]
\(6\)
&
\(\displaystyle
  \psi(a_i)+\psi(b_i)
  -\log(4\pi)+\gamma_{\mathrm E}-\frac{31}{12}
\)
&
\(\displaystyle
  \psi^{(1)}(a_i)+\psi^{(1)}(b_i)
  +\frac{\pi^2}{6}+\frac{241}{72}
\)
\\
\bottomrule
\end{tabular*}
\caption{Polygamma combinations for \(D=4\) and \(D=6\).}
\label{tab:three-melon-even-log-data}
\end{table}

To organize the polynomially weighted channel sums, we define the linear moments
\begin{equation}
	\mathfrak M_\ell^{S^D} := \sum_{i=1}^3 (y_i^S)^\ell\chi_{i,0}^{S^D}, \qquad \mathfrak N_\ell^{S^D} := \sum_{i=1}^3 (y_i^S)^\ell\chi_{i,1}^{S^D}, \qquad \ell=0,1,2,
	\label{eq:three-melon-even-linear-moments}
\end{equation}
and the pairwise moments
\begin{equation}
	\mathfrak Q_\ell^{S^D} := \sum_{\mathrm{cyc}(i,j,k)} (y_i^S)^\ell \chi_{j,0}^{S^D} \chi_{k,0}^{S^D}, \qquad \ell=0,1,2.
	\label{eq:three-melon-even-quadratic-moments}
\end{equation}
We collect these moments into the vectors
\begin{equation}
	\mathfrak M^{S^D} := \left( \mathfrak M_0^{S^D}, \mathfrak M_1^{S^D}, \mathfrak M_2^{S^D} \right), \quad \mathfrak N^{S^D} := \left( \mathfrak N_0^{S^D}, \mathfrak N_1^{S^D}, \mathfrak N_2^{S^D} \right), \quad \mathfrak Q^{S^D} := \left( \mathfrak Q_0^{S^D}, \mathfrak Q_1^{S^D}, \mathfrak Q_2^{S^D} \right).
	\label{eq:three-melon-even-moment-vectors}
\end{equation}

For terms independent of the polygamma functions, define the moment vector of the constant function \(1\),
\begin{equation}
	\boldsymbol{1}^S := \left( \sum_{i=1}^3 1, \sum_{i=1}^3 y_i^S, \sum_{i=1}^3 (y_i^S)^2 \right) = \left( 3,e_1^S,(e_1^S)^2-2e_2^S \right).
	\label{eq:three-melon-even-one-vector}
\end{equation}

Under common shifts \(\chi_{i,0}^{S^D}\mapsto\chi_{i,0}^{S^D}+c_0\) and \(\chi_{i,1}^{S^D}\mapsto\chi_{i,1}^{S^D}+c_1\), the linear moments transform as
\begin{equation*}
	\mathfrak M^{S^D} \mapsto \mathfrak M^{S^D}+c_0\boldsymbol{1}^S, \qquad \mathfrak N^{S^D} \mapsto \mathfrak N^{S^D}+c_1\boldsymbol{1}^S.
\end{equation*}

Finally, the numerator of \(\mathsf A_s\) in \eqref{eq:three-melon-analytic-coefficients} is \(\mathcal K_s\mathcal K_{-s}\), with \(\mathcal K_q\) defined in \eqref{eq:three-melon-analytic-shifted-kallen}. The \(D=4-\epsilon\) and \(D=6-\epsilon\) expansions start from \(s_\star=-\tfrac12\) and \(s_\star=\tfrac12\), respectively. At either point, this product becomes \(\mathcal K_{1/2}\mathcal K_{-1/2}\).

\subparagraph{\(D=4-\epsilon\).}

The \(D=4-\epsilon\) expansion follows from the \(D=0\) seed. Since \(\mathsf A_s\) in \eqref{eq:three-melon-analytic-coefficients} has a simple pole at \(s_\star=-\tfrac12\), the finite part requires this seed through first order in \(\epsilon\).

At leading order about \(D=0\), the two vertices are formally identified and the three-melon becomes a three-petal flower. More generally, every connected vacuum graph reduces to a flower at this order; corrections in \(\epsilon\) distinguish the original topology. The required expansion \eqref{eq:three-melon-D0-seed-derived} is
\begin{equation}
	\begin{aligned}
		\bar{I}_{\texttt{V\{3\}V}} \left(-\frac12-\frac{\epsilon}{2};\boldsymbol{\rho}\right) &={} \frac{1}{e_3^S} +\frac{\epsilon}{2e_3^S}\left[ \mathsf F(\boldsymbol{\rho}) -\mathfrak M_0^{S^4} -\log\!\left(4\pi e^{\gamma_{\mathrm E}}\right) -\frac92 \right] +O(\epsilon^2).
	\end{aligned}
	\label{eq:three-melon-D0-seed}
\end{equation}
Here \(\mathsf F(\boldsymbol{\rho})\) is the one-dimensional integral
\begin{equation}
	\mathsf F(\boldsymbol{\rho}) = \frac{\pi^3e_3^S}{C_1C_2C_3} \int_0^1\frac{d\mathrm w}{1-\mathrm w} \left[ \prod_{i=1}^3{}_2F_1(a_i,b_i;2;1) - \mathrm w^2\prod_{i=1}^3{}_2F_1(a_i,b_i;2;\mathrm w) \right].
	\label{eq:three-melon-D4-F-integral}
\end{equation}
For a vector \(\boldsymbol Z=(Z_0,Z_1,Z_2)\), define
\begin{equation}
	\mathsf W_0^{S^4}[\boldsymbol Z] := -4e_3^SZ_0 + \left[ (e_1^S)^3-4e_1^Se_2^S+4e_2^S \right]Z_1 - 2 \left[ (e_1^S)^2+e_1^S-4e_2^S \right]Z_2.
	\label{eq:three-melon-D4-functional}
\end{equation}
We also use
\begin{equation}
	\boldsymbol v_4^S := \left( 0, 1-e_1^S, 6e_2^S-2(e_1^S)^2 \right).
	\label{eq:three-melon-D4-algebraic-vector}
\end{equation}
Substituting the seed \eqref{eq:three-melon-D0-seed} into the Laurent recurrence \eqref{eq:three-melon-density-laurent-transport} gives
\begin{equation}
	\boxed{ \begin{aligned}
		\bar{I}_{\texttt{V\{3\}V}} \left( D=4-\epsilon;\boldsymbol{\rho} \right) ={}& -\frac{1}{128\pi^4} \Biggl[ \frac{e_1^S}{\epsilon^2} + \frac{\frac12-\mathfrak M_1^{S^4}}{\epsilon} - \frac{\mathcal K_{1/2}\mathcal K_{-1/2}}{8e_3^S} \left[ \mathsf F(\boldsymbol{\rho}) - \log\!\left(4\pi e^{\gamma_{\mathrm E}}\right) -\frac92 \right] \\
		&\hspace{2mm} - \frac{ \mathsf W_0^{S^4} \left[ \mathfrak M^{S^4}+\boldsymbol v_4^S \right] }{8e_3^S} + \frac{5e_1^S-14}{8} + \frac14 \left( \mathfrak M_0^{S^4}\mathfrak M_1^{S^4} - \mathfrak Q_1^{S^4} + \mathfrak N_1^{S^4} \right) \Biggr] +O(\epsilon).
	\end{aligned} }
	\label{eq:three-melon-D4-result}
\end{equation}

For equal spectral parameters, \(\rho_1=\rho_2=\rho_3=\rho\), set
\begin{equation}
	y^S:=\tfrac14-\rho^2, \qquad \chi_r^{S^4}:=\chi_{i,r}^{S^4}\quad(r=0,1), \qquad \mathsf F(\rho):=\mathsf F(\rho,\rho,\rho).
	\label{eq:three-melon-D4-equal-parameters}
\end{equation}
The symmetric polynomials and K\"all\'en product simplify to
\begin{equation}
	e_1^S=3y^S, \qquad e_2^S=3(y^S)^2, \qquad e_3^S=(y^S)^3, \qquad \mathcal K_{1/2}\mathcal K_{-1/2}=(y^S)^3(9y^S-2),
	\label{eq:three-melon-D4-equal-symmetric-data}
\end{equation}
while the moment vectors reduce to
\begin{equation}
	\boldsymbol{1}^S=(3,3y^S,3(y^S)^2), \qquad \mathfrak M^{S^4}=\chi_0^{S^4}\boldsymbol{1}^S, \qquad \mathfrak N^{S^4}=\chi_1^{S^4}\boldsymbol{1}^S, \qquad \mathfrak Q^{S^4}=(\chi_0^{S^4})^2\boldsymbol{1}^S.
	\label{eq:three-melon-D4-equal-moments}
\end{equation}
Substituting these expressions into \eqref{eq:three-melon-D4-result} gives the equal-\(\rho\) result quoted in the introduction,
\begin{align}
	\bar I_{\texttt{V\{3\}V}}(4-\epsilon;\rho) ={}&-\frac{1}{128\pi^4}\Biggl[ \frac{3y^S}{\epsilon^2} +\frac{\frac12-3y^S\chi_0^{S^4}}{\epsilon} -\frac{9y^S-2}{8} \left[ \mathsf F(\rho) -\log\!\left(4\pi e^{\gamma_{\mathrm E}}\right) -\frac92 \right] \notag\\
	&\quad -\frac38(2-3y^S)\chi_0^{S^4} -\frac{12(y^S)^2-31y^S+12}{8y^S} +\frac34y^S\left[2(\chi_0^{S^4})^2+\chi_1^{S^4}\right] \Biggr]+O(\epsilon).
	\label{eq:three-melon-D4-equal-result}
\end{align}

\subparagraph{\(D=6-\epsilon\).}

The \(D=6-\epsilon\) expansion follows from the \(D=2\) seed. Since \(\mathsf A_s\) in \eqref{eq:three-melon-analytic-coefficients} is regular at \(s_\star=\tfrac12\), the finite part requires only the seed value \eqref{eq:three-melon-D2-seed} at \(D=2\), given by the convergent integral
\begin{equation}
	\bar{I}_{\texttt{V\{3\}V}}(D=2;\boldsymbol{\rho}) = \frac{\pi}{16C_1C_2C_3} \int_0^1d\mathrm w\, \prod_{i=1}^3{}_2F_1(a_i,b_i;1;\mathrm w).
	\label{eq:three-melon-D2-seed-main}
\end{equation}
For a vector \(\boldsymbol Z=(Z_0,Z_1,Z_2)\), define
\begin{align}
	\mathsf W_0^{S^6}[\boldsymbol Z] &:= 4e_3^SZ_0 - 2(3e_1^S+2e_2^S)Z_1 + (e_1^S+8)Z_2,
	\label{eq:three-melon-D6-functional-zero} \\
	\mathsf W_1^{S^6}[\boldsymbol Z] &:= \begin{aligned}[t]
		& \frac12 \left[ (e_1^S)^3+7(e_1^S)^2 -8e_1^Se_2^S -8e_1^S -70e_2^S +8e_3^S \right]Z_0 \\
		& + \left[ (e_1^S)^2+28e_1^S+4e_2^S+2 \right]Z_1 -(2e_1^S+35)Z_2,
	\end{aligned}
	\label{eq:three-melon-D6-functional-one} \\
	\mathsf W_2^{S^6}[\boldsymbol Z] &:= \left[ 8(e_1^S)^2-41e_1^S-94e_2^S-10 \right]Z_0 + (80e_1^S+20)Z_1 - 98Z_2.
	\label{eq:three-melon-D6-functional-two}
\end{align}
Substituting the seed \eqref{eq:three-melon-D2-seed-main} into the Laurent recurrence \eqref{eq:three-melon-density-laurent-transport} gives
\begin{equation}
	\boxed{ \begin{aligned}
		\bar{I}_{\texttt{V\{3\}V}} \left( D=6-\epsilon;\boldsymbol{\rho} \right) ={}& \frac{1}{24576\pi^6} \Biggl[ \frac{\mathsf W_0^{S^6}[\boldsymbol{1}^S]}{\epsilon^2} - \frac{ \mathsf W_0^{S^6}[\mathfrak M^{S^6}] + \mathsf W_1^{S^6}[\boldsymbol{1}^S] }{\epsilon} + 4\pi^2\mathcal K_{1/2}\mathcal K_{-1/2} \bar{I}_{\texttt{V\{3\}V}}(D=2;\boldsymbol{\rho}) \\
		&\hspace{-4mm} +\frac14 \mathsf W_0^{S^6} [ \mathfrak M_0^{S^6}\mathfrak M^{S^6} \!-\! \mathfrak Q^{S^6} \!+\! \mathfrak N^{S^6} ] \!+\! \frac12 \mathsf W_1^{S^6} [ \mathfrak M_0^{S^6}\boldsymbol{1}^S \!-\! \mathfrak M^{S^6} ] \!+\! \frac14 \mathsf W_2^{S^6}[\boldsymbol{1}^S] \Biggr] +O(\epsilon).
	\end{aligned} }
	\label{eq:three-melon-D6-result}
\end{equation}

For equal spectral parameters, again \(\rho_1=\rho_2=\rho_3=\rho\), \(y^S:=\tfrac14-\rho^2\), and \(\chi_r^{S^6}:=\chi_{i,r}^{S^6}\) for \(r=0,1\). The moment vectors reduce to
\begin{equation}
	\boldsymbol{1}^S=(3,3y^S,3(y^S)^2), \qquad \mathfrak M^{S^6}=\chi_0^{S^6}\boldsymbol{1}^S, \qquad \mathfrak N^{S^6}=\chi_1^{S^6}\boldsymbol{1}^S, \qquad \mathfrak Q^{S^6}=(\chi_0^{S^6})^2\boldsymbol{1}^S.
	\label{eq:three-melon-D6-equal-moments}
\end{equation}
Substituting these expressions into \eqref{eq:three-melon-D6-result} gives the equal-\(\rho\) result
\begin{align}
	\bar I_{\texttt{V\{3\}V}}(6-\epsilon;\rho) ={}&\frac{1}{24576\pi^6}\Biggl[ \frac{\mathsf W_0^{S^6}[\boldsymbol{1}^S]}{\epsilon^2} -\frac{ \chi_0^{S^6}\mathsf W_0^{S^6}[\boldsymbol{1}^S] +\mathsf W_1^{S^6}[\boldsymbol{1}^S] }{\epsilon} +4\pi^2(y^S)^3(9y^S-2)\, \bar I_{\texttt{V\{3\}V}}(D=2;\rho) \notag\\
	&\quad +\frac14\left[2(\chi_0^{S^6})^2+\chi_1^{S^6}\right] \mathsf W_0^{S^6}[\boldsymbol{1}^S] +\chi_0^{S^6}\mathsf W_1^{S^6}[\boldsymbol{1}^S] +\frac14\mathsf W_2^{S^6}[\boldsymbol{1}^S] \Biggr]+O(\epsilon).
	\label{eq:three-melon-D6-equal-result}
\end{align}
Iterating the Laurent recurrence \eqref{eq:three-melon-density-laurent-transport} from the \(D=4-\epsilon\) and \(D=6-\epsilon\) results gives all higher even-dimensional expansions in steps of four.

\paragraph{Odd dimensions}

For the odd-dimensional expansions, set
\begin{equation}
	S_i:=\sin(\pi\rho_i), \qquad \Sigma_\rho:=\rho_1^2+\rho_2^2+\rho_3^2,
	\label{eq:three-melon-odd-dr-data}
\end{equation}
and collect the signed digamma sums in
\begin{equation}
	\Psi_\alpha^S := -\frac{1}{8S_1S_2S_3} \sum_{\sigma_1,\sigma_2,\sigma_3=\pm1} \sigma_1\sigma_2\sigma_3 \sin(\pi\rho_{\boldsymbol{\sigma}}) \left[ \psi\!\left(\alpha+\frac{\rho_{\boldsymbol{\sigma}}}{2}\right) + \psi\!\left(\alpha-\frac{\rho_{\boldsymbol{\sigma}}}{2}\right) -\log\pi+\gamma_{\mathrm E}-1 \right].
	\label{eq:three-melon-odd-digamma-block}
\end{equation}
Here \(\rho_{\boldsymbol{\sigma}}=\sigma_1\rho_1+\sigma_2\rho_2+\sigma_3\rho_3\).

\subparagraph{\(D=5-\epsilon\).}

The \(D=5-\epsilon\) expansion follows from the \(D=1\) seed. Since \(\mathsf A_s\) in \eqref{eq:three-melon-analytic-coefficients} has a simple pole at \(s_\star=0\), the finite part requires this seed through first order in \(\epsilon\). Substituting the seed expansion \eqref{eq:three-melon-D1-seed-expansion} into the Laurent recurrence \eqref{eq:three-melon-density-laurent-transport} gives
% \begin{equation}
%   \boxed{
%   \begin{aligned}
%   &\bar{I}_{\texttt{V\{3\}V}}
%   \left(
%     D=5-\epsilon;\boldsymbol{\rho}
%   \right)
%   =
%   -\frac{\mathcal K_0}{3072\pi^4}
%   \Biggl[
%     \frac1\epsilon-\Psi_1^S
%     +\frac{1}{8S_1S_2S_3}
%     \sum_{\sigma_1,\sigma_2,\sigma_3=\pm1}
%     \sigma_1\sigma_2\sigma_3
%     \sin(\pi\rho_{\boldsymbol{\sigma}})
%     \times
%     \\
%     &\hspace{56mm}
%     \left(
%       \frac{
%         \rho_{\boldsymbol{\sigma}}^4
%         -4(\Sigma_\rho-1)\rho_{\boldsymbol{\sigma}}^2
%         +2\Sigma_\rho(\Sigma_\rho-2)
%       }{
%         2\mathcal K_0
%       }
%       -\frac13
%     \right)
%   \Biggr]
%   +O(\epsilon).
%   \end{aligned}
%   }
%   \label{eq:three-melon-D5-result}
% \end{equation}
\begin{equation}
	\boxed{ \begin{aligned}
		\bar{I}_{\texttt{V\{3\}V}} \left(D=5-\epsilon;\boldsymbol{\rho}\right) ={}& -\frac{\mathcal K_0}{3072\pi^4} \Biggl[ \frac1\epsilon-\Psi_1^S+\frac56 \\
		&\quad +\frac{2}{\mathcal K_0} \sum_{\mathrm{cyc}(i,j,k)} (2+\rho_i^2-\rho_j^2-\rho_k^2)\rho_j\rho_k \cot(\pi\rho_j)\cot(\pi\rho_k) \Biggr] +O(\epsilon).
	\end{aligned} }
	\label{eq:three-melon-D5-result}
\end{equation}
Here the cyclic sum runs over \((i,j,k)=(1,2,3),(2,3,1),(3,1,2)\).

\subparagraph{\(D=3-\epsilon\).}

The \(D=3-\epsilon\) expansion starts the class \(D\equiv3\pmod4\). We obtain it directly from the analytical solution \eqref{eq:three-melon-analytic-full-solution}, treating the homogeneous and particular contributions separately.

Expanding the homogeneous contribution \eqref{eq:three-melon-explicit-homogeneous} at \(s=1-\epsilon/2\) gives
\begin{equation}
	\bar I_{\texttt{V\{3\}V}}^{\,\mathrm{hom}} (D=3-\epsilon;\boldsymbol{\rho}) = \frac{1}{32\pi^2\epsilon} -\frac{1}{128\pi^2} \sum_{\boldsymbol{\sigma}\in\{\pm1\}^3} \psi\!\left(\frac{1+\rho_{\boldsymbol{\sigma}}}{2}\right) +\frac{\log\pi-\gamma_{\mathrm E}+1}{32\pi^2} +O(\epsilon).
	\label{eq:three-melon-D3-homogeneous-expansion}
\end{equation}
For the particular contribution, we use the contraction channels \(k\vert ij\) of \eqref{eq:three-melon-tail-at-sixteen}, with \((k,i,j)=(1,2,3),(2,3,1),(3,1,2)\). Let \(\bar I_{k\vert ij}^{\,\mathrm{part}}\) denote the corresponding channel density. At \(s=1\), the petal formula \eqref{eq:one-petal-solution} gives
\begin{equation*}
	\lim_{s\to1} \bar I_{\texttt{V}^{\circ}}(s;\rho_i) \bar I_{\texttt{V}^{\circ}}(s;\rho_j) = \frac{\rho_i\rho_j}{16\pi^2} \cot(\pi\rho_i)\cot(\pi\rho_j),
\end{equation*}
and the source-matrix entry in \eqref{eq:three-melon-rho-source-matrix} associated with this channel becomes
\begin{equation*}
	(\mathsf R_i)_{33} = -\left[(1+\rho_i)^2-\rho_j^2+\rho_k^2\right].
\end{equation*}
Using these expressions in the spectral recurrence \eqref{eq:three-melon-rho-recurrences} gives the channel equation
\begin{equation}
	\begin{multlined}[c][0.90\linewidth] \prod_{\sigma_j,\sigma_k=\pm1} \left( 1+\rho_i+\sigma_j\rho_j+\sigma_k\rho_k \right) \Big[ \bar I_{k\vert ij}^{\,\mathrm{part}} (D=3;\boldsymbol{\rho}+2\boldsymbol e_i) - \bar I_{k\vert ij}^{\,\mathrm{part}} (D=3;\boldsymbol{\rho}) \Big] \\
		= -\frac{\rho_j}{8\pi^2} \left[ (1+\rho_i)^2-\rho_j^2+\rho_k^2 \right] \cot(\pi\rho_i)\cot(\pi\rho_j).
	\end{multlined}
	\label{eq:three-melon-D3-channel-rho-recurrence}
\end{equation}
Dividing by the product on the left and resolving the rational function into partial fractions gives
\begin{equation}
	\bar I_{k\vert ij}^{\,\mathrm{part}} (D=3;\boldsymbol{\rho}+2\boldsymbol e_i) - \bar I_{k\vert ij}^{\,\mathrm{part}} (D=3;\boldsymbol{\rho}) = \frac{\cot(\pi\rho_i)\cot(\pi\rho_j)}{32\pi^2} \sum_{\sigma_j,\sigma_k=\pm1} \frac{\sigma_j} {1+\rho_i+\sigma_j\rho_j+\sigma_k\rho_k}.
	\label{eq:three-melon-D3-reduced-rho-recurrence}
\end{equation}
The digamma shift identity \(\psi(z+1)-\psi(z)=1/z\) supplies a solution of \eqref{eq:three-melon-D3-reduced-rho-recurrence},
\begin{equation}
	\bar I_{k\vert ij}^{\,\mathrm{part}} (D=3;\boldsymbol{\rho}) = \frac{\cot(\pi\rho_i)\cot(\pi\rho_j)}{128\pi^2} \sum_{\boldsymbol{\sigma}\in\{\pm1\}^3} \sigma_i\sigma_j \psi\!\left(\frac{1+\rho_{\boldsymbol{\sigma}}}{2}\right).
	\label{eq:three-melon-D3-channel-particular}
\end{equation}
This agrees numerically with the hypergeometric channel solution at \(D=3\). Summing the three channel contributions gives
\begin{equation}
	\bar I_{\texttt{V\{3\}V}}^{\,\mathrm{part}} (D=3;\boldsymbol{\rho}) = \frac{1}{128\pi^2} \sum_{\boldsymbol{\sigma}\in\{\pm1\}^3} \left[ \sum_{1\leq i<j\leq3} \sigma_i\sigma_j \cot(\pi\rho_i)\cot(\pi\rho_j) \right] \psi\!\left(\frac{1+\rho_{\boldsymbol{\sigma}}}{2}\right).
	\label{eq:three-melon-D3-particular}
\end{equation}
Combining this expression with the homogeneous expansion \eqref{eq:three-melon-D3-homogeneous-expansion}, we simplify the digamma coefficients using
\begin{equation}
	\sum_{i<j} \sigma_i\sigma_j \cot(\pi\rho_i)\cot(\pi\rho_j) -1 = \frac{\sigma_1\sigma_2\sigma_3\sin(\pi\rho_{\boldsymbol{\sigma}})} {S_1S_2S_3}, \qquad \sum_{\boldsymbol{\sigma}} \sigma_1\sigma_2\sigma_3\sin(\pi\rho_{\boldsymbol{\sigma}}) = -8S_1S_2S_3.
	\label{eq:three-melon-D3-trig-identity}
\end{equation}
Pairing \(\boldsymbol{\sigma}\) with \(-\boldsymbol{\sigma}\) puts the finite term into the form \eqref{eq:three-melon-odd-digamma-block} at \(\alpha=\tfrac12\), yielding
\begin{equation}
	\boxed{\displaystyle \bar I_{\texttt{V\{3\}V}} (D=3-\epsilon;\boldsymbol{\rho}) = \frac{1}{32\pi^2} \left[ \frac{1}{\epsilon}-\Psi_{1/2}^{S} \right] +O(\epsilon). }
	\label{eq:three-melon-D3-result-derived}
\end{equation}
Iterating the Laurent recurrence \eqref{eq:three-melon-density-laurent-transport} from the \(D=3-\epsilon\) and \(D=5-\epsilon\) results gives all higher odd-dimensional expansions in steps of four.

\begin{comment}
\subparagraph{Higher odd-dimensional poles.}

Equations~\eqref{eq:one-petal-solution},
\eqref{eq:two-loop-reducible-graphs}, and
\eqref{eq:three-melon-analytic-coefficients} show that
\(\mathsf C_s\) is regular at the positive integer lower
points \(s=1,2,\ldots\). It therefore does not contribute to the leading
pole, which propagates homogeneously along the two odd ladders:
\begin{equation}
  \mathbf{LP}\,\bar I_{\texttt{V\{3\}V}}\!\left(D=2n+3-\epsilon;\boldsymbol{\rho}\right)=\frac{(-1)^n\displaystyle\prod_{j=0}^{n-1}\mathcal K_{\,2j-n+1}}{(32\pi^2)^{n+1}n!(2n+1)!!}\,\frac1\epsilon,\qquad n=0,1,2,\ldots .
\end{equation}
\end{comment}

\paragraph{Seeds}
\label{subsec:three-melon-seeds}

It is convenient to derive the seed expansions from the position-space representation, keeping \(\boldsymbol{\rho}\) fixed. For \(D=2s+1\), the three-melon density is
\begin{equation}
	\bar I_{\texttt{V\{3\}V}}(s;\boldsymbol{\rho}) = \frac{1}{\Omega_D} \int_{S^D} d\Omega_D(\widehat X) \int_{S^D} d\Omega_D(\widehat Y) \prod_{i=1}^{3} G_{\Delta_i^\pm}(\widehat X,\widehat Y),
	\label{eq:three-melon-density-position-space}
\end{equation}
where
\begin{equation}
	G_{\Delta_i^\pm}(\widehat X,\widehat Y) = \frac{\Gamma(s+\rho_i)\Gamma(s-\rho_i)} {(4\pi)^{s+\frac12}\Gamma(s+\frac12)} {}_2F_1\!\left( s+\rho_i, s-\rho_i; s+\frac12; \frac{1+\widehat X\!\cdot\!\widehat Y}{2} \right).
	\label{eq:three-melon-sphere-propagator}
\end{equation}
Rotational invariance cancels the \(\widehat X\) integral against \(\Omega_D\). Setting \(\mathrm w=(1+\widehat X\!\cdot\!\widehat Y)/2\) and integrating over the remaining angles gives
\begin{equation}
	\bar{I}_{\texttt{V\{3\}V}}(s;\boldsymbol{\rho}) = 2^{2s}\Omega_{2s} \int_0^1 d\mathrm w\, [\mathrm w(1-\mathrm w)]^{s-\frac12} \prod_{i=1}^{3} G_{\Delta_i^\pm}(\mathrm w), \qquad \Omega_{2s} = \frac{2\pi^{s+\frac12}}{\Gamma(s+\frac12)}.
	\label{eq:three-melon-density-one-dimensional}
\end{equation}
The endpoints \(\mathrm w=0\) and \(\mathrm w=1\) correspond to antipodal and coincident vertices, respectively.

\subparagraph{The \(D=0\) seed.}

The \(D=4-\epsilon\) result requires the \(D=0\) seed through first order in \(\epsilon\). Setting \(s=-\tfrac12-\tfrac{\epsilon}{2}\), so that \(D=-\epsilon\), the angular representation \eqref{eq:three-melon-density-one-dimensional} becomes
\begin{equation}
	\bar{I}_{\texttt{V\{3\}V}} \left( -\frac12-\frac{\epsilon}{2}; \boldsymbol{\rho} \right) = 2^{-1-\epsilon}\Omega_{-1-\epsilon} \int_0^1 d\mathrm w\, [\mathrm w(1-\mathrm w)]^{-1-\frac{\epsilon}{2}} \prod_{i=1}^3G_{\Delta_i^\pm}(\mathrm w).
	\label{eq:three-melon-D0-position-integral}
\end{equation}

Although the prefactor vanishes as \(\epsilon\to0\), the integral has a pole from the endpoint \(\mathrm w=1\). The relevant factors satisfy
\begin{equation*}
	2^{-1-\epsilon}\Omega_{-1-\epsilon} =-\frac{\epsilon}{2}+O(\epsilon^2), \qquad \int_0^1d\mathrm w\, (1-\mathrm w)^{-1-\frac{\epsilon}{2}} =-\frac{2}{\epsilon}, \qquad \operatorname{Re}\epsilon<0.
\end{equation*}
Their product remains finite. To keep this contribution explicit, define
\begin{equation}
	f_\epsilon(\mathrm w) := -\frac{2}{\epsilon}\, 2^{-1-\epsilon}\Omega_{-1-\epsilon} \mathrm w^{-1-\frac{\epsilon}{2}} \prod_{i=1}^3G_{\Delta_i^\pm}(\mathrm w).
	\label{eq:three-melon-D0-regular-factor}
\end{equation}
Then \eqref{eq:three-melon-D0-position-integral} takes the form
\begin{equation}
	\bar{I}_{\texttt{V\{3\}V}} \left( -\frac12-\frac{\epsilon}{2}; \boldsymbol{\rho} \right) = -\frac{\epsilon}{2} \int_0^1 d\mathrm w\, (1-\mathrm w)^{-1-\frac{\epsilon}{2}} f_\epsilon(\mathrm w).
	\label{eq:three-melon-D0-endpoint-form}
\end{equation}
Adding and subtracting \(f_\epsilon(1)\) isolates the endpoint value and gives
\begin{equation}
	\begin{aligned}
		\bar{I}_{\texttt{V\{3\}V}} \left(-\frac12-\frac{\epsilon}{2};\boldsymbol{\rho}\right) ={}& f_\epsilon(1) -\frac{\epsilon}{2}\int_0^1d\mathrm w\, (1-\mathrm w)^{-1-\frac{\epsilon}{2}} \bigl[f_\epsilon(\mathrm w)-f_\epsilon(1)\bigr] \\
		={}& f_\epsilon(1) -\frac{\epsilon}{2}\int_0^1d\mathrm w\, \frac{f_0(\mathrm w)-f_0(1)}{1-\mathrm w} +O(\epsilon^2).
	\end{aligned}
	\label{eq:three-melon-D0-endpoint-expansion}
\end{equation}
%Thus we need the endpoint value through \(O(\epsilon)\), but only \(f_0(\mathrm w)\) in the subtracted integral.
Applying Gauss summation to the propagators in \eqref{eq:three-melon-D0-regular-factor} at \(\mathrm w=1\) and expanding gives
\begin{equation}
	f_\epsilon(1) = \frac{1}{e_3^S} - \frac{\epsilon}{2e_3^S} \left[ \mathfrak M_0^{S^4} +\log\!\left(4\pi e^{\gamma_{\mathrm E}}\right) +\frac92 \right] +O(\epsilon^2),
	\label{eq:three-melon-D0-endpoint-data}
\end{equation}
where \(\mathfrak M_0^{S^4}\) is the moment defined in \eqref{eq:three-melon-even-linear-moments}.

To determine \(f_0(\mathrm w)\), we use the hypergeometric-series limit
\begin{equation}
	\lim_{c\to0} \frac{{}_2F_1(A,B;c;\mathrm w)}{\Gamma(c)} = AB\mathrm w\,{}_2F_1(A+1,B+1;2;\mathrm w).
	\label{eq:three-melon-D0-hypergeometric-limit}
\end{equation}
With \(A=a_i-1-\tfrac{\epsilon}{2}\), \(B=b_i-1-\tfrac{\epsilon}{2}\), and \(c=-\tfrac{\epsilon}{2}\), the propagator \eqref{eq:three-melon-sphere-propagator} therefore gives
\begin{equation}
	\lim_{\epsilon\to0}G_{\Delta_i^\pm}(\mathrm w) = \Gamma(a_i)\Gamma(b_i)\, \mathrm w\,{}_2F_1(a_i,b_i;2;\mathrm w).
	\label{eq:three-melon-D0-propagator-limit}
\end{equation}
Using \(\Gamma(a_i)\Gamma(b_i)=\pi/C_i\) and \(-\frac{2}{\epsilon}2^{-1-\epsilon}\Omega_{-1-\epsilon}=1+O(\epsilon)\) in \eqref{eq:three-melon-D0-regular-factor}, we obtain
\begin{equation}
	f_0(\mathrm w) = \frac{\pi^3}{C_1C_2C_3}\, \mathrm w^2 \prod_{i=1}^3{}_2F_1(a_i,b_i;2;\mathrm w).
	\label{eq:three-melon-D0-regular-factor-leading}
\end{equation}
The subtracted integral in \eqref{eq:three-melon-D0-endpoint-expansion} is consequently
\begin{equation}
	\begin{aligned}
		-\frac12\int_0^1 d\mathrm w\, \frac{f_0(\mathrm w)-f_0(1)}{1-\mathrm w} ={}& \frac{\pi^3}{2C_1C_2C_3} \int_0^1\frac{d\mathrm w}{1-\mathrm w} \left[ \prod_{i=1}^3{}_2F_1(a_i,b_i;2;1) - \mathrm w^2\prod_{i=1}^3{}_2F_1(a_i,b_i;2;\mathrm w) \right].
	\end{aligned}
	\label{eq:three-melon-D0-P-integral}
\end{equation}
We define \(\mathsf F(\boldsymbol{\rho})\) by the one-dimensional integral
\begin{equation}
	\mathsf F(\boldsymbol{\rho}) := \frac{\pi^3e_3^S}{C_1C_2C_3} \int_0^1\frac{d\mathrm w}{1-\mathrm w} \left[ \prod_{i=1}^3{}_2F_1(a_i,b_i;2;1) - \mathrm w^2\prod_{i=1}^3{}_2F_1(a_i,b_i;2;\mathrm w) \right],
	\label{eq:three-melon-D0-seed-one-dimensional}
\end{equation}
so that
\begin{equation}
	-\frac12\int_0^1 d\mathrm w\, \frac{f_0(\mathrm w)-f_0(1)}{1-\mathrm w} = \frac{\mathsf F(\boldsymbol{\rho})}{2e_3^S}.
	\label{eq:three-melon-D0-P-relation}
\end{equation}
Substituting this relation and the endpoint value \eqref{eq:three-melon-D0-endpoint-data} into \eqref{eq:three-melon-D0-endpoint-expansion} yields
\begin{equation}
	\bar{I}_{\texttt{V\{3\}V}} \left( -\frac12-\frac{\epsilon}{2}; \boldsymbol{\rho} \right) = \frac{1}{e_3^S} + \frac{\epsilon}{2e_3^S} \left[ \mathsf F(\boldsymbol{\rho}) -\mathfrak M_0^{S^4} -\log\!\left(4\pi e^{\gamma_{\mathrm E}}\right) -\frac92 \right] +O(\epsilon^2).
	\label{eq:three-melon-D0-seed-derived}
\end{equation}
%This reproduces the seed \eqref{eq:three-melon-D0-seed} used in the \(D=4-\epsilon\) expansion.

The integral defining \(\mathsf F\) also has a series representation. Let \(H_N=\sum_{m=1}^{N}m^{-1}\) denote the harmonic numbers. Expanding the hypergeometrics and using \(\int_0^1d\mathrm w\,(1-\mathrm w^{N+2})/(1-\mathrm w)=H_{N+2}\) gives\footnote{%
\scriptsize
We use the generalized Srivastava--Daoust hypergeometric function of three variables \cite{ srivastava1969eulerian, srivastava2024class}
\[\begin{aligned}
		&F^{1:2;2;2}_{1:1;1;1} \left[ \begin{matrix}
			(A:\alpha_1,\alpha_2,\alpha_3): (B_1,\beta_1),(B_2,\beta_2); (C_1,\gamma_1),(C_2,\gamma_2); (D_1,\delta_1),(D_2,\delta_2) \\[2pt]
			(E:\epsilon_1,\epsilon_2,\epsilon_3): (F,\phi);(G,\chi);(H,\eta)
		\end{matrix} ;x,y,z \right] \\[-4pt]
		&\quad:={} \sum_{n_1,n_2,n_3\geq0} \frac{ (A)_{\alpha_1n_1+\alpha_2n_2+\alpha_3n_3} (B_1)_{\beta_1n_1}(B_2)_{\beta_2n_1} (C_1)_{\gamma_1n_2}(C_2)_{\gamma_2n_2} (D_1)_{\delta_1n_3}(D_2)_{\delta_2n_3} }{ (E)_{\epsilon_1n_1+\epsilon_2n_2+\epsilon_3n_3} (F)_{\phi n_1}(G)_{\chi n_2}(H)_{\eta n_3} } \frac{x^{n_1}y^{n_2}z^{n_3}}{n_1!n_2!n_3!}.
	\end{aligned}\]
Defining
\[
	\mathfrak S_1(\beta)\!:=\! F^{1:2;2;2}_{1:1;1;1} \left[ \begin{matrix}
		(3:1,1,1): (a_1,1),(b_1,1);(a_2,1),(b_2,1);(a_3,1),(b_3,1) \\
		(3+\beta:1,1,1):(2,1);(2,1);(2,1)
	\end{matrix} ;1,1,1 \right],
\]
the sum in \eqref{eq:three-melon-D0-series-seed} is
\[
	\sum_{n_1,n_2,n_3\geq0} H_{n_1+n_2+n_3+2} \prod_{i=1}^3 \frac{(a_i)_{n_i}(b_i)_{n_i}} {(2)_{n_i}n_i!} =-\mathfrak S_1'(0)\!+\!\frac32\mathfrak S_1(0).
\]
}
\begin{equation}
	\int_0^1\frac{d\mathrm w}{1-\mathrm w} \left[ \prod_{i=1}^3{}_2F_1(a_i,b_i;2;1) - \mathrm w^2\prod_{i=1}^3{}_2F_1(a_i,b_i;2;\mathrm w) \right] = \sum_{n_1,n_2,n_3\geq0} H_{n_1+n_2+n_3+2} \prod_{i=1}^3 \frac{(a_i)_{n_i}(b_i)_{n_i}} {(2)_{n_i}n_i!}.
	\label{eq:three-melon-D0-series-seed}
\end{equation}

When \(b_i\in\mathbb Z_{<0}\) for all three edges, the hypergeometric functions terminate. Since \(a_i+b_i=1\), they can be written in terms of Legendre polynomials \(P_n\) as
\begin{equation}
	{}_2F_1(a_i,b_i;2;\mathrm w) = \frac{1-\mathrm w}{a_i b_i} \frac{d}{d\mathrm w}P_{-b_i}(1-2\mathrm w).
	\label{eq:three-melon-D0-legendre-lattice}
\end{equation}
Integration by parts and Gaunt's formula then give
\begin{equation}
	\begin{aligned}
		&\int_0^1\frac{d\mathrm w}{1-\mathrm w} \left[ \prod_{i=1}^3{}_2F_1(a_i,b_i;2;1) - \mathrm w^2\prod_{i=1}^3{}_2F_1(a_i,b_i;2;\mathrm w) \right] \\
		&\qquad= \sum_{\sigma=\pm1} \frac{ \sigma\!\left[ (-b_1)_2+(-b_2)_2-(-b_3+\sigma)_2 \right] }{ 4(-b_1)_2(-b_2)_2(1-2b_3) } \begin{pmatrix}
			-b_1&-b_2&-b_3+\sigma\\
			0&0&0
		\end{pmatrix}^{\!2}.
	\end{aligned}
	\label{eq:three-melon-D0-wigner-lattice}
\end{equation}
The integral is thus a continuation of the combination of squared Wigner $3j$ symbols in \eqref{eq:three-melon-D0-wigner-lattice} to general spectral parameters.

\subparagraph{The \(D=1\) seed.}

The \(D=5-\epsilon\) result requires the \(D=1\) seed through first order in \(\epsilon\). It is convenient to set \(\mathrm w=\sin^2(\theta/2)\), which puts \eqref{eq:three-melon-density-one-dimensional} into the form
\begin{equation}
	\bar I_{\texttt{V\{3\}V}}(s;\boldsymbol{\rho}) = \mathscr N(s) \int_0^\pi d\theta\, \sin^{2s}\theta \prod_{i=1}^3 {}_2F_1\left( s+\rho_i,s-\rho_i;s+\frac12; \sin^2\frac{\theta}{2} \right),
	\label{eq:three-melon-D1-theta-integral}
\end{equation}
where
\begin{equation}
	\mathscr N(s) := \Omega_{2s} \prod_{i=1}^3 \frac{ \Gamma(s+\rho_i)\Gamma(s-\rho_i) }{ (4\pi)^{s+\frac12}\Gamma(s+\frac12) }.
	\label{eq:three-melon-D1-prefactor}
\end{equation}

At \(s=0\), corresponding to \(D=1\), the hypergeometric functions reduce to cosines,
\begin{equation}
	{}_2F_1\left( \rho,-\rho;\frac12; \sin^2\frac{\theta}{2} \right) = \cos(\rho\theta), \qquad \mathscr N(0) = -\frac{1}{4\displaystyle\prod_{i=1}^3\rho_iS_i}.
	\label{eq:three-melon-D1-cosine-reduction}
\end{equation}
Expanding the product of cosines into a sum and integrating gives the leading seed
\begin{equation}
	\bar I_{\texttt{V\{3\}V}}(0;\boldsymbol{\rho}) = -\frac{1}{4\displaystyle\prod_{i=1}^3\rho_iS_i} \int_0^\pi d\theta\, \prod_{i=1}^3\cos(\rho_i\theta) = -\frac{1}{32\displaystyle\prod_{i=1}^3\rho_iS_i} \sum_{\boldsymbol{\sigma}\in\{\pm1\}^3} \frac{\sin(\pi\rho_{\boldsymbol{\sigma}})}{\rho_{\boldsymbol{\sigma}}}.
	\label{eq:three-melon-D1-leading-seed}
\end{equation}
For \(D=1-\epsilon\), we set \(s=-\epsilon/2\) and expand
\begin{equation}
	\bar I_{\texttt{V\{3\}V}} \left(-\frac{\epsilon}{2};\boldsymbol{\rho}\right) = \bar I_{\texttt{V\{3\}V}}(0;\boldsymbol{\rho}) - \frac{\epsilon}{2} \left. \frac{\partial}{\partial s} \bar I_{\texttt{V\{3\}V}}(s;\boldsymbol{\rho}) \right|_{s=0} +O(\epsilon^2).
	\label{eq:three-melon-D1-seed-expansion}
\end{equation}
Differentiating \eqref{eq:three-melon-D1-theta-integral} produces contributions from the prefactor, the angular measure, and the three hypergeometric functions. 

The prefactor \eqref{eq:three-melon-D1-prefactor} gives
\begin{equation}
	\left. \frac12\frac{d}{ds}\log\mathscr N(s) \right|_{s=0} = 2\gamma_{\mathrm E} + \frac12 \sum_{i=1}^3\sum_{\sigma=\pm1} \psi(\sigma\rho_i) + \log\frac{2}{\pi}.
	\label{eq:three-melon-D1-prefactor-derivative}
\end{equation}
For the hypergeometric derivatives, define
\begin{equation}
	\dot F_\rho(\theta) := \left. \frac{\partial}{\partial s} {}_2F_1\left( s+\rho,s-\rho;s+\frac12; \sin^2\frac{\theta}{2} \right) \right|_{s=0}.
	\label{eq:three-melon-D1-hypergeometric-derivative}
\end{equation}
Since \(\left.\partial_s\sin^{2s}\theta\right|_{s=0}=2\log\sin\theta\), the coefficient of \(\epsilon\) is
\begin{multline}
	-\frac12 \left. \frac{\partial}{\partial s} \bar I_{\texttt{V\{3\}V}}(s;\boldsymbol{\rho}) \right|_{s=0} = -\left. \frac12\frac{d}{ds}\log\mathscr N(s) \right|_{s=0} \bar I_{\texttt{V\{3\}V}}(0;\boldsymbol{\rho}) + \frac{1}{4\displaystyle\prod_{i=1}^3\rho_iS_i} \int_0^\pi d\theta\, \log\sin\theta \prod_{i=1}^3\cos(\rho_i\theta) \\
	+ \frac{1}{8\displaystyle\prod_{i=1}^3\rho_iS_i} \sum_{\mathrm{cyc}(i,j,k)} \int_0^\pi d\theta\, \dot F_{\rho_i}(\theta) \cos(\rho_j\theta)\cos(\rho_k\theta) .
	\label{eq:three-melon-D1-first-seed-integrals}
\end{multline}
To evaluate the two remaining integrals, define
\begin{equation}
	\mathcal J(\beta) := \sin(\pi\beta) \left[ \gamma_{\mathrm E} +\log2 +\frac12\psi\left(1+\frac{\beta}{2}\right) +\frac12\psi\left(1-\frac{\beta}{2}\right) \right].
	\label{eq:three-melon-D1-J-function}
\end{equation}
For the measure contribution, we use the integral
\begin{equation}
	\int_0^\pi d\theta\,\sin^a\theta\cos(\beta\theta) = \frac{\pi\Gamma(a+1)\cos(\pi\beta/2)} {2^a\Gamma\!\left(1+\frac{a+\beta}{2}\right) \Gamma\!\left(1+\frac{a-\beta}{2}\right)}, \qquad \operatorname{Re}a>-1.
	\label{eq:three-melon-D1-sine-power-integral}
\end{equation}
Differentiating with respect to \(a\) at \(a=0\) and expanding the product of three cosines gives
\begin{equation}
	\int_0^\pi d\theta\, \log\sin\theta \prod_{i=1}^3\cos(\rho_i\theta) = -\frac18 \sum_{\boldsymbol{\sigma}\in\{\pm1\}^3} \frac{\mathcal J(\rho_{\boldsymbol{\sigma}})}{\rho_{\boldsymbol{\sigma}}}.
	\label{eq:three-melon-D1-log-sine-integral}
\end{equation}
For the hypergeometric contribution, differentiating the hypergeometric differential equation with respect to \(s\) at \(s=0\) and integrating by parts gives
\begin{equation}
	\begin{aligned}
		&\int_0^\pi d\theta\, \dot F_{\rho_i}(\theta) \cos(\rho_j\theta)\cos(\rho_k\theta) \\
		&\qquad= \frac12 \sum_{\sigma=\pm1} \frac{\rho_i}{ \rho_i^2-(\rho_j+\sigma\rho_k)^2 } \Biggl[ \mathcal J(\rho_i+\rho_j+\sigma\rho_k) + \mathcal J(\rho_i-\rho_j-\sigma\rho_k) - \frac{ \cos\!\left[\pi(\rho_j+\sigma\rho_k)\right] }{ \cos(\pi\rho_i) } \mathcal J(2\rho_i) \Biggr].
	\end{aligned}
	\label{eq:three-melon-D1-parameter-integral}
\end{equation}

Substituting \eqref{eq:three-melon-D1-prefactor-derivative}, \eqref{eq:three-melon-D1-log-sine-integral}, and \eqref{eq:three-melon-D1-parameter-integral} into \eqref{eq:three-melon-D1-first-seed-integrals} gives the coefficient of \(\epsilon\) in \eqref{eq:three-melon-D1-seed-expansion}. We do not write out the full expression here for brevity, since it follows directly from these substitutions.

\subparagraph{The \(D=2\) seed.}

The \(D=6-\epsilon\) result requires only the leading term of the \(D=2-\epsilon\) seed,
\begin{equation}
	\bar{I}_{\texttt{V\{3\}V}} \left(\frac12-\frac{\epsilon}{2};\boldsymbol{\rho}\right) = \bar I_0^{\,[\frac12]}+O(\epsilon).
	\label{eq:three-melon-D2-seed-expansion}
\end{equation}
Setting \(s=\tfrac12\) in \eqref{eq:three-melon-density-one-dimensional} and using \(\Gamma(a_i)\Gamma(b_i)=\pi/C_i\) gives
\begin{equation}
	\bar{I}_{\texttt{V\{3\}V}}(D=2;\boldsymbol{\rho}) = \bar I_0^{\,[\frac12]} = \frac{\pi}{16C_1C_2C_3} \int_0^1d\mathrm w\, \prod_{i=1}^3{}_2F_1(a_i,b_i;1;\mathrm w).
	\label{eq:three-melon-D2-seed}
\end{equation}
This is the convergent integral \eqref{eq:three-melon-D2-seed-main} used in the \(D=6-\epsilon\) expansion. 

Expanding the three hypergeometric functions and integrating term by term gives the series representation\footnote{%
\scriptsize
Again, using the generalized Srivastava--Daoust hypergeometric function \cite{srivastava1969eulerian,srivastava2024class}, we define
\[\begin{aligned}
		\mathfrak S_2(\beta) &:={} F^{1:2;2;2}_{1:1;1;1} \left[ \begin{matrix}
			(1:1,1,1): (a_1,1),(b_1,1);(a_2,1),(b_2,1);(a_3,1),(b_3,1) \\
			(2+\beta:1,1,1):(1,1);(1,1);(1,1)
		\end{matrix} ;1,1,1 \right].
	\end{aligned}\]
Setting \(\beta=0\) reproduces the sum in \eqref{eq:three-melon-D2-series-seed},
\[
	\mathfrak S_2(0) = \sum_{n_1,n_2,n_3\geq0} \frac{1}{n_1+n_2+n_3+1} \prod_{i=1}^3 \frac{(a_i)_{n_i}(b_i)_{n_i}}{(n_i!)^2}.
\]
}
\begin{equation}
	\int_0^1d\mathrm w\, \prod_{i=1}^3{}_2F_1(a_i,b_i;1;\mathrm w) = \sum_{n_1,n_2,n_3\geq0} \frac{1}{n_1+n_2+n_3+1} \prod_{i=1}^3 \frac{(a_i)_{n_i}(b_i)_{n_i}}{(n_i!)^2}.
	\label{eq:three-melon-D2-series-seed}
\end{equation}
Both the integral and the series are symmetric under permutations of the three spectral parameters.

The same integral also has a Legendre representation. With \(y=1-2\mathrm w\) and \(a_i=1-b_i\), the hypergeometric factors become Legendre functions,
\begin{equation}
	{}_2F_1(a_i,b_i;1;\mathrm w) = P_{-b_i}(y).
	\label{eq:three-melon-D2-legendre-seed}
\end{equation}
When \(b_i\in\mathbb Z_{\leq0}\) for all three edges, these functions reduce to Legendre polynomials, and Gaunt's formula gives
\begin{equation}
	\int_0^1d\mathrm w\, \prod_{i=1}^3{}_2F_1(a_i,b_i;1;\mathrm w) = \frac12\int_{-1}^1dy\, \prod_{i=1}^3P_{-b_i}(y) = \begin{pmatrix}
		-b_1&-b_2&-b_3\\
		0&0&0
	\end{pmatrix}^{\!2}.
	\label{eq:three-melon-D2-wigner-seed}
\end{equation}
The integral factor in \eqref{eq:three-melon-D2-seed} thus continues these squared Wigner \(3j\) symbols to general spectral parameters.

\subsubsection{Flat-space limit}
\label{sec:three-melon-flat-space-contraction}

We have so far worked on the unit sphere. To take the flat-space limit, restore its radius \(R\) and use the principal-series scaling
\begin{equation}
	\rho_i=\mathrm{i}Rm_i, \qquad R\longrightarrow\infty, \qquad m_i\ \text{fixed}.
	\label{eq:three-melon-flat-scaling}
\end{equation}
With the radius restored, \(M_i^2=R^{-2}[(D-1)^2/4-\rho_i^2]\longrightarrow m_i^2\). Here \(m_i\) is the flat-space mass. Denote \(\boldsymbol m=(m_1,m_2,m_3)\). In the sphere density \eqref{eq:three-melon-recurrence-density}, the integration measure contributes \(R^D\) and the three propagators contribute \(R^{3(2-D)}\). Thus
\begin{equation}
	\bar I^{\,\mathbb R}_{\texttt{V\{3\}V}}(D;\boldsymbol m) = \lim_{R\to\infty} R^{6-2D} \bar I_{\texttt{V\{3\}V}}(D;\mathrm{i}R\boldsymbol m).
	\label{eq:three-melon-flat-density}
\end{equation}
Below, we take the flat-space limit of the sphere recurrences \eqref{eq:three-melon-s-recurrence} and \eqref{eq:three-melon-rho-recurrences} to recover the dimensional recurrence and mass differential equations for the flat-space three-melon. The sphere channel variables defined in \eqref{eq:three-melon-channel-data} become
\begin{equation}
	\mathsf z_k^{\mathbb R} := \lim_{R\to\infty}R^{-2}\mathsf z_k =m_i^2+m_j^2-m_k^2, \qquad \{i,j,k\}=\{1,2,3\},
	\label{eq:three-melon-flat-channel-data}
\end{equation}
with the corresponding matrix
\begin{equation}
	\mathsf Z^{\mathbb R} = \operatorname{diag} \left( \mathsf z_1^{\mathbb R}, \mathsf z_2^{\mathbb R}, \mathsf z_3^{\mathbb R} \right).
	\label{eq:three-melon-flat-channel-matrix}
\end{equation}
For fixed \(q\), the shifted K\"all\'en polynomial \(\mathcal K_q\) in \eqref{eq:three-melon-analytic-shifted-kallen} similarly has the flat-space limit
\begin{equation}
	\mathcal K^{\mathbb R} := \lim_{R\to\infty}R^{-4}\mathcal K_q = \prod_{\sigma_2,\sigma_3=\pm1} \left( m_1+\sigma_2m_2+\sigma_3m_3 \right).
	\label{eq:three-melon-flat-kallen}
\end{equation}
After volume normalization, the petal and petal matrix \eqref{eq:three-melon-petal-matrix} appearing in the sphere source terms become the flat-space petal density and petal-density matrix
\begin{equation}
	\bar I_{\texttt{V}^{\circ}}^{\,\mathbb R}(s;m) = \frac{\Gamma(\frac12-s)}{(4\pi)^{s+\frac12}}m^{2s-1}, \qquad \overline{\mathsf P}^{\mathbb R}(s;\alpha,\beta,\gamma) = \operatorname{diag} \left( \bar I_{\texttt{V}^{\circ}}^{\,\mathbb R}(s;\alpha), \bar I_{\texttt{V}^{\circ}}^{\,\mathbb R}(s;\beta), \bar I_{\texttt{V}^{\circ}}^{\,\mathbb R}(s;\gamma) \right).
	\label{eq:three-melon-flat-petal-matrix}
\end{equation}
Substituting these limits into the density form of the dimensional recurrence \eqref{eq:three-melon-s-recurrence}, the resulting recurrence equation can be written as
\begin{equation}
	\boxed{ 32\pi^2s(2s+1) \bar I^{\,\mathbb R}_{\texttt{V\{3\}V}}(s+1;\boldsymbol m) + \mathcal K^{\mathbb R} \bar I^{\,\mathbb R}_{\texttt{V\{3\}V}}(s;\boldsymbol m) = -\operatorname{tr} \left[ \mathsf Z^{\mathbb R} \operatorname{adj}\overline{\mathsf P}^{\mathbb R}(s;m_1,m_2,m_3) \right]. }
	\label{eq:three-melon-flat-dimension-recurrence}
\end{equation}
The spectral recurrences \eqref{eq:three-melon-rho-recurrences} reduce to mass derivatives after Taylor expansion. For each cyclic permutation \((i,j,k)\), the source matrix \eqref{eq:three-melon-rho-source-matrix} becomes
\begin{equation}
	\mathsf R_i^{\mathbb R} := \lim_{R\to\infty}R^{-2}\mathsf R_i = \operatorname{diag} \left( -2m_i^2, \mathsf z_k^{\mathbb R}, \mathsf z_j^{\mathbb R} \right).
	\label{eq:three-melon-flat-mass-source-matrix}
\end{equation}
The mass equations then take the form
\begin{equation}
	\boxed{\displaystyle \mathcal K^{\mathbb R} m_i\partial_{m_i} \bar I^{\,\mathbb R}_{\texttt{V\{3\}V}}(s;\boldsymbol m) +4(s-1)m_i^2\mathsf z_i^{\mathbb R} \bar I^{\,\mathbb R}_{\texttt{V\{3\}V}}(s;\boldsymbol m) =(2s-1) \operatorname{tr} \left[ \mathsf R_i^{\mathbb R} \operatorname{adj} \overline{\mathsf P}^{\mathbb R}(s;m_i,m_j,m_k) \right]. }
	\label{eq:three-melon-flat-mass-recurrence}
\end{equation}
The known flat-space result \cite{cacciatori2023banana, davydychev1993two} satisfies both the dimensional recurrence \eqref{eq:three-melon-flat-dimension-recurrence} \cite{davydychev1996connection} and the mass equations \eqref{eq:three-melon-flat-mass-recurrence}. In fact, the dimensional recurrence can be solved as in Section~\ref{sec:three-melon-analytical-solution} to recover this result, but since the result is already known, we do not carry it out here.

For dimensional regularization, the analytical solution is not required. The sphere recurrence \eqref{eq:three-melon-analytic-recurrence} advances \(D\) by four, but the flat-space recurrence \eqref{eq:three-melon-flat-dimension-recurrence} advances it by two, leaving only the \(D\equiv1\pmod 2\) and \(D\equiv2\pmod 2\) ladders. So we only need \(D=1\) and \(D=2\) seed expansions for the flat space case. Our aim here, however, is not to obtain the regularized results from the flat space recurrences but to show that the regularized flat-space densities are obtained by taking the limit \eqref{eq:three-melon-flat-density} of the sphere results in Section~\ref{sec:three-melon-dimensional-regularization}; the flat-space recurrence then \eqref{eq:three-melon-flat-dimension-recurrence} serves as an additional check. We take this limit first in odd dimensions and then in even dimensions.

\paragraph{Odd dimensions}

Set
\begin{equation}
	\widetilde e_1^{\mathbb R}:=m_1+m_2+m_3, \qquad \widetilde e_2^{\mathbb R}:=m_1m_2+m_1m_3+m_2m_3, \qquad \widetilde e_3^{\mathbb R}:=m_1m_2m_3.
	\label{eq:three-melon-flat-mass-symmetric}
\end{equation}
Under \(\rho_i=\mathrm{i}Rm_i\), with \(m_i>0\), only the sign choices \(\boldsymbol{\sigma}=\pm(1,1,1)\) contribute at leading order to the normalized sum in \eqref{eq:three-melon-odd-digamma-block}; the mixed-sign terms are exponentially suppressed. Using \(\psi(z)\sim\log z\) gives the limit
\begin{equation}
	\Psi_\alpha^S-2\log R\longrightarrow\Psi^{\mathbb R}, \qquad \Psi^{\mathbb R}:=2\log\widetilde e_1^{\mathbb R} -\log(4\pi)+\gamma_{\mathrm E}-1,
	\label{eq:three-melon-flat-odd-digamma}
\end{equation}
which is independent of the fixed shift \(\alpha\).

\subparagraph{\(D=3-\epsilon\).}

The radius factor in \eqref{eq:three-melon-flat-density} is \(R^{2\epsilon}=1+2\epsilon\log R+O(\epsilon^2)\). Multiplying the bracket in the sphere result \eqref{eq:three-melon-D3-result-derived} by this factor adds \(2\log R\) to its finite part. This cancels the radius dependence of \(-\Psi_{1/2}^S\), giving
\begin{equation}
	\boxed{\displaystyle \bar I^{\,\mathbb R}_{\texttt{V\{3\}V}}(3-\epsilon;\boldsymbol m) = \frac{1}{32\pi^2} \left[ \frac1\epsilon-\Psi^{\mathbb R} \right] +O(\epsilon). }
	\label{eq:three-melon-flat-D3}
\end{equation}
This agrees with \cite{cacciatori2023banana}, which uses \(d=3-2\varepsilon\), after identifying \(\varepsilon=\epsilon/2\).

\subparagraph{\(D=5-\epsilon\).}

The radius factor is now \(R^{-4+2\epsilon}\). Using \eqref{eq:three-melon-flat-kallen}, the factor \(R^{-4}\) converts the prefactor \(\mathcal K_0\) in the sphere result \eqref{eq:three-melon-D5-result} to \(\mathcal K^{\mathbb R}\), while \(R^{2\epsilon}\) produces the same logarithmic cancellation in \(-\Psi_1^S\). Since \(\rho_i\cot(\pi\rho_i)\sim Rm_i\), the cyclic sum in \eqref{eq:three-melon-D5-result} gives
\begin{equation}
	\boxed{ \bar I^{\,\mathbb R}_{\texttt{V\{3\}V}}(5-\epsilon;\boldsymbol m) = -\frac{\mathcal K^{\mathbb R}}{3072\pi^4} \Biggl[ \frac1\epsilon-\Psi^{\mathbb R}+\frac56 + \frac{ 2(\widetilde e_1^{\mathbb R})^2\widetilde e_2^{\mathbb R} -4(\widetilde e_2^{\mathbb R})^2 -4\widetilde e_1^{\mathbb R}\widetilde e_3^{\mathbb R} }{ \mathcal K^{\mathbb R} } \Biggr] +O(\epsilon). }
	\label{eq:three-melon-flat-D5}
\end{equation}

Applying the flat-space recurrence \eqref{eq:three-melon-flat-dimension-recurrence} to the \(D=3-\epsilon\) result \eqref{eq:three-melon-flat-D3} reproduces \eqref{eq:three-melon-flat-D5}, checking both the pole and the finite part.

\paragraph{Even dimensions}

Under \(\rho_i=\mathrm{i}Rm_i\), the variables \eqref{eq:three-melon-even-basic-data} and their elementary symmetric polynomials \eqref{eq:three-melon-even-symmetric-data} satisfy
\begin{equation}
	R^{-2}y_i^S\longrightarrow y_i^{\mathbb R}:=m_i^2, \qquad \left( R^{-2}e_1^S, R^{-4}e_2^S, R^{-6}e_3^S \right) \longrightarrow \left( e_1^{\mathbb R},e_2^{\mathbb R},e_3^{\mathbb R} \right).
	\label{eq:three-melon-flat-even-basic-limits}
\end{equation}
Here \(e_a^{\mathbb R}\) denotes the elementary symmetric polynomial of degree \(a\) in the squared masses. The Källén product has the limit
\begin{equation}
	R^{-8}\mathcal K_{1/2}\mathcal K_{-1/2} \longrightarrow (\mathcal K^{\mathbb R})^2, \qquad \mathcal K^{\mathbb R}=(e_1^{\mathbb R})^2-4e_2^{\mathbb R}.
	\label{eq:three-melon-flat-even-kallen-limit}
\end{equation}
Applying the large-argument asymptotics of the polygamma functions to Table~\ref{tab:three-melon-even-log-data} gives
\begin{align}
	\chi_{i,0}^{S^4}-2\log R &\longrightarrow \chi_{i,0}^{\mathbb R^4} := \log y_i^{\mathbb R}-\log(4\pi)+\gamma_{\mathrm E}-\frac32, & \chi_{i,1}^{S^4} &\longrightarrow \chi_{i,1}^{\mathbb R^4} := \frac{\pi^2}{6}, \notag\\
	\chi_{i,0}^{S^6}-2\log R &\longrightarrow \chi_{i,0}^{\mathbb R^6} := \log y_i^{\mathbb R}-\log(4\pi)+\gamma_{\mathrm E}-\frac{31}{12}, & \chi_{i,1}^{S^6} &\longrightarrow \chi_{i,1}^{\mathbb R^6} := \frac{\pi^2}{6}+\frac{241}{72}.
	\label{eq:three-melon-flat-even-log-data}
\end{align}
Define the flat-space moment vectors \(\mathfrak M^{\mathbb R^D}\), \(\mathfrak N^{\mathbb R^D}\), and \(\mathfrak Q^{\mathbb R^D}\) by \eqref{eq:three-melon-even-linear-moments}-- \eqref{eq:three-melon-even-moment-vectors}, using \(y_i^{\mathbb R}\) and \(\chi_{i,r}^{\mathbb R^D}\) in place of their sphere counterparts. The corresponding moment vector for the constant function is
\begin{equation}
	\boldsymbol{1}^{\mathbb R} := \left( 3,e_1^{\mathbb R},(e_1^{\mathbb R})^2-2e_2^{\mathbb R} \right).
	\label{eq:three-melon-flat-even-one-vector}
\end{equation}

The normalization \eqref{eq:three-melon-flat-density} contributes \(R^{-2+2\epsilon}\) at \(D=4-\epsilon\) and \(R^{-6+2\epsilon}\) at \(D=6-\epsilon\). Since both sphere results have double poles, the finite parts require expanding \(R^{2\epsilon}\) through second order in \(\epsilon\). After all contributions are combined, the \(\log R\) terms cancel. We write both results in terms of the common flat-space density \(\bar I_{\texttt{V\{3\}V}}^{\,\mathbb R}(D=2;\boldsymbol m)\), whose reduction from the sphere seeds is derived after these expansions.

\subparagraph{\(D=4-\epsilon\).}

Taking the flat-space limit of \eqref{eq:three-melon-D4-result} gives
\begin{equation}
	\boxed{ \begin{aligned}
		\bar I_{\texttt{V\{3\}V}}^{\,\mathbb R} (4-\epsilon;\boldsymbol m) ={}& -\frac{1}{128\pi^4} \Bigg[ \frac{e_1^{\mathbb R}}{\epsilon^2} -\frac{\mathfrak M_1^{\mathbb R^4}}{\epsilon} +4\pi^2\mathcal K^{\mathbb R} \bar I_{\texttt{V\{3\}V}}^{\,\mathbb R}(D=2;\boldsymbol m) +\frac58e_1^{\mathbb R} \\
		&\hspace{3cm} +\frac14 \left( \mathfrak M_0^{\mathbb R^4}\mathfrak M_1^{\mathbb R^4} -\mathfrak Q_1^{\mathbb R^4} +\mathfrak N_1^{\mathbb R^4} \right) \Bigg] +O(\epsilon).
	\end{aligned} }
	\label{eq:three-melon-flat-D4}
\end{equation}
The result agrees with \cite{cacciatori2023banana, davydychev1993two, ford1992standard}.

\subparagraph{\(D=6-\epsilon\).}

The term \(\mathsf W_2^{S^6}[\boldsymbol{1}^S]\) in \eqref{eq:three-melon-D6-result} is \(O(R^4)\) and therefore vanishes after multiplication by \(R^{-6}\). To express the surviving polynomial terms, define, for \(\boldsymbol Z=(Z_0,Z_1,Z_2)\),
\begin{align}
	\mathsf W_0^{\mathbb R^6}[\boldsymbol Z] &:={} 4e_3^{\mathbb R}Z_0-4e_2^{\mathbb R}Z_1+e_1^{\mathbb R}Z_2, \notag\\
	\mathsf W_1^{\mathbb R^6}[\boldsymbol Z] &:={} \frac12 \left[(e_1^{\mathbb R})^3-8e_1^{\mathbb R}e_2^{\mathbb R}+8e_3^{\mathbb R}\right]Z_0 +\left[(e_1^{\mathbb R})^2+4e_2^{\mathbb R}\right]Z_1 -2e_1^{\mathbb R}Z_2.
	\label{eq:three-melon-flat-W6}
\end{align}
The limit of \eqref{eq:three-melon-D6-result} is
\begin{equation}
	\boxed{ \begin{aligned}
		\bar I_{\texttt{V\{3\}V}}^{\,\mathbb R} (6-\epsilon;\boldsymbol m) ={}& \frac{1}{24576\pi^6} \Bigg[ \frac{\mathsf W_0^{\mathbb R^6}[\boldsymbol{1}^{\mathbb R}]}{\epsilon^2} -\frac{ \mathsf W_0^{\mathbb R^6}[\mathfrak M^{\mathbb R^6}] +\mathsf W_1^{\mathbb R^6}[\boldsymbol{1}^{\mathbb R}] }{\epsilon} +4\pi^2(\mathcal K^{\mathbb R})^2 \bar I_{\texttt{V\{3\}V}}^{\,\mathbb R}(D=2;\boldsymbol m) \\
		&\hspace{14mm} +\frac14 \mathsf W_0^{\mathbb R^6} [ \mathfrak M_0^{\mathbb R^6}\mathfrak M^{\mathbb R^6} -\mathfrak Q^{\mathbb R^6} +\mathfrak N^{\mathbb R^6} ] +\frac12 \mathsf W_1^{\mathbb R^6} [ \mathfrak M_0^{\mathbb R^6}\boldsymbol{1}^{\mathbb R} -\mathfrak M^{\mathbb R^6} ] \Bigg] +O(\epsilon).
	\end{aligned} }
	\label{eq:three-melon-flat-D6}
\end{equation}
%The flat recurrence \eqref{eq:three-melon-flat-dimension-recurrence} reproduces both even-dimensional expansions.

\subparagraph{The \(D=2\) seed.}

The common flat-space seed follows from the sphere \(D=2\) melon density. Using \eqref{eq:three-melon-D2-seed}, \eqref{eq:three-melon-D2-series-seed}, and \(\Gamma(a_i)\Gamma(b_i)=\pi/C_i\), we obtain
\begin{equation}
	\bar I_{\texttt{V\{3\}V}} (D=2;\mathrm{i}R\boldsymbol m) = \frac{1}{16\pi^2} \sum_{n_1,n_2,n_3\geq0} \frac{1}{n_1+n_2+n_3+1} \prod_{i=1}^3 \frac{ \Gamma(n_i+\frac12+\mathrm{i}Rm_i) \Gamma(n_i+\frac12-\mathrm{i}Rm_i) }{ \Gamma(n_i+1)^2 }.
	\label{eq:three-melon-flat-D2-gamma-series}
\end{equation}
Set \(n_i=(Rm_i)^2x_i\). At fixed \(x_i>0\), Stirling's formula gives
\begin{equation}
	\frac{ \Gamma(n_i+\frac12+\mathrm{i}Rm_i) \Gamma(n_i+\frac12-\mathrm{i}Rm_i) }{ \Gamma(n_i+1)^2 } = \frac{e^{-1/x_i}}{(Rm_i)^2x_i} \left[1+O(R^{-2})\right].
	\label{eq:three-melon-flat-D2-stirling}
\end{equation}
The triple sum therefore has the Riemann limit
\begin{equation}
	\bar I_{\texttt{V\{3\}V}} (D=2;\mathrm{i}R\boldsymbol m) = \frac{1}{16\pi^2R^2} \int_{(0,\infty)^3} \frac{ e^{-1/x_1-1/x_2-1/x_3} \,dx_1\,dx_2\,dx_3 }{ x_1x_2x_3 (m_1^2x_1+m_2^2x_2+m_3^2x_3) } +O(R^{-4}).
	\label{eq:three-melon-flat-D2-riemann-limit}
\end{equation}
To integrate out the common scale, set \(x=x_3/x_1\), \(y=x_3/x_2\), and \(t=1/x_3\). Then
\begin{equation}
	x_1=\frac1{tx}, \qquad x_2=\frac1{ty}, \qquad x_3=\frac1t, \qquad dx_1\,dx_2\,dx_3=\frac{dt\,dx\,dy}{t^4x^2y^2}.
	\label{eq:three-melon-flat-scale-ratios}
\end{equation}
The Jacobian cancels the factor \(t^{-4}x^{-2}y^{-2}\) in the denominator. Taking the flat-space limit and integrating over \(t\) gives
\begin{equation}
	\begin{aligned}
		\bar I_{\texttt{V\{3\}V}}^{\,\mathbb R}(D=2;\boldsymbol m) &={} \frac1{16\pi^2} \int_0^\infty dx\,dy\, \frac{\displaystyle\int_0^\infty dt\,e^{-t(1+x+y)}} {m_1^2y+m_2^2x+m_3^2xy} \\
		&={} \frac1{16\pi^2} \int_0^\infty \frac{dx\,dy} {(1+x+y)\bigl[m_2^2x+(m_1^2+m_3^2x)y\bigr]}.
	\end{aligned}
	\label{eq:three-melon-flat-D2-scale-reduction}
\end{equation}
The \(y\)-integration uses the elementary identity
\begin{equation}
	\int_0^\infty\frac{dy}{(y+c)(Ay+B)} = \frac{\log(Ac/B)}{Ac-B}, \qquad A,B,c>0,
	\label{eq:three-melon-flat-elementary-integral}
\end{equation}
with the value at \(Ac=B\) understood by continuity. Taking \(c=1+x\), \(A=m_1^2+m_3^2x\), and \(B=m_2^2x\) yields
\begin{equation}
	\bar I_{\texttt{V\{3\}V}}^{\,\mathbb R}(D=2;\boldsymbol m) = \frac{1}{16\pi^2} \int_0^\infty \frac{dx}{q_{12|3}(x)} \log\!\left[ \frac{(1+x)(m_1^2+m_3^2x)}{m_2^2x} \right],
	\label{eq:three-melon-flat-D2-hyperlog}
\end{equation}
where
\begin{equation}
	q_{12|3}(x) = m_3^2x^2+(m_1^2+m_3^2-m_2^2)x+m_1^2 =m_3^2(x-r_+)(x-r_-), \qquad r_\pm = \frac{ -(m_1^2+m_3^2-m_2^2)\pm\sqrt{\mathcal K^{\mathbb R}} }{2m_3^2}.
	\label{eq:three-melon-flat-hyperlog-roots}
\end{equation}
The partial-fraction decomposition
\begin{equation}
	\frac{1}{q_{12|3}(x)} = \frac{1}{\sqrt{\mathcal K^{\mathbb R}}} \left( \frac{1}{x-r_+}-\frac{1}{x-r_-} \right)
	\label{eq:three-melon-flat-hyperlog-partial-fraction}
\end{equation}
reduces the remaining integral to dilogarithms\footnote{The dilogarithm is defined by \(\operatorname{Li}_2(z)=-\int_0^z \frac{\log(1-t)}{t}\,dt\), using the principal logarithm and an integration path that avoids the cut \([1,\infty)\).},
\begin{equation}
	\begin{aligned}
		\bar I_{\texttt{V\{3\}V}}^{\,\mathbb R}(D=2;\boldsymbol m) ={}& \frac{1}{16\pi^2\sqrt{\mathcal K^{\mathbb R}}} \Bigg[ \operatorname{Li}_2(1+r_+)-\operatorname{Li}_2(1+r_-) +\operatorname{Li}_2(1+r_-^{-1}) -\operatorname{Li}_2(1+r_+^{-1}) \\
		&\quad +\frac12\log^2(-r_-)-\frac12\log^2(-r_+) +\log\!\left(\frac{m_3^2}{m_2^2}\right) \log\!\left(\frac{r_-}{r_+}\right) \Bigg].
	\end{aligned}
	\label{eq:three-melon-flat-D2-dilogarithm}
\end{equation}
Equation \eqref{eq:three-melon-flat-D2-dilogarithm} is obtained by evaluating \eqref{eq:three-melon-flat-D2-hyperlog} for distinct negative roots \(r_\pm\), where \(\mathcal K^{\mathbb R}>0\). For positive masses, symmetry allows us to label the largest mass as \(m_1\). With this labeling, we use the principal branches of the square root, logarithms, and dilogarithms in \eqref{eq:three-melon-flat-D2-dilogarithm}. This prescription also covers \(\mathcal K^{\mathbb R}<0\), where the roots are complex conjugates, while the value at \(\mathcal K^{\mathbb R}=0\) is obtained by continuity. %Although the expression uses the \(12|3\) channel, the result is symmetric in the three masses.

\begin{comment}
For \(\mathcal K^{\mathbb R}<0\), the masses form the sides of a triangle.
Set \(\kappa:=\sqrt{-\mathcal K^{\mathbb R}}\), and let \(\theta_{ij|k}\)
denote the angle opposite \(m_k\), so that
\begin{equation}
  \cos\theta_{ij|k}
  =
  \frac{m_i^2+m_j^2-m_k^2}{2m_im_j},
  \qquad
  \sin\theta_{ij|k}
  =
  \frac{\kappa}{2m_im_j},
  \qquad
  \theta_{12|3}+\theta_{23|1}+\theta_{31|2}=\pi.
  \label{eq:three-melon-flat-triangle-angles}
\end{equation}
The roots \(r_\pm\) are then complex conjugates. In terms of the Clausen
function
\begin{equation}
  \operatorname{Cl}_2(\phi)
  :=
  \operatorname{Im}\operatorname{Li}_2(e^{\mathrm{i}\phi})
  =
  \sum_{n=1}^{\infty}\frac{\sin(n\phi)}{n^2},
  \label{eq:clausen-function}
\end{equation}
pairing the conjugate dilogarithms gives
\begin{equation}
  \boxed{\displaystyle
  \bar I_{\texttt{V\{3\}V}}^{\,\mathbb R}(D=2;\boldsymbol m)
  =
  \frac{1}{8\pi^2\kappa}
  \sum_{\mathrm{cyc}(i,j,k)}
  \operatorname{Cl}_2(2\theta_{ij|k}).
  }
  \label{eq:three-melon-flat-D2-clausen}
\end{equation}
\end{comment}

The sphere \(D=2\) seed enters the \(D=6-\epsilon\) result directly. Combining its limit with \eqref{eq:three-melon-flat-even-kallen-limit} gives
\begin{equation}
	R^{-6} \left[ 4\pi^2\mathcal K_{1/2}\mathcal K_{-1/2} \bar I_{\texttt{V\{3\}V}} (D=2;\mathrm{i}R\boldsymbol m) \right] \longrightarrow 4\pi^2(\mathcal K^{\mathbb R})^2 \bar I_{\texttt{V\{3\}V}}^{\,\mathbb R}(D=2;\boldsymbol m).
	\label{eq:three-melon-flat-D6-seed-limit}
\end{equation}

For \(D=4-\epsilon\), the sphere \(D=0\) seed must be combined with the accompanying terms. Equations \eqref{eq:three-melon-D0-seed-one-dimensional} and \eqref{eq:three-melon-D0-series-seed} express \(\mathsf F\) as
\begin{equation}
	\mathsf F(\mathrm{i}R\boldsymbol m) = e_3^S \sum_{n_1,n_2,n_3\geq0} H_{N+2} \prod_{i=1}^3 \frac{ \Gamma(n_i+\frac12+\mathrm{i}Rm_i) \Gamma(n_i+\frac12-\mathrm{i}Rm_i) }{ \Gamma(n_i+2)\Gamma(n_i+1) }, \qquad N=n_1+n_2+n_3.
	\label{eq:three-melon-flat-D0-gamma-series}
\end{equation}
On the same scale \(n_i=(Rm_i)^2x_i\), the summand has the asymptotics
\begin{equation}
	\frac{ \Gamma(n_i+\frac12+\mathrm{i}Rm_i) \Gamma(n_i+\frac12-\mathrm{i}Rm_i) }{ \Gamma(n_i+2)\Gamma(n_i+1) } = \frac{e^{-1/x_i}}{(Rm_i)^4x_i^2} \left[1+O(R^{-2})\right],
	\label{eq:three-melon-flat-D0-asymptotics}
\end{equation}
\begin{equation}
	\text{and}\quad H_{N+2} = 2\log R +\log\!\left( m_1^2x_1+m_2^2x_2+m_3^2x_3 \right) +\gamma_{\mathrm E} +O(R^{-2}).
	\label{eq:three-melon-flat-D0-harmonic-asymptotics}
\end{equation}
The Riemann limit therefore gives
\begin{equation}
	\mathsf F(\mathrm{i}R\boldsymbol m) = 2\log R +\gamma_{\mathrm E} +\mathcal I_0(\boldsymbol m) +O(R^{-2}\log R),
	\label{eq:three-melon-flat-D0-riemann-limit}
\end{equation}
where
\begin{equation}
	\mathcal I_0(\boldsymbol m) := \int_{(0,\infty)^3} \prod_{i=1}^3 \frac{e^{-1/x_i}\,dx_i}{x_i^2} \log\!\left( m_1^2x_1+m_2^2x_2+m_3^2x_3 \right).
	\label{eq:three-melon-flat-D0-log-integral}
\end{equation}
Stokes' theorem reduces this logarithmic integral to the same \(D=2\) density,
\begin{equation}
	\mathcal I_0(\boldsymbol m) = \gamma_{\mathrm E} +\sum_{i=1}^3\alpha_i^\star\log m_i^2 +\frac{32\pi^2(m_1m_2m_3)^2}{-\mathcal K^{\mathbb R}} \bar I_{\texttt{V\{3\}V}}^{\,\mathbb R}(D=2;\boldsymbol m), \qquad \alpha_i^\star = \frac{m_i^2(m_i^2-m_j^2-m_k^2)}{\mathcal K^{\mathbb R}}.
	\label{eq:three-melon-flat-D0-reduction}
\end{equation}
Combining this reduction with the accompanying terms in \eqref{eq:three-melon-D4-result} gives
\begin{equation}
	\lim_{R\to\infty}R^{-2} \Bigg\{ -\frac{\mathcal K_{1/2}\mathcal K_{-1/2}}{8e_3^S} \left[ \mathsf F(\mathrm{i}R\boldsymbol m) -\log\!\left(4\pi e^{\gamma_{\mathrm E}}\right) -\frac92 \right] -\frac{ \mathsf W_0^{S^4} [\mathfrak M^{S^4}+\boldsymbol v_4^S] }{8e_3^S} \Bigg\} = 4\pi^2\mathcal K^{\mathbb R} \bar I_{\texttt{V\{3\}V}}^{\,\mathbb R}(D=2;\boldsymbol m).
	\label{eq:three-melon-flat-D4-seed-limit}
\end{equation}

\section{Higher loops and outlook}
\label{sec:conclusion}

We demonstrated a lattice framework that combines total-derivative identities and edge-contraction relations into a coupled linear system for scalar vacuum integrals on \(S^D\). We solved this system for all scalar vacuum graphs upto two loops. For the two-loop three-melon, we constructed a meromorphic solution for general dimension and spectral parameters, obtained the poles and finite terms at singular integer dimensions, and recovered the corresponding known flat-space limits too. The next steps are to extend these calculations to higher loops and to spinning fields.

Nothing in our construction is tied to two loops. Edge contractions organize the graph lattices into a hierarchy. Once a graph has been solved, its solution and lattice relations are allowed source data for the graphs that contract to it. Higher-loop melons provide a direct extension of the two-loop calculation. Contracting any edge of an \(n\)-melon leaves an \((n-1)\)-petal flower, so all contraction sources are fixed by the known petal factorization. For general higher-loop graphs, the contraction sources need not factorize and can involve other nontrivial integrals at the same loop order.

\begin{figure}[ht]
  \centering
  % Smaller nodes, with the same stroke width as Figures 1--4.
  \renewcommand{\GraphNodeDiameter}{2.625pt}
  \renewcommand{\GraphLineWidth}{0.75pt}
  \begin{tikzpicture}[
    graph edge/.style={edge},
    graph vertex/.style={vertex},
    contraction/.style={graph arrow},
    graph label/.style={text=black,font=\small}
  ]
    % (a) Match the top, middle, and bottom rows of panel (b).
    \begin{scope}[shift={(-3.8,0)}]
    % Tetrahedron.
    \begin{scope}[shift={(-2.3,2.175)}]
      \coordinate (a) at (-0.7,-0.45);
      \coordinate (b) at (0.7,-0.45);
      \coordinate (c) at (0,0.7);
      \coordinate (d) at (0,0);
      \draw[graph edge] (a)--(b)--(c)--(a);
      \draw[graph edge] (d)--(a) (d)--(b);
      \draw[contracted edge] (d)--(c);
      \node[graph vertex] at (a) {};
      \node[graph vertex] at (b) {};
      \node[contracted vertex] at (c) {};
      \node[contracted vertex] at (d) {};
    \end{scope}

    % Square with opposite doubled edges.
    \begin{scope}[shift={(2.3,2.3)},rotate=-90]
      \coordinate (a) at (-0.7,-0.7);
      \coordinate (b) at (0.7,-0.7);
      \coordinate (c) at (0.7,0.7);
      \coordinate (d) at (-0.7,0.7);
      \draw[graph edge] (a) to[bend left=10] (b);
      \draw[graph edge] (a) to[bend right=10] (b);
      \draw[graph edge] (d) to[bend left=10] (c);
      \draw[graph edge] (d) to[bend right=10] (c);
      \draw[contracted edge] (a)--(d);
      \draw[graph edge] (b)--(c);
      \node[contracted vertex] at (a) {};
      \node[graph vertex] at (b) {};
      \node[graph vertex] at (c) {};
      \node[contracted vertex] at (d) {};
    \end{scope}

    % Doubled triangle.
    \begin{scope}[shift={(0,-0.125)}]
      \coordinate (a) at (-0.7,-0.45);
      \coordinate (b) at (0.7,-0.45);
      \coordinate (c) at (0,0.7);
      \draw[contracted edge] (a)--(b);
      \draw[graph edge] (b) to[bend left=10] (c);
      \draw[graph edge] (b) to[bend right=10] (c);
      \draw[graph edge] (c) to[bend left=10] (a);
      \draw[graph edge] (c) to[bend right=10] (a);
      \node[contracted vertex] at (a) {};
      \node[contracted vertex] at (b) {};
      \node[graph vertex] at (c) {};
    \end{scope}

    % Four-melon.
    \begin{scope}[shift={(0,-2.3)}]
      \coordinate (a) at (-0.7,0);
      \coordinate (b) at (0.7,0);
      \draw[graph edge] (a) to[bend left=42] (b);
      \draw[graph edge] (a) to[bend left=14] (b);
      \draw[graph edge] (a) to[bend right=14] (b);
      \draw[graph edge] (a) to[bend right=42] (b);
      \node[graph vertex] at (a) {};
      \node[graph vertex] at (b) {};
    \end{scope}

    \draw[contraction] (-1.10,1.10) ++(135:0.45) -- ++(-45:0.9);
    \draw[contraction] (1.10,1.10) ++(45:0.45) -- ++(225:0.9);
    \draw[contraction] (0,-1.30) ++(90:0.45) -- ++(-90:0.9);
    \end{scope}

    % (b) One spanning tree per graph; the three other edges become petals.
    \begin{scope}[shift={(3.8,0)}]
      % Upper left: the three interior edges span the tetrahedron.
      \begin{scope}[shift={(-2.3,2.175)}]
        \coordinate (a) at (-0.7,-0.45);
        \coordinate (b) at (0.7,-0.45);
        \coordinate (c) at (0,0.7);
        \coordinate (d) at (0,0);
        \draw[graph edge] (a)--(b)--(c)--(a);
        \draw[contracted edge] (d)--(a) (d)--(b) (d)--(c);
        \foreach \vertex in {a,b,c,d}
          \node[contracted vertex] at (\vertex) {};
      \end{scope}

      % Upper right: two single edges and one doubled-side edge form a path.
      \begin{scope}[shift={(2.3,2.3)},rotate=-90]
        \coordinate (a) at (-0.7,-0.7);
        \coordinate (b) at (0.7,-0.7);
        \coordinate (c) at (0.7,0.7);
        \coordinate (d) at (-0.7,0.7);
        \draw[contracted edge] (a) to[bend left=10] (b);
        \draw[graph edge] (a) to[bend right=10] (b);
        \draw[graph edge] (d) to[bend left=10] (c);
        \draw[graph edge] (d) to[bend right=10] (c);
        \draw[contracted edge] (a)--(d) (b)--(c);
        \foreach \vertex in {a,b,c,d}
          \node[contracted vertex] at (\vertex) {};
      \end{scope}

      % Lower right: one edge from each doubled side spans the triangle.
      \begin{scope}[shift={(2.3,-2.425)}]
        \coordinate (a) at (-0.7,-0.45);
        \coordinate (b) at (0.7,-0.45);
        \coordinate (c) at (0,0.7);
        \draw[graph edge] (a)--(b);
        \draw[contracted edge] (b) to[bend left=10] (c);
        \draw[graph edge] (b) to[bend right=10] (c);
        \draw[contracted edge] (c) to[bend left=10] (a);
        \draw[graph edge] (c) to[bend right=10] (a);
        \foreach \vertex in {a,b,c}
          \node[contracted vertex] at (\vertex) {};
      \end{scope}

      % Lower left: any one edge spans the four-melon.
      \begin{scope}[shift={(-2.3,-2.3)}]
        \coordinate (a) at (-0.7,0);
        \coordinate (b) at (0.7,0);
        \draw[contracted edge] (a) to[bend left=42] (b);
        \draw[graph edge] (a) to[bend left=14] (b);
        \draw[graph edge] (a) to[bend right=14] (b);
        \draw[graph edge] (a) to[bend right=42] (b);
        \node[contracted vertex] at (a) {};
        \node[contracted vertex] at (b) {};
      \end{scope}

      % The common three-petal flower.
      \foreach \petalangle in {90,210,330} {
        \begin{scope}[rotate=\petalangle]
          \draw[graph edge] (0,0)
            .. controls (1.00,0.62) and (1.00,-0.62) .. (0,0);
        \end{scope}
      }
      \node[graph vertex] at (0,0) {};

      % Equal-length arrows, centered in the gaps around the flower.
      \draw[contraction] (-1.10,1.10) ++(135:0.45) -- ++(-45:0.9);
      \draw[contraction] (1.10,1.10) ++(45:0.45) -- ++(225:0.9);
      \draw[contraction] (1.20,-1.20) ++(-45:0.45) -- ++(135:0.9);
      \draw[contraction] (-1.20,-1.20) ++(225:0.45) -- ++(45:0.9);
    \end{scope}

    \node[font=\normalfont\normalsize] at (-3.8,3.65) {(a) Successive edge contractions};
    \node[font=\normalfont\normalsize] at (3.8,3.65) {(b) Spanning tree contractions};
  \end{tikzpicture}
  \caption{(a) The nontrivial three-loop contraction network. Each arrow denotes a representative one-edge contraction. Symmetry-related arrows and contractions to petal-decorated graphs are omitted; those sources are fixed by the lower-loop results. (b) Representative spanning trees are shown dashed. Contracting each tree gives the central three-petal flower.}
  \label{fig:three-loop-contraction-network}
\end{figure}
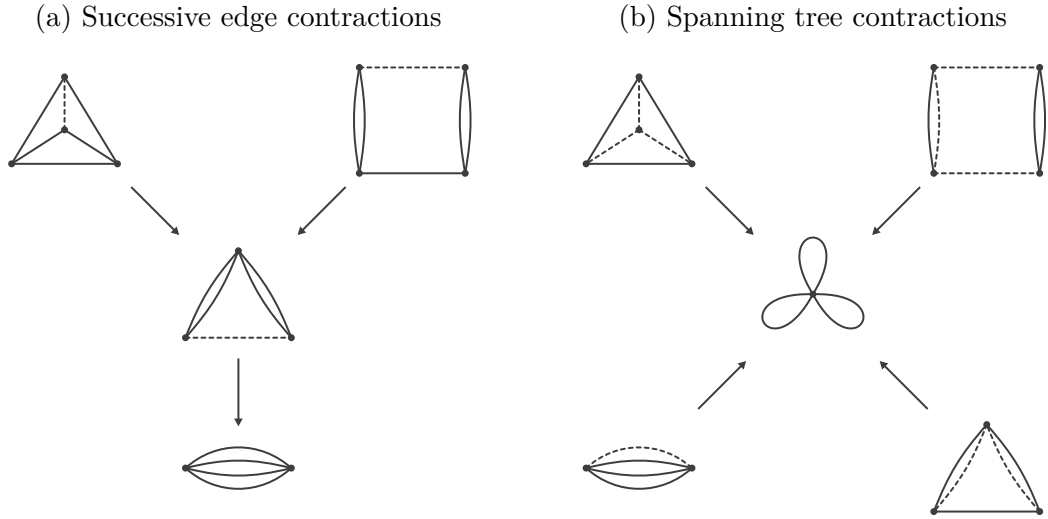

At three loops, the reduction rules for trees, chains, and petals leave four nontrivial graphs: the four-melon, the triangle with two doubled sides, the square with two opposite doubled sides, and the tetrahedron. As shown in Figure~\ref{fig:three-loop-contraction-network}(a), the tetrahedron and doubled square contract to the doubled triangle, which in turn contracts to the four-melon. The remaining one-edge contractions are related by symmetry or give petal-decorated graphs fixed by lower-loop results.

Tree and forest contractions provide direct routes through this hierarchy. Their equations can be constructed using the contraction identities in Appendix~\ref{app:contact-contractions}. For example, contracting two disjoint edges of the tetrahedron, or the two undoubled sides of the doubled square, reaches the four-melon directly, while contracting a spanning tree takes any connected graph directly to a one-vertex flower, as illustrated in Figure~\ref{fig:three-loop-contraction-network}(b). The flower integral factorizes into petal integrals, so petals lie at the bottom of the contraction hierarchy. In ongoing work, we are studying how best to combine these contractions with total-derivative identities to close higher-loop systems.

\enlargethispage{\baselineskip}
Altogether, the graph relations form sparse linear systems, but the rapid growth of their rational coefficients during symbolic elimination poses a computational challenge at higher loops. Finite-field reconstruction addresses this problem by solving the same system of equations at numerical parameter values modulo primes, and then reconstructing the rational coefficients. This strategy is well established in flat-space integral reduction \cite{von2015novel,peraro2016scattering,peraro2019finiteflow}. It avoids the intermediate symbolic expression growth, but solving the sparse systems efficiently would still require algorithms that control the loss of sparsity during elimination or avoid elimination altogether \cite{wiedemann1986solving}. Using modular calculations, we have obtained four-melon recurrences for general spectral parameters and recurrences for the other three-loop graphs on special spectral slices. These results will be presented separately.

Spinning fields are the next logical step. Once their vacuum integrals are written in Schwinger-parameter form \cite{bandaru2024quantum}, our lattice framework can be applied to them. These calculations would supply the integral input needed to obtain the renormalized sphere partition function for interacting theories with general field content and determine the resulting higher-loop corrections to de Sitter entropy.

\section*{Acknowledgements}
We thank Zimo Sun and Sebastian Mizera for very helpful discussions and suggestions. VL acknowledges the assistance of OpenAI’s Codex with C++ programming, proofreading, numerical checks, and language polishing during the preparation of this manuscript.

\newpage

\appendix

\section{Graph Nomenclature}
\label{app:nomenclature}

We use short strings to label the Feynman graphs\footnote{We use the term \emph{graph} and write \(I_g\) for its integral. Calling it a \emph{diagram} would instead give \(I_d\) or \(I_D\), conflicting with the standard symbols \(d\) and \(D\) for dimension.} that recur in the paper. The symbol \(\texttt{V}\) denotes a vertex, and a single edge between successive vertices is implicit, so \(\texttt{VV}\) denotes one edge. A superscript \(\circ\) denotes a self-loop. In either case, a multiplicity \(n>1\) is marked by \(\texttt{\{n\}}\); thus, \(\texttt{V\{n\}V}\) denotes \(n\) parallel edges between two vertices, while \(\texttt{V}^{\circ\texttt{\{n\}}}\) denotes \(n\) self-loops at one vertex. Longer cycles are written by following their vertices and marking the closing edge with matching integers; for example, \(\texttt{V1VV1}\) is the triangle. Table~\ref{tab:graph-nomenclature} lists the graph names used in the paper.

\tikzset{
  naming diagram/.style={
    baseline=(current bounding box.center)
  },
  naming vertex/.style={vertex},
  naming edge/.style={edge},
  naming omitted edge/.style={omitted edge}
}

\begin{table}[ht]
\centering
\renewcommand{\GraphLineWidth}{0.75pt}
\renewcommand{\GraphNodeDiameter}{3.75pt}
\small
\begin{tabular*}{\textwidth}{@{\extracolsep{\fill}} c c c @{}}
\toprule
\multicolumn{3}{c}{\textbf{Nomenclature}}\\
\midrule
\multicolumn{3}{c}{\textit{One loop}}\\[7pt]
\begin{tikzpicture}[naming diagram,scale=0.75]
  \useasboundingbox (-1.15,-0.18) rectangle (1.15,1.42);
  \coordinate (a) at (0,0);
  \draw[naming edge] (a)
    .. controls +(-0.78,1.20) and +(0.78,1.20) .. (a);
  \node[naming vertex] at (a) {};
\end{tikzpicture}
&
\begin{tikzpicture}[naming diagram,scale=0.75]
  \useasboundingbox (-1.15,-0.92) rectangle (1.15,0.92);
  \coordinate (a) at (-0.82,0);
  \coordinate (b) at (0.82,0);
  \draw[naming edge] (a)
    arc[start angle=180,end angle=0,radius=0.82];
  \draw[naming edge] (a)
    arc[start angle=180,end angle=360,radius=0.82];
  \node[naming vertex] at (a) {};
  \node[naming vertex] at (b) {};
\end{tikzpicture}
&
\begin{tikzpicture}[naming diagram,scale=0.75]
  \useasboundingbox (-1.25,-1.05) rectangle (1.25,1.05);
  \coordinate (x1) at (0,1.00);
  \coordinate (x2) at (0.71,0.71);
  \coordinate (x3) at (1.00,0);
  \coordinate (x4) at (0.71,-0.71);
  \coordinate (x5) at (0,-1.00);
  \coordinate (x6) at (-0.71,-0.71);
  \coordinate (x7) at (-1.00,0);
  \coordinate (xn) at (-0.71,0.71);
  \draw[naming edge] (xn)--(x1)--(x2)--(x3)--(x4)--(x5);
  \draw[naming omitted edge] (x5)--(x6);
  \draw[naming edge] (x6)--(x7)--(xn);
  \foreach \p in {x1,x2,x3,x4,x5,x6,x7,xn}
    \node[naming vertex] at (\p) {};
\end{tikzpicture}
\\[24pt]
\(\texttt{V}^{\circ}\)
&
\(\texttt{V1V1}\equiv\texttt{V\{2\}V}\)
&
\(\texttt{V1}\,\texttt{V}^{\,n-2}\texttt{V1},\ n\geq3\)
\\[10pt]
\midrule
\multicolumn{3}{c}{\textit{Two loops}}\\[7pt]
\begin{tikzpicture}[naming diagram,scale=0.75]
  \useasboundingbox (-1.45,-0.82) rectangle (1.45,0.82);
  \coordinate (a) at (0,0);
  \draw[naming edge] (a)
    .. controls +(-1.20,0.78) and +(-1.20,-0.78) .. (a);
  \draw[naming edge] (a)
    .. controls +(1.20,0.78) and +(1.20,-0.78) .. (a);
  \node[naming vertex] at (a) {};
\end{tikzpicture}
&
\begin{tikzpicture}[naming diagram,scale=0.75]
  \useasboundingbox (-1.72,-0.72) rectangle (1.72,0.72);
  \coordinate (a) at (-0.54,0);
  \coordinate (b) at (0.54,0);
  \draw[naming edge] (a)--(b);
  \draw[naming edge] (a)
    .. controls +(-1.20,0.78) and +(-1.20,-0.78) .. (a);
  \draw[naming edge] (b)
    .. controls +(1.20,0.78) and +(1.20,-0.78) .. (b);
  \node[naming vertex] at (a) {};
  \node[naming vertex] at (b) {};
\end{tikzpicture}
&
\begin{tikzpicture}[naming diagram,scale=0.75]
  \useasboundingbox (-1.15,-1.02) rectangle (1.15,1.02);
  \coordinate (a) at (-0.82,0);
  \coordinate (b) at (0.82,0);
  \draw[naming edge] (a)--(b);
  \draw[naming edge] (a)
    arc[start angle=180,end angle=0,radius=0.82];
  \draw[naming edge] (a)
    arc[start angle=180,end angle=360,radius=0.82];
  \node[naming vertex] at (a) {};
  \node[naming vertex] at (b) {};
\end{tikzpicture}
\\[24pt]
\(\texttt{V}^{\circ\texttt{\{2\}}}\)
&
\(\texttt{V}^{\circ}\texttt{V}^{\circ}\)
&
\(\texttt{V\{3\}V}\)
\\[10pt]
\bottomrule
\end{tabular*}
\caption{One- and two-loop graph names used in the paper. Wavy segments indicate omitted parts.}
\label{tab:graph-nomenclature}
\end{table}

\section{Edge Kernels}
\label{app:embedding-propagator}
In this appendix, we lift the massive scalar field on \(S^D\) to the ambient space \(\mathbb R^{D+1}\) and derive a scale-redundant ambient-space representation of its propagator. Each edge of a scalar Feynman graph carries this propagator. We call it the physical edge kernel. We then introduce a family of generalized edge kernels and derive their descent equations. We obtain their hypergeometric and spectral forms and recover the corresponding expressions for the physical propagator.
\subsection{Physical edge}
Consider a free massive scalar field \(\phi(\widehat X)\) on \(S^D\) with external source \(j(\widehat X)\). It satisfies
\begin{equation}
	\left( -\nabla_{S^D}^2+M^2 \right) \phi(\widehat X) = j(\widehat X).
	\label{eq:massive-scalar-equation}
\end{equation}
We lift \(\phi(\widehat X)\) to an ambient field \(\Phi(X)\) on \(\mathbb R^{D+1}\) with fixed radial weight,
\begin{equation}
	\Phi(X) = \mathscr r^{-\Delta^+}\phi(\widehat X), \qquad \mathscr r>0, \quad \text{and} \quad X = \mathscr r \widehat X,
	\label{eq:propagator-homogeneous-lift}
\end{equation}
and take \(\Phi\) to satisfy the massless equation
\begin{equation}
	-\partial_X^2\Phi(X) = \mathsf J(X).
	\label{eq:propagator-ambient-equation}
\end{equation}
In radial coordinates, the Laplacian on \(\mathbb R^{D+1}\) takes the form
\begin{equation}
	\partial_X^2 = \partial_{\mathscr r}^2 + \frac{D}{\mathscr r}\partial_{\mathscr r} + \frac{1}{\mathscr r^2}\nabla_{S^D}^2.
\end{equation}
Applying it to \eqref{eq:propagator-homogeneous-lift} and defining \(\Delta^-=D-1-\Delta^+\) gives
\begin{equation}
	-\partial_X^2\Phi(\mathscr r\widehat X) = \mathscr r^{-\Delta^+-2} \left[ -\nabla_{S^D}^2 + \Delta^+\Delta^- \right] \phi(\widehat X).
	\label{eq:propagator-radial-reduction}
\end{equation}
Lifting the sphere source as
\begin{equation}
	\mathsf J(\mathscr r\widehat X) = \mathscr r^{-\Delta^+-2}j(\widehat X),
	\label{eq:propagator-source-lift}
\end{equation}
reduces the ambient equation \eqref{eq:propagator-ambient-equation} to
\begin{equation}
	\left( -\nabla_{S^D}^2+\Delta^+\Delta^- \right) \phi(\widehat X) = j(\widehat X).
	\label{eq:propagator-sphere-equation}
\end{equation}
This reproduces the massive scalar equation \eqref{eq:massive-scalar-equation} when
\begin{equation}
	M^2=\Delta^+\Delta^-, \qquad \Delta^++\Delta^-=D-1.
	\label{eq:mass-delta-parameterization}
\end{equation}
We define the sphere propagator \(G_{\Delta^\pm}(\widehat X,\widehat Y)\) by
\begin{equation}
	\left( -\nabla_{S^D}^2+\Delta^+\Delta^- \right) G_{\Delta^\pm}(\widehat X,\widehat Y) = \delta_{S^D}(\widehat X,\widehat Y),
	\label{eq:propagator-target-equation}
\end{equation}
where the delta function is normalized by
\begin{equation}
	\int_{S^D} d\Omega_D(\widehat X)\, \delta_{S^D}(\widehat X,\widehat Y) = 1.
	\label{eq:sphere-delta-normalization}
\end{equation}
The solution to \eqref{eq:propagator-sphere-equation} is then
\begin{equation}
	\phi(\widehat X) = \int_{S^D} d\Omega_D(\widehat Y)\, G_{\Delta^\pm}(\widehat X,\widehat Y)\, j(\widehat Y).
	\label{eq:propagator-solution}
\end{equation}

We obtain the sphere propagator from the massless scalar propagator on \(\mathbb R^{D+1}\), given by
\begin{equation}
	G_{\mathbb R}(X,Y) = \int \frac{d^{D+1}K}{(2\pi)^{D+1}}\, \frac{e^{\mathrm{i}K\cdot(X-Y)}}{K^2}.
	\label{eq:propagator-flat-green}
\end{equation}
Under a common rescaling of \(X\) and \(Y\), the propagator transforms homogeneously as
\begin{equation}
	G_{\mathbb R}(\mathscr q X,\mathscr q Y) = \mathscr q^{-(D-1)} G_{\mathbb R}(X,Y).
	\label{eq:propagator-flat-homogeneity}
\end{equation}
The solution to the ambient equation \eqref{eq:propagator-ambient-equation} is
\begin{equation}
	\Phi(X) = \int_{\mathbb R^{D+1}} d^{D+1}Y\,G_{\mathbb R}(X,Y)\mathsf J(Y).
	\label{eq:propagator-ambient-solution}
\end{equation}
Using the radial coordinates and measure
\begin{equation}
	X=\mathscr r\widehat X, \qquad Y=\mathscr s\widehat Y, \qquad d^{D+1}Y = \mathscr s^D d\mathscr s\,d\Omega_D(\widehat Y),
\end{equation}
together with the lifted source \eqref{eq:propagator-source-lift}, the ambient solution \eqref{eq:propagator-ambient-solution} becomes
\begin{align}
	\Phi(\mathscr r\widehat X) &= \int_{S^D} d\Omega_D(\widehat Y)\, j(\widehat Y) \int_0^\infty \frac{d\mathscr s}{\mathscr s}\, \mathscr s^{D - 1 - \Delta^+} G_{\mathbb R} (\mathscr r\widehat X,\mathscr s\widehat Y).
	\label{eq:propagator-before-mellin-change}
\end{align}
Setting \(\lambda=\mathscr r/\mathscr s\) and using the homogeneity relation \eqref{eq:propagator-flat-homogeneity} gives
\begin{align}
	\Phi(\mathscr r\widehat X) &= \mathscr r^{-\Delta^+} \int_{S^D} d\Omega_D(\widehat Y)\, j(\widehat Y) \int_0^\infty \frac{d\lambda}{\lambda}\, \lambda^{\Delta^+} G_{\mathbb R} (\lambda\widehat X,\widehat Y).
\end{align}
Comparing this result with the sphere solution \eqref{eq:propagator-solution} and the homogeneous lift \eqref{eq:propagator-homogeneous-lift} identifies the sphere propagator as
\begin{equation}
	G_{\Delta^\pm}(\widehat X,\widehat Y) = \int_0^\infty \frac{d\lambda}{\lambda}\, \lambda^{\Delta^+} G_{\mathbb R} (\lambda\widehat X,\widehat Y).
	\label{eq:propagator-one-sided-mellin}
\end{equation}
The integral in \eqref{eq:propagator-one-sided-mellin} is a Mellin transform. Introducing a second Mellin scale gives the symmetric, scale-redundant form
\begin{align}
	G_{\Delta^\pm} (\widehat X,\widehat Y) &= \frac{1}{\operatorname{Vol}(\mathbb R_+)} \int_0^\infty \frac{d\lambda}{\lambda} \int_0^\infty \frac{d\mu}{\mu}\, \lambda^{\Delta^+} \mu^{\Delta^-} G_{\mathbb R} (\lambda\widehat X,\mu\widehat Y).
	\label{eq:propagator-radial-quotient}
\end{align}
The common rescaling \((\lambda,\mu)\mapsto(\mathscr q\lambda,\mathscr q\mu)\) leaves the integrand invariant: the radial weights contribute \(\mathscr q^{D-1}\), which cancels the homogeneity \(\mathscr q^{-(D-1)}\) of \(G_{\mathbb R}\). The factor
\begin{equation}
	\operatorname{Vol}(\mathbb R_+) = \int_0^\infty\frac{d\mathscr q}{\mathscr q}
\end{equation}
divides out this redundant scale. Fixing \(\mu=1\) recovers the one-sided Mellin representation \eqref{eq:propagator-one-sided-mellin}.

For later convenience, set \(K=-2P\) in \eqref{eq:propagator-flat-green}, which gives
\begin{equation}
	G_{\mathbb R}(X,Y) = \frac{1}{4\pi^{D+1}} \int d^{D+1}P\, \frac{e^{-2\mathrm{i}P\cdot(X-Y)}}{P^2}.
	\label{eq:propagator-flat-rescaled}
\end{equation}
Substituting \eqref{eq:propagator-flat-rescaled} into the scale-redundant representation \eqref{eq:propagator-radial-quotient}, we get
\begin{align}
	G_{\Delta^\pm}(\widehat X,\widehat Y) &= \frac{1}{\operatorname{Vol}(\mathbb R_+)} \frac{1}{4\pi^{D+1}} \int_0^\infty \frac{d\lambda}{\lambda} \int_0^\infty \frac{d\mu}{\mu}\, \lambda^{\Delta^+} \mu^{\Delta^-} \int d^{D+1}P\, \frac{ e^{-2\mathrm{i}P\cdot (\lambda\widehat X-\mu\widehat Y)} }{P^2}.
	\label{eq:propagator-embedding-form}
\end{align}
For \(X=\mathscr r\widehat X\) and \(Y=\mathscr s\widehat Y\), we assign \(\Delta^+\) to the first endpoint and \(\Delta^-\) to the second. Endpoint homogeneity in \eqref{eq:propagator-embedding-form} gives
\begin{equation}
	G_{\Delta^\pm} (\mathscr r\widehat X,\mathscr s\widehat Y) = \mathscr r^{-\Delta^+}\mathscr s^{-\Delta^-} G_{\Delta^\pm}(\widehat X,\widehat Y).
	\label{eq:propagator-physical-homogeneity}
\end{equation}
\subsection{General edge}
We define a general edge by extending the physical edge \eqref{eq:propagator-embedding-form} to
\begin{align}
	G_{\eta,\Delta^+} (\widehat X,\widehat Y) &= \frac{1}{\operatorname{Vol}(\mathbb R_+)} \frac{1}{4^\eta\pi^{D+1}} \int_0^\infty \frac{d\lambda}{\lambda} \int_0^\infty \frac{d\mu}{\mu}\, \lambda^{\Delta^+} \mu^{\Delta^-} \int d^{D+1}P\, \frac{ e^{-2\mathrm{i}P\cdot (\lambda\widehat X-\mu\widehat Y)} }{(P^2)^\eta}.
	\label{eq:propagator-generalized-line}
\end{align}
Here \(\Delta^+\) and \(\Delta^-\) are the radial weights at the two endpoints. The general edge is likewise homogeneous at both endpoints, as in \eqref{eq:propagator-physical-homogeneity}. Scale invariance under the common rescaling \((\lambda,\mu,P)\mapsto (\mathscr q\lambda,\mathscr q\mu,\mathscr q^{-1}P)\) requires
\begin{equation}
	\Delta^++\Delta^-+2\eta = D+1.
	\label{eq:propagator-generalized-balance}
\end{equation}
At fixed \(D\), specifying \(\eta\) and \(\Delta^+\) fixes \(\Delta^-=D+1-2\eta-\Delta^+\), which we therefore omit from the kernel label. For \(\eta=1\), this recovers \(\Delta^++\Delta^-=D-1\), the physical-edge relation. For general \(\eta\), we interpret the edge through its descent equations and then derive its position-space and spectral forms.
\paragraph{Descent equations.}
Define the flat kernel appearing in \eqref{eq:propagator-generalized-line} by
\begin{equation}
	G_{\mathbb R,\eta}(X,Y) = \frac{1}{4^\eta\pi^{D+1}} \int d^{D+1}P\, \frac{e^{-2\mathrm{i}P\cdot(X-Y)}}{(P^2)^\eta}.
	\label{eq:propagator-flat-general-line}
\end{equation}
It obeys the descent equation
\begin{equation}
	-\partial_X^2 G_{\mathbb R,\eta}(X,Y) = G_{\mathbb R,\eta-1}(X,Y).
	\label{eq:propagator-flat-descent}
\end{equation}
The flat-space physical and contact kernels correspond to \(\eta=1\) and \(\eta=0\), respectively. Each descent sends \(\eta\) to \(\eta-1\), while the operator \(-\partial_X^2\) remains unchanged. Thus, positive integers \(\eta>1\) give successive inverses of the flat Laplacian, while negative integers give its successive action on the contact kernel.
% At \(\eta=0\),
% \begin{equation}
%   G_{\mathbb R,0}(X,Y)
%   =
%   \delta^{(D+1)}(X-Y).
%   \label{eq:propagator-flat-zero-line}
% \end{equation}
% Equivalently, the flat kernel is the Green function of a power of the
% Laplacian:
% \begin{equation}
%   \left(-\partial_X^2\right)^\eta
%   G_{\mathbb R,\eta}(X,Y)
%   =
%   \delta^{(D+1)}(X-Y).
%   \label{eq:propagator-flat-fractional-green}
% \end{equation}

To derive the corresponding descent equation on the sphere, replace \(\widehat X\) with \(X=\mathscr r\widehat X\) in \eqref{eq:propagator-generalized-line} and apply \(-\partial_X^2\) at the first endpoint. Differentiating the plane wave supplies \(4\lambda^2P^2\), which shifts \((\eta,\Delta^+)\) to \((\eta-1,\Delta^++2)\). Thus
\begin{equation}
	-\partial_X^2 G_{\eta,\Delta^+}(X,\widehat Y) = G_{\eta-1,\Delta^++2}(X,\widehat Y).
	\label{eq:propagator-ambient-descent}
\end{equation}
Applying the same radial reduction as in \eqref{eq:propagator-radial-reduction} to the first endpoint gives
\begin{equation}
	-\partial_X^2 G_{\eta,\Delta^+}(\mathscr r\widehat X,\widehat Y) = \mathscr r^{-\Delta^+-2} \left[ -\nabla_{S^D}^2 + \Delta^+(D-1-\Delta^+) \right] G_{\eta,\Delta^+}(\widehat X,\widehat Y).
	\label{eq:propagator-ambient-left-radial}
\end{equation}
The first-endpoint homogeneity of the shifted kernel in \eqref{eq:propagator-ambient-descent} gives
\begin{equation}
	G_{\eta-1,\Delta^++2}(X,\widehat Y) = \mathscr r^{-\Delta^+-2} G_{\eta-1,\Delta^++2}(\widehat X,\widehat Y).
	\label{eq:propagator-shifted-homogeneity}
\end{equation}
Substituting \eqref{eq:propagator-ambient-left-radial} and \eqref{eq:propagator-shifted-homogeneity} into \eqref{eq:propagator-ambient-descent} and canceling \(\mathscr r^{-\Delta^+-2}\) gives
\begin{equation}
	\left[ -\nabla_{S^D}^2 + \Delta^+(D-1-\Delta^+) \right] G_{\eta,\Delta^+}(\widehat X,\widehat Y) = G_{\eta-1,\Delta^++2}(\widehat X,\widehat Y).
	\label{eq:propagator-left-descent}
\end{equation}
The physical and contact edges correspond to \(\eta=1\) and \(\eta=0\), respectively. Each descent sends \((\eta,\Delta^+)\) to \((\eta-1,\Delta^++2)\), preserving the scale-invariance condition \(\Delta^++\Delta^-+2\eta=D+1\). The ambient operator \(-\partial_X^2\) remains the same, but its reduction on the sphere depends on the shifted radial weight. After \(k\) descents, the next sphere operator is \(-\nabla_{S^D}^2+(\Delta^++2k)(D-1-\Delta^+-2k)\). Thus, positive integers \(\eta>1\) give successive inverses of different shifted Laplacians, while negative integers give their successive action on the contact edge.
\paragraph{Hypergeometric form.}
Starting from the Mellin representation \eqref{eq:propagator-generalized-line}, we fix the common radial scale by setting \(\mu=1\) and evaluate the momentum integral to obtain
\begin{equation}
	G_{\eta,\Delta^+}(\widehat X,\widehat Y) = \frac{ \Gamma\!\left(\frac{D+1}{2}-\eta\right) }{ 4^\eta\pi^{\frac{D+1}{2}}\Gamma(\eta) } \int_0^\infty\frac{d\lambda}{\lambda}\, \frac{\lambda^{\Delta^+}} {\left( 1+\lambda^2-2\lambda\,\widehat X\cdot\widehat Y \right)^{\frac{D+1}{2}-\eta}}.
\end{equation}
Setting \(\lambda=u/(1-u)\) and using the scale-invariance condition \eqref{eq:propagator-generalized-balance}, we obtain
\begin{equation}
	G_{\eta,\Delta^+}(\widehat X,\widehat Y) = \frac{\Gamma\!\left(\frac{D+1}{2}-\eta\right)}{4^\eta\pi^{\frac{D+1}{2}}\Gamma(\eta)} \int_0^1du\,u^{\Delta^+-1}(1-u)^{\Delta^--1} \left[1-2\left(1+\widehat X\cdot\widehat Y\right)u(1-u)\right]^{-\frac{D+1}{2}+\eta}.
\end{equation}
Expanding the last factor and evaluating the beta integral term by term gives
\begin{equation}
	G_{\eta,\Delta^+}(\widehat X,\widehat Y) = \frac{\Gamma(\Delta^+)\Gamma(\Delta^-)}{(4\pi)^{D/2}\Gamma(\eta)\Gamma\!\left(\frac{D+2}{2}-\eta\right)} {}_2F_1\!\left(\Delta^+,\Delta^-;\frac{D+2}{2}-\eta;\frac{1+\widehat X\cdot\widehat Y}{2}\right).
\end{equation}
These formulas are understood by analytic continuation and, where necessary, as distributions. For the physical edge, \(\eta=1\) and \(\Delta^++\Delta^-=D-1\). The hypergeometric form becomes
\begin{equation}
	G_{\Delta^\pm}(\widehat X,\widehat Y) := G_{1,\Delta^+}(\widehat X,\widehat Y) = \frac{ \Gamma(\Delta^+)\Gamma(\Delta^-) }{ (4\pi)^{D/2}\Gamma(D/2) } {}_2F_1\!\left( \Delta^+,\Delta^-; \frac D2; \frac{1+\widehat X\cdot\widehat Y}{2} \right).
	\label{eq:propagator-hypergeometric-form}
\end{equation}
\paragraph{Spectral form.}
Expanding the general edge in an orthonormal basis of spherical harmonics gives
\begin{equation}
	G_{\eta,\Delta^+}(\widehat X,\widehat Y) = \sum_{\ell=0}^{\infty}\sum_{\boldsymbol{m}} \mathsf g_\ell(\eta,\Delta^+) Y_{\ell\boldsymbol{m}}(\widehat X)Y_{\ell\boldsymbol{m}}^*(\widehat Y).
\end{equation}
Using the left descent equation \eqref{eq:propagator-left-descent} and \(-\nabla_{S^D}^2Y_{\ell\boldsymbol{m}}=\ell(\ell+D-1)Y_{\ell\boldsymbol{m}}\), we find
\begin{equation}
	(\ell+\Delta^+)(\ell+D-1-\Delta^+)\, \mathsf g_\ell(\eta,\Delta^+) = \mathsf g_\ell(\eta-1,\Delta^++2).
\end{equation}
At \(\eta=0\), completeness gives the contact edge and hence the initial condition
\begin{equation*}
	G_{0,\Delta^+}(\widehat X,\widehat Y) = \delta_{S^D}(\widehat X,\widehat Y) \quad\Longrightarrow\quad \mathsf g_\ell(0,\Delta^+)=1.
\end{equation*}
Iterating the recurrence and analytically continuing in \(\eta\) gives the spectral form
\begin{equation}
	G_{\eta,\Delta^+}(\widehat X,\widehat Y) = \frac{1}{4^\eta}\sum_{\ell=0}^{\infty}\sum_{\boldsymbol{m}} \frac{\Gamma\!\left(\frac{\ell+\Delta^+}{2}\right)\Gamma\!\left(\frac{\ell+\Delta^-}{2}\right)} {\Gamma\!\left(\frac{\ell+\Delta^+}{2}+\eta\right)\Gamma\!\left(\frac{\ell+\Delta^-}{2}+\eta\right)} Y_{\ell\boldsymbol{m}}(\widehat X)Y_{\ell\boldsymbol{m}}^*(\widehat Y).
\end{equation}
For the physical edge, \(\eta=1\), so the spectral form becomes
\begin{equation}
	G_{\Delta^\pm}(\widehat X,\widehat Y) = G_{1,\Delta^+}(\widehat X,\widehat Y) = \sum_{\ell=0}^{\infty}\sum_{\boldsymbol{m}} \frac{ Y_{\ell\boldsymbol{m}}(\widehat X)Y_{\ell\boldsymbol{m}}^*(\widehat Y) }{ (\ell+\Delta^+)(\ell+\Delta^-) }.
	\label{eq:propagator-spectral-form}
\end{equation}
\section{Vacuum Graph Integral}
\label{app:graph-integrals}

In this appendix, we derive the Schwinger-parameter representation used in Section~\ref{sec:graph-lattices} for a general vacuum graph integral.

\subsection{Scale-redundant form}
Let \(g\) be a vacuum graph with \(E\) oriented general edges and \(V\) vertices. Assign \(\widehat X_v\in S^D\) to each vertex, and write \(\boldsymbol{\eta}=(\eta_1,\ldots,\eta_E)\) and \(\boldsymbol{\Delta}^{+}=(\Delta_1^+,\ldots,\Delta_E^+)\). The associated vacuum graph integral is
\begin{equation}
	I_g(D;\boldsymbol{\eta},\boldsymbol{\Delta}^{+}) = \int_{(S^D)^V} \prod_{v=1}^{V} d\Omega_D(\widehat X_v) \prod_{i=1}^{E} G_{\eta_i,\Delta_i^+} (\widehat X_{\mathsf a(i)},\widehat X_{\mathsf b(i)}).
	\label{eq:graph-position-integral}
\end{equation}
Here \(\Delta_i^-=D+1-2\eta_i-\Delta_i^+\), and \(\mathsf a,\mathsf b:\{1,\ldots,E\}\longrightarrow\{1,\ldots,V\}\) are the endpoint maps, with edge \(i\) oriented from \(\mathsf a(i)\) to \(\mathsf b(i)\). In the intermediate steps below, we write \(I_g\) for \(I_g(D;\boldsymbol{\eta},\boldsymbol{\Delta}^{+})\).

Lifting each \(\widehat X_v\) in \eqref{eq:graph-position-integral} from \(S^D\) to a nonzero vector \(X_v\in\mathbb R^{D+1}\) gives
\begin{equation}
	I_g = \frac{1}{\operatorname{Vol}((\mathbb R_+)^V)} \int \prod_{v=1}^{V} \frac{d^{D+1}X_v}{(X_v^2)^{(D+1)/2}} \prod_{i=1}^{E} G_{\eta_i,\Delta_i^+} \left( \frac{X_{\mathsf a(i)}}{|X_{\mathsf a(i)}|}, \frac{X_{\mathsf b(i)}}{|X_{\mathsf b(i)}|} \right).
	\label{eq:graph-lifted-position-integral}
\end{equation}
Applying endpoint homogeneity to each edge in \eqref{eq:graph-lifted-position-integral} gives
\begin{equation}
	I_g = \frac{1}{\operatorname{Vol}((\mathbb R_+)^V)} \int \prod_{v=1}^{V} \frac{d^{D+1}X_v}{(X_v^2)^{w_v}} \prod_{i=1}^{E} G_{\eta_i,\Delta_i^+} (X_{\mathsf a(i)},X_{\mathsf b(i)}),
	\label{eq:graph-homogeneous-position-integral}
\end{equation}
where the vertex weight \(w_v\) depends on the weighted degree \(B_v\) of \(v\) through
\begin{equation}
	w_v = \frac{D+1-B_v}{2}, \qquad B_v = \sum_{\mathsf a(i)=v}\Delta_i^+ + \sum_{\mathsf b(i)=v}\Delta_i^-.
	\label{eq:graph-vertex-weight}
\end{equation}
Introducing a Schwinger parameter \(\alpha_v\) for each denominator \((X_v^2)^{w_v}\) gives the vertex-Schwinger representation
\begin{equation}
	I_g=\frac{1}{\operatorname{Vol}((\mathbb R_+)^V)\prod_{v=1}^{V}\Gamma(w_v)} \int_0^\infty\prod_{v=1}^{V}\left[\frac{d\alpha_v}{\alpha_v}\,\alpha_v^{w_v}\right] \int\prod_{v=1}^{V} d^{D+1}X_v\,e^{-\sum_v\alpha_vX_v^2} \prod_{i=1}^{E}G_{\eta_i,\Delta_i^+}(X_{\mathsf a(i)},X_{\mathsf b(i)}).
	\label{eq:graph-vertex-Schwinger-integral}
\end{equation}
With a Schwinger parameter \(\tau_i\) for each denominator \((P_i^2)^{\eta_i}\) in \eqref{eq:propagator-generalized-line}, the edge kernels become
\begin{align}
	G_{\eta_i,\Delta_i^+} (X_{\mathsf a(i)},X_{\mathsf b(i)}) &={} \frac{1}{\operatorname{Vol}(\mathbb R_+)} \frac{1}{4^{\eta_i}\pi^{D+1}\Gamma(\eta_i)} \int_0^\infty \frac{d\lambda_i}{\lambda_i} \int_0^\infty \frac{d\mu_i}{\mu_i} \int_0^\infty \frac{d\tau_i}{\tau_i}\, \lambda_i^{\Delta_i^+} \mu_i^{\Delta_i^-} \tau_i^{\eta_i}\notag\\
	&\hspace{3cm}\times \int d^{D+1}P_i\, \exp\left[ -\tau_iP_i^2 -2\mathrm{i}P_i\cdot \left( \lambda_iX_{\mathsf a(i)} -\mu_iX_{\mathsf b(i)} \right) \right].
	\label{eq:general-line-Schwinger}
\end{align}
Substituting the edge kernels into the vertex-Schwinger representation leaves Gaussian integrals over \(P_i\) and \(X_v\), with exponent
\begin{equation}
	-\sum_{i=1}^{E}\tau_iP_i^2 -\sum_{v=1}^{V}\alpha_vX_v^2 -2\mathrm{i}\sum_{i=1}^{E} P_i\cdot \left( \lambda_iX_{\mathsf a(i)} -\mu_iX_{\mathsf b(i)} \right).
	\label{eq:graph-unfixed-exponent}
\end{equation}
Write the ambient momenta and vertex vectors as \(W^T=(P_1,\ldots,P_E,X_1,\ldots,X_V)\), and define the \(E\times V\) matrix \(L\) by
\begin{equation}
	L_{iv} = \lambda_i\delta_{v,\mathsf a(i)} - \mu_i\delta_{v,\mathsf b(i)}.
	\label{eq:graph-incidence-matrix}
\end{equation}
The exponent is then \(-W^T\widetilde U_gW\), where
\begin{equation}
	\widetilde U_g = \begin{pmatrix}
		\operatorname{diag}(\tau_1,\ldots,\tau_E) & \mathrm{i}L \\
		\mathrm{i}L^T & \operatorname{diag}(\alpha_1,\ldots,\alpha_V)
	\end{pmatrix}.
	\label{eq:graph-unfixed-matrix}
\end{equation}
Multiplying the vertex prefactor in \eqref{eq:graph-vertex-Schwinger-integral} by the \(E\) edge prefactors in \eqref{eq:general-line-Schwinger} gives the normalization
\begin{equation}
	\widetilde N_g = \frac{1}{\operatorname{Vol}((\mathbb R_+)^{E+V})} \frac{1}{ \pi^{E(D+1)} \prod_{i=1}^{E}4^{\eta_i}\Gamma(\eta_i) \prod_{v=1}^{V}\Gamma(w_v) },
	\label{eq:graph-unfixed-normalization}
\end{equation}
and then the graph integral takes the scale-redundant form
\begin{equation}
	I_g = \widetilde N_g \int_0^\infty \prod_{i=1}^{E} \left[ \frac{d\lambda_i}{\lambda_i} \frac{d\mu_i}{\mu_i} \frac{d\tau_i}{\tau_i}\, \lambda_i^{\Delta_i^+} \mu_i^{\Delta_i^-} \tau_i^{\eta_i} \right] \int_0^\infty \prod_{v=1}^{V} \left[ \frac{d\alpha_v}{\alpha_v}\, \alpha_v^{w_v} \right] \int d^{(D+1)(E+V)}W\, e^{-W^T\widetilde U_gW}.
	\label{eq:graph-unfixed-matrix-integral}
\end{equation}
In this form, the integral \eqref{eq:graph-unfixed-matrix-integral} has one independent positive rescaling for each edge and vertex. Table~\ref{tab:graph-gauge-fixing} gives their action on the integration variables.
\begin{table}[H]
\centering
\small
\begin{tabular*}{\textwidth}{
  @{\extracolsep{\fill}}
  p{0.11\textwidth}
  p{0.82\textwidth}
  @{}
}
\toprule
Scale & Transformation\\
\midrule
Edge $i$
&
$P_i\mapsto \mathscr q_i^{-1}P_i$,
$\tau_i\mapsto \mathscr q_i^2\tau_i$,
$\lambda_i\mapsto \mathscr q_i\lambda_i$,
$\mu_i\mapsto \mathscr q_i\mu_i$
\\[1mm]
Vertex $v$
&
$X_v\mapsto \mathscr r_v^{-1}X_v$,
$\alpha_v\mapsto \mathscr r_v^2\alpha_v$;
$\lambda_i\mapsto \mathscr r_v\lambda_i$ if $\mathsf a(i)=v$ and
$\mu_i\mapsto \mathscr r_v\mu_i$ if $\mathsf b(i)=v$
\\
\bottomrule
\end{tabular*}
\caption{Edge and vertex scale redundancies.}
\label{tab:graph-gauge-fixing}
\end{table}
For edge \(i\), invariance follows from \(\Delta_i^++\Delta_i^-+2\eta_i=D+1\); for vertex \(v\), it follows from \(2w_v+B_v=D+1\). The factor \(\operatorname{Vol}((\mathbb R_+)^{E+V})^{-1}\) in \(\widetilde N_g\) therefore divides out their scale volume.
\subsection{Scale-fixed Schwinger-parameter form}
To fix these scales, insert the following Faddeev--Popov identities into the integral \eqref{eq:graph-unfixed-matrix-integral}, one for every edge \(i\) and vertex \(v\):
\begin{equation}
	1 = 2\int_0^\infty\frac{d\mathscr q_i}{\mathscr q_i}\, \delta(\mathscr q_i^2\tau_i-1), \qquad 1 = 2\int_0^\infty\frac{d\mathscr r_v}{\mathscr r_v}\, \delta(\mathscr r_v^2\alpha_v-1).
	\label{eq:graph-faddeev-popov}
\end{equation}
We use the rescalings in Table~\ref{tab:graph-gauge-fixing} to absorb \(\mathscr q_i\) and \(\mathscr r_v\) into the integration variables. The delta functions then impose \(\tau_i=1\) and \(\alpha_v=1\), while the \(E+V\) factors of \(2\) give \(2^{E+V}\). The scale integrals cancel the factor \(\operatorname{Vol}((\mathbb R_+)^{E+V})^{-1}\) in \eqref{eq:graph-unfixed-normalization}, leaving
\begin{equation}
	I_g = \frac{2^{E+V-2\sum_{i=1}^{E}\eta_i}}{ \pi^{E(D+1)} \prod_{i=1}^{E}\Gamma(\eta_i) \prod_{v=1}^{V}\Gamma(w_v) } \int_0^\infty \prod_{i=1}^{E} \left[ \frac{d\lambda_i}{\lambda_i} \frac{d\mu_i}{\mu_i}\, \lambda_i^{\Delta_i^+} \mu_i^{\Delta_i^-} \right] \int d^{(D+1)(E+V)}W\, e^{-W^TU_gW},
	\label{eq:graph-gauge-fixed-matrix-integral}
\end{equation}
where
\begin{equation}
	U_g := \left.\widetilde U_g\right|_{ \boldsymbol{\tau}=\boldsymbol{1},\, \boldsymbol{\alpha}=\boldsymbol{1} } = \begin{pmatrix}
		I_E & \mathrm{i}L\\
		\mathrm{i}L^T & I_V
	\end{pmatrix}.
\end{equation}
Define the graph polynomial by
\begin{equation}
	A_g := \det U_g = \det(I_V+L^TL) = \det(I_E+LL^T).
	\label{eq:graph-polynomial}
\end{equation}
Performing the Gaussian integral over \(W\) in \eqref{eq:graph-gauge-fixed-matrix-integral} gives the Euler form
\begin{equation}
	I_g = \frac{ 2^{E+V-2\sum_i\eta_i} \pi^{\frac{D+1}{2}(V-E)} }{ \prod_{i=1}^{E}\Gamma(\eta_i) \prod_{v=1}^{V}\Gamma(w_v) } \int_0^\infty \prod_{i=1}^{E} \left[ \frac{d\lambda_i}{\lambda_i} \frac{d\mu_i}{\mu_i}\, \lambda_i^{\Delta_i^+} \mu_i^{\Delta_i^-} \right] A_g^{-(D+1)/2}.
\end{equation}
Finally, introducing a Schwinger parameter \(z\) for \(A_g^{-(D+1)/2}\) gives
\begin{align}
    I_g(D;\boldsymbol{\eta},\boldsymbol{\Delta}^{+}) &= \frac{ 2^{E+V-2\sum_i\eta_i} \pi^{\frac{D+1}{2}(V-E)}}{ \Gamma\!\left(\frac{D+1}{2}\right) \prod_{i=1}^{E}\Gamma(\eta_i) \prod_{v=1}^{V}\Gamma(w_v) }\times\notag\\
    & \hspace{1.5cm}\int_0^\infty \frac{dz}{z}\, z^{\frac{D+1}{2}} \prod_{i=1}^{E} \frac{d\lambda_i}{\lambda_i} \frac{d\mu_i}{\mu_i}\, \lambda_i^{\Delta_i^+} \mu_i^{\Delta_i^-} e^{-zA_g}.
	\label{eq:graph-final-z-form}
\end{align}
\paragraph{Spectral parameterization.}
For each edge, \(D\) and \(\eta_i\) determine the sum of \(\Delta_i^\pm\), and we introduce the spectral parameter \(\rho_i\) through their difference:
\begin{equation}
	\frac{\Delta_i^++\Delta_i^-}{2}= \frac{D+1}{2} - \eta_i, \qquad \frac{\Delta_i^+-\Delta_i^-}{2} = \rho_i.
\end{equation}

A convenient parameterization of the dimension and spectral data is then
\begin{equation}
	D=2t+1,\qquad \Delta_i^+=u_i+\rho_i, \qquad \Delta_i^-=u_i-\rho_i, \qquad \eta_i=t+1-u_i.
	\label{eq:graph-line-parameters}
\end{equation}
In terms of \(\boldsymbol{x}=(t,\boldsymbol{u},\boldsymbol{\rho})\), write \(I_g(\boldsymbol{x}):=I_g(D;\boldsymbol{\eta},\boldsymbol{\Delta}^{+})\). The gauge-fixed graph integral \eqref{eq:graph-final-z-form} can then be written as
\begin{equation}
	I_g(\boldsymbol{x}) = N_g(\boldsymbol{x})J_g(\boldsymbol{x}),
	\label{eq:graph-final-I-NJ}
\end{equation}
\begin{equation}
	\text{where}\quad J_g(\boldsymbol{x}) = \int_0^\infty \frac{dz}{z}\, z^{t+1} \prod_{i=1}^{E} \frac{d\lambda_i}{\lambda_i} \frac{d\mu_i}{\mu_i}\, \lambda_i^{u_i+\rho_i} \mu_i^{u_i-\rho_i} e^{-zA_g(\boldsymbol{\lambda},\boldsymbol{\mu})},
	\label{eq:graph-final-Euler-integral}
\end{equation}
\begin{equation}
	\text{and}\quad N_g(\boldsymbol{x}) = \frac{ 2^{E+V-2\sum_i\eta_i(\boldsymbol{x})} \pi^{(t+1)(V-E)} }{ \Gamma(t+1) \prod_{i=1}^{E}\Gamma\!\bigl(\eta_i(\boldsymbol{x})\bigr) \prod_{v=1}^{V}\Gamma\!\bigl(w_v(\boldsymbol{x})\bigr) }.
	\label{eq:graph-final-normalization}
\end{equation}

A physical edge has \(\eta_i=1\), so \(u_i=t\). When every edge is physical, denote the common value of \(t\) and all \(u_i\) by \(s\). The physical point is \(\bar{\boldsymbol{x}}:=(s,s\boldsymbol{1}_E,\boldsymbol{\rho})\), for which
\begin{equation}
	D=2s+1, \qquad \Delta_i^+=s+\rho_i, \qquad \Delta_i^-=s-\rho_i.
\end{equation}
Specializing \eqref{eq:graph-final-I-NJ}--\eqref{eq:graph-final-normalization} to \(\bar{\boldsymbol{x}}\) gives
\begin{equation}
	I_g(\bar{\boldsymbol{x}}) = N_g(\bar{\boldsymbol{x}}) J_g(\bar{\boldsymbol{x}}),
	\label{eq:graph-physical-I-NJ}
\end{equation}
\begin{equation}
	\text{where}\quad J_g(\bar{\boldsymbol{x}}) = \int_0^\infty \frac{dz}{z}\, z^{s+1} \prod_{i=1}^{E} \frac{d\lambda_i}{\lambda_i} \frac{d\mu_i}{\mu_i}\, \lambda_i^{s+\rho_i} \mu_i^{s-\rho_i} e^{-zA_g(\boldsymbol{\lambda},\boldsymbol{\mu})},
	\label{eq:graph-physical-Euler-integral}
\end{equation}
\begin{equation}
	\text{and}\quad N_g(\bar{\boldsymbol{x}}) = \frac{ 2^{V-E} \pi^{(s+1)(V-E)} }{ \Gamma(s+1) \prod_{v=1}^{V}\Gamma\!\bigl(w_v(\bar{\boldsymbol{x}})\bigr) }.
	\label{eq:graph-physical-normalization}
\end{equation}
\section{Graph Contractions and Cuts}
\label{app:contact-contractions}
Contracting a connected subgraph merges its vertices into a single vertex. We enforce this identification in the graph integral using insertions that turn the selected physical edge kernels into delta functions. We first treat a single edge, then apply the same construction to a forest. Finally, we introduce spectral cuts: residues that select a harmonic degree on a chosen edge. The zero-mode cut deletes the edge while retaining all vertices.

\subsection{Edge contractions}
Choose an oriented edge \(i\) with distinct endpoints and write
\begin{equation}
	a:=\mathsf a(i), \qquad b:=\mathsf b(i).
\end{equation}
Let \(g/i\) be the graph obtained by contracting edge \(i\) and identifying its endpoints \(a\) and \(b\). Write \(\pi_i\boldsymbol{x}\) for \(\boldsymbol{x}\) with the parameters of edge \(i\) removed, so that the contracted integral is \(I_{g/i}(\pi_i\boldsymbol{x})\). For any \(\xi\), define the insertion
\begin{equation}
	\mathcal O_{i,\xi} := \frac{ \lambda_i^{2\xi}\mu_i^{2(1-\xi)} }{ \alpha_a^{\xi}\alpha_b^{1-\xi} } P_i^2.
	\label{eq:contact-general-operator}
\end{equation}
This insertion is invariant under the edge and vertex rescalings in Table~\ref{tab:graph-gauge-fixing}. We denote the scale-redundant graph integral with the insertion \eqref{eq:contact-general-operator} by \(\langle\mathcal O_{i,\xi}\rangle_g\). We evaluate \(\langle\mathcal O_{i,\xi}\rangle_g\) in two different ways to relate \(I_g(\boldsymbol{x})\) to \(I_{g/i}(\pi_i\boldsymbol{x})\). The intermediate formulas keep \(\eta_i\) general; the contraction follows by specializing only edge \(i\) to the physical value \(\eta_i=1\).

First, inserting \(\lambda_i^{2\xi}\mu_i^{2(1-\xi)}P_i^2\) into the Mellin representation \eqref{eq:propagator-generalized-line} for edge \(i\) gives
\begin{align}
	& \frac{1}{\operatorname{Vol}(\mathbb R_+)} \frac{1}{4^{\eta_i}\pi^{D+1}} \int_0^\infty\frac{d\lambda_i}{\lambda_i} \int_0^\infty\frac{d\mu_i}{\mu_i}\, \lambda_i^{\Delta_i^++2\xi} \mu_i^{\Delta_i^-+2(1-\xi)} \int d^{D+1}P_i\, \frac{P_i^2}{(P_i^2)^{\eta_i}} e^{-2\mathrm{i}P_i\cdot (\lambda_iX_a-\mu_iX_b)} \notag\\
	&\hspace{3cm} = \frac14\, G_{\eta_i-1,\Delta_i^++2\xi} (X_a,X_b).
	\label{eq:contact-edge-descent}
\end{align}
Substituting this result into the vertex-Schwinger representation \eqref{eq:graph-vertex-Schwinger-integral} gives
\begin{align}
	\left\langle\mathcal O_{i,\xi}\right\rangle_g ={}& \frac{1}{ 4\,\operatorname{Vol}((\mathbb R_+)^V) \prod_{v=1}^{V}\Gamma(w_v) } \int_0^\infty \prod_{v=1}^{V} \left[ \frac{d\alpha_v}{\alpha_v}\,\alpha_v^{w_v} \right] \frac{1}{\alpha_a^{\xi}\alpha_b^{1-\xi}} \notag\\
	&\times \int\prod_{v=1}^{V}d^{D+1}X_v\, e^{-\sum_{v=1}^{V}\alpha_vX_v^2}\, G_{\eta_i-1,\Delta_i^++2\xi}(X_a,X_b) \prod_{j\ne i} G_{\eta_j,\Delta_j^+} (X_{\mathsf a(j)},X_{\mathsf b(j)}).
	\label{eq:contact-vertex-Schwinger-integral}
\end{align}
The shifts in \(\Delta_i^\pm\) change the endpoint weights to \(w_a-\xi\) and \(w_b+\xi-1\). The \(\alpha\)-factor in \(\mathcal O_{i,\xi}\) changes the two vertex measures to the same weights.
\begin{multline}
	\left[ \frac{d\alpha_a}{\alpha_a}\, \frac{\alpha_a^{w_a}}{\Gamma(w_a)} \right] \left[ \frac{d\alpha_b}{\alpha_b}\, \frac{\alpha_b^{w_b}}{\Gamma(w_b)} \right] \frac{1}{\alpha_a^{\xi}\alpha_b^{1-\xi}} \\
	= \frac{ \Gamma(w_a-\xi)\Gamma(w_b+\xi-1) }{ \Gamma(w_a)\Gamma(w_b) } \left[ \frac{d\alpha_a}{\alpha_a}\, \frac{\alpha_a^{w_a-\xi}} {\Gamma(w_a-\xi)} \right] \left[ \frac{d\alpha_b}{\alpha_b}\, \frac{\alpha_b^{w_b+\xi-1}} {\Gamma(w_b+\xi-1)} \right] .
\end{multline}
Since all other vertex measures and edges are unchanged, the first evaluation becomes
\begin{align}
	\left\langle\mathcal O_{i,\xi}\right\rangle_g ={}& \frac14 \frac{ \Gamma(w_a-\xi)\Gamma(w_b+\xi-1) }{ \Gamma(w_a)\Gamma(w_b) } \notag\\
	&\times \int_{(S^D)^V} \prod_{v=1}^{V}d\Omega_D(\widehat X_v)\, G_{\eta_i-1,\Delta_i^++2\xi} (\widehat X_a,\widehat X_b) \prod_{j\neq i} G_{\eta_j,\Delta_j^+} (\widehat X_{\mathsf a(j)},\widehat X_{\mathsf b(j)}).
\end{align}
At \(\eta_i=1\), the shifted kernel on edge \(i\) is the contact edge
\begin{equation}
	G_{0,\Delta_i^++2\xi} (\widehat X_a,\widehat X_b) = \delta_{S^D}(\widehat X_a,\widehat X_b).
	\label{eq:contact-edge}
\end{equation}
The delta function performs the \(\widehat X_b\) integral, contracting edge \(i\) and identifying its endpoints. The remaining integral is \(I_{g/i}(\pi_i\boldsymbol{x})\), so
\begin{equation}
	\left\langle\mathcal O_{i,\xi}\right\rangle_g = \frac14 \frac{ \Gamma(w_a-\xi)\Gamma(w_b+\xi-1) }{ \Gamma(w_a)\Gamma(w_b) } I_{g/i}(\pi_i\boldsymbol{x}).
	\label{eq:contact-general-boundary-evaluation}
\end{equation}

Second, write \(\langle\mathcal O_{i,\xi}\rangle_g\) in the scale-redundant Gaussian form as
\begin{align}
	\left\langle \mathcal O_{i,\xi} \right\rangle_g &={} \widetilde N_g \int_0^\infty \prod_{j=1}^{E} \left[ \frac{d\lambda_j}{\lambda_j} \frac{d\mu_j}{\mu_j} \frac{d\tau_j}{\tau_j}\, \lambda_j^{\Delta_j^+} \mu_j^{\Delta_j^-} \tau_j^{\eta_j} \right] \notag\\
	&\qquad\times \int_0^\infty \prod_{v=1}^{V} \left[ \frac{d\alpha_v}{\alpha_v}\, \alpha_v^{w_v} \right] \frac{ \lambda_i^{2\xi}\mu_i^{2(1-\xi)} }{ \alpha_a^{\xi}\alpha_b^{1-\xi} } \int d^{(D+1)(E+V)}W\, P_i^2e^{-W^T\widetilde U_gW}.
	\label{eq:contact-general-unfixed-integral}
\end{align}
Since \(\tau_i\) multiplies \(P_i^2\) in \(W^T\widetilde U_gW\), the \(W\)-integral in \eqref{eq:contact-general-unfixed-integral} is
\begin{equation}
	\int d^{(D+1)(E+V)}W\, P_i^2e^{-W^T\widetilde U_gW} = -\pi^{(D+1)(E+V)/2} \frac{\partial}{\partial\tau_i} (\det\widetilde U_g)^{-(D+1)/2}.
	\label{eq:contact-Gaussian-moment}
\end{equation}
Applying the chain rule gives
\begin{equation}
	-\frac{\partial}{\partial\tau_i}(\det\widetilde U_g)^{-(D+1)/2} = \frac{D+1}{2} \frac{\partial_{\tau_i}\det\widetilde U_g} {(\det\widetilde U_g)^{(D+1)/2+1}}.
	\label{eq:contact-det-derivative}
\end{equation}
Now fix the edge and vertex scales by setting \(\boldsymbol{\tau}=\boldsymbol{1}\) and \(\boldsymbol{\alpha}=\boldsymbol{1}\), and define the polynomial \(Q_i\) by
\begin{equation}
	Q_i := \left. \frac{\partial}{\partial\tau_i}\det\widetilde U_g \right|_{\boldsymbol{\tau}=\boldsymbol{1},\,\boldsymbol{\alpha}=\boldsymbol{1}} = \det(I_V+L_{\,\widehat i}^{T}L_{\,\widehat i}),
	\label{eq:app-contact-Q-definition}
\end{equation}
where \(L_{\,\widehat i}\) is \(L\) with row \(i\) removed. Equation \eqref{eq:contact-det-derivative} then gives
\begin{equation}
	-\left. \frac{\partial}{\partial\tau_i} (\det\widetilde U_g)^{-(D+1)/2} \right|_{\boldsymbol{\tau}=\boldsymbol{1},\,\boldsymbol{\alpha}=\boldsymbol{1}} = \frac{D+1}{2}\frac{Q_i}{A_g^{(D+1)/2+1}}.
	\label{eq:contact-gauge-fixed-derivative}
\end{equation}
Introducing the \(z\) parameter used in \eqref{eq:graph-final-z-form} rewrites this as
\begin{equation}
	\frac{D+1}{2}\frac{Q_i}{A_g^{(D+1)/2+1}} = \frac{1}{\Gamma((D+1)/2)} \int_0^\infty\frac{dz}{z}\, z^{(D+1)/2}(zQ_i)e^{-zA_g}.
	\label{eq:contact-z-insertion}
\end{equation}
At the same gauge-fixed point, the remaining factor from \(\mathcal O_{i,\xi}\) is \(\lambda_i^{2\xi}\mu_i^{2(1-\xi)}\). Using \(I_g(\boldsymbol{x})=N_g(\boldsymbol{x})J_g(\boldsymbol{x})\) and writing \(J_g[\mathcal M;\boldsymbol{x}]\) for the insertion of a monomial \(\mathcal M\) in \(J_g\), the second evaluation is
\begin{equation}
	\left\langle\mathcal O_{i,\xi}\right\rangle_g = N_g(\boldsymbol{x}) J_g[ z\lambda_i^{2\xi}\mu_i^{2(1-\xi)}Q_i; \boldsymbol{x} ].
	\label{eq:contact-Gaussian-evaluation}
\end{equation}

Equating the two evaluations at the physical value \(\eta_i=1\) gives the contraction identity
\begin{equation}
	4 \frac{\Gamma(w_a)\Gamma(w_b)} {\Gamma(w_a-\xi)\Gamma(w_b+\xi-1)} N_g(\boldsymbol{x}) J_g[ z\lambda_i^{2\xi}\mu_i^{2(1-\xi)}Q_i; \boldsymbol{x} ] = I_{g/i}(\pi_i\boldsymbol{x}).
	\label{eq:app-contact-fixed-equations}
\end{equation}
\subsection{Forest contractions}
The same construction contracts a forest \(F\) of physical edges, with \(\eta_i=1\) for every \(i\in F\). Let \(V(F)\) denote its incident vertices, \(g/F\) the graph obtained by contracting all edges in \(F\), and \(\pi_F\boldsymbol{x}\) the parameters remaining after those edges are removed. For arbitrary \(\xi_i\), use the scale-invariant insertion
\begin{equation}
	\mathcal O_{F,\boldsymbol{\xi}} := \prod_{i\in F}\mathcal O_{i,\xi_i} = \prod_{i\in F} \left[ \frac{\lambda_i^{2\xi_i}\mu_i^{2(1-\xi_i)}} {\alpha_{\mathsf a(i)}^{\xi_i}\alpha_{\mathsf b(i)}^{1-\xi_i}} P_i^2 \right].
	\label{eq:forest-contact-operator}
\end{equation}
The vertex weights shift from \(w_v\) to \(w_v-\kappa_v\), where
\begin{equation}
	\kappa_v := \sum_{i\in F} \left[ \xi_i\delta_{v,\mathsf a(i)} +(1-\xi_i)\delta_{v,\mathsf b(i)} \right].
	\label{eq:forest-contact-vertex-power}
\end{equation}
First, in the vertex-Schwinger representation, applying \eqref{eq:contact-edge-descent} and \eqref{eq:contact-edge} to each selected edge replaces its kernel by a sphere delta function, with a factor \(1/4\). Since \(F\) is a forest, these constraints are independent and contract all its edges. Including the shifted vertex measures gives
\begin{equation}
	\left\langle\mathcal O_{F,\boldsymbol{\xi}}\right\rangle_g = \frac{1}{4^{|F|}} \prod_{v\in V(F)} \frac{\Gamma(w_v-\kappa_v)}{\Gamma(w_v)} I_{g/F}(\pi_F\boldsymbol{x}).
	\label{eq:forest-contact-boundary}
\end{equation}

Second, in the scale-redundant Gaussian representation, each factor \(P_i^2\) is generated by \(-\partial_{\tau_i}\), as in \eqref{eq:contact-Gaussian-moment}. The forest insertion therefore requires one derivative for every \(i\in F\). Repeated application of the chain rule to \((\det\widetilde U_g)^{-(D+1)/2}\) produces mixed derivatives of the determinant. After fixing \(\boldsymbol{\tau}=\boldsymbol{1}\) and \(\boldsymbol{\alpha}=\boldsymbol{1}\), define these polynomials by
\begin{equation}
	Q_B := \left. \left(\prod_{i\in B}\frac{\partial}{\partial\tau_i}\right) \det\widetilde U_g \right|_{\boldsymbol{\tau}=\boldsymbol{1},\,\boldsymbol{\alpha}=\boldsymbol{1}} = \det\!\left(I_V+L_{\widehat B}^{T}L_{\widehat B}\right), \qquad \varnothing\ne B\subseteq F,
	\label{eq:forest-contact-minor}
\end{equation}
where \(L_{\widehat B}\) is obtained by deleting the rows indexed by \(B\).

The chain rule groups the derivatives into partitions of \(F\). Let \(\Pi(F)\) be the set of all partitions of \(F\). Each partition \(\mathfrak p\in\Pi(F)\) consists of disjoint nonempty subsets \(B\subseteq F\), called blocks, whose union is \(F\). We write \(|\mathfrak p|\) for the number of blocks. Each block contributes \(Q_B\), while introducing the Schwinger parameter as in \eqref{eq:contact-z-insertion} gives a factor of \(z\) per block. The resulting insertion polynomial is
\begin{equation}
	\mathcal P_F(z) := \sum_{\mathfrak p\in\Pi(F)} (-1)^{|F|-|\mathfrak p|} z^{|\mathfrak p|} \prod_{B\in\mathfrak p}Q_B.
	\label{eq:forest-contact-polynomial}
\end{equation}
For two and three selected edges, this gives
\begin{equation}
	\begin{aligned}
		\mathcal P_{\{i,j\}}(z) &=z^2Q_iQ_j-zQ_{ij}, \\
		\mathcal P_{\{i,j,k\}}(z) &=z^3Q_iQ_jQ_k -z^2\left(Q_{ij}Q_k+Q_{ik}Q_j+Q_{jk}Q_i\right) +zQ_{ijk},
	\end{aligned}
	\label{eq:forest-contact-first-depths}
\end{equation}
where \(Q_{ij}:=Q_{\{i,j\}}\), and similarly for \(Q_{ijk}\).

Restoring the remaining \(\lambda_i,\mu_i\) factors from the insertion and equating the two evaluations gives
\begin{equation}
	4^{|F|} \prod_{v\in V(F)} \frac{\Gamma(w_v)}{\Gamma(w_v-\kappa_v)} \,N_g(\boldsymbol{x})\, J_g\!\left[ \left(\prod_{i\in F}\lambda_i^{2\xi_i}\mu_i^{2(1-\xi_i)}\right) \mathcal P_F(z); \boldsymbol{x} \right] = I_{g/F}(\pi_F\boldsymbol{x}),
	\label{eq:forest-contact-fixed-equations}
\end{equation}
where \(\xi_i\in\{0,\tfrac12,1\}\). For \(F=\{i\}\), \(\mathcal P_F=zQ_i\), recovering \eqref{eq:app-contact-fixed-equations}.

\subsection{Spectral cuts}
\label{app:spectral-cuts}
Consider a general graph integral \(I_g(\boldsymbol{x})\) and choose a physical edge \(i\), so \(\eta_i=1\) and \(u_i=t\). Inserting its spectral form \eqref{eq:propagator-spectral-form} into the position-space graph integral \eqref{eq:graph-position-integral} gives
\begin{equation}
	I_g(\boldsymbol{x}) = \sum_{\ell=0}^{\infty} \frac{ I_g^{(i,\ell)}(\pi_i\boldsymbol{x}) }{ (t+\ell)^2-\rho_i^2 },
	\label{eq:graph-spectral-expansion}
\end{equation}
where
\begin{equation}
	I_g^{(i,\ell)}(\pi_i\boldsymbol{x}) := \int_{(S^D)^V} \prod_{v=1}^{V}d\Omega_D(\widehat X_v)\, \mathrm P_\ell(\widehat X_{\mathsf a(i)},\widehat X_{\mathsf b(i)}) \prod_{j\ne i} G_{\eta_j,\Delta_j^+} (\widehat X_{\mathsf a(j)},\widehat X_{\mathsf b(j)}),
	\label{eq:graph-spectral-cut-integral}
\end{equation}
and
\begin{equation}
	\mathrm P_\ell(\widehat X,\widehat Y) := \sum_{\boldsymbol{m}} Y_{\ell\boldsymbol{m}}(\widehat X)Y_{\ell\boldsymbol{m}}^{*}(\widehat Y)
	\label{eq:graph-harmonic-projector}
\end{equation}
is the projector onto spherical harmonics of degree \(\ell\). The selected edge kernel is now specified by \(\ell\), while \(\pi_i\boldsymbol{x}\) contains the remaining parameters.

Since \(I_g^{(i,\ell)}(\pi_i\boldsymbol{x})\) is independent of \(\rho_i\), taking the normalized residue gives
\begin{equation}
	\lim_{\rho_i\to\pm(t+\ell)} \bigl[(t+\ell)^2-\rho_i^2\bigr] I_g(\boldsymbol{x}) = I_g^{(i,\ell)}(\pi_i\boldsymbol{x}).
	\label{eq:graph-spectral-cut}
\end{equation}
We call this operation a spectral cut of edge \(i\). It replaces the selected propagator by \(\mathrm P_\ell\), which generally retains a coupling between its endpoints.

For the zero mode, \(\mathrm P_0=1/\Omega_{2t+1}\), so the spectral cut gives
\begin{equation}
	\lim_{\rho_i\to\pm t} \Omega_{2t+1}(t^2-\rho_i^2)\, I_g(\boldsymbol{x}) = I_{g\setminus i}(\pi_i\boldsymbol{x}).
	\label{eq:graph-zero-mode-cut}
\end{equation}
Here \(g\setminus i\) is obtained by deleting edge \(i\) while retaining all vertices, including any left isolated.

The zero-mode cut also follows from the Schwinger representation \eqref{eq:graph-final-Euler-integral}. Set \(\epsilon=t-\rho_i\), so the selected physical edge contributes \(\lambda_i^{2t-\epsilon}\mu_i^\epsilon\). Together with the measure \(d\mu_i/\mu_i\), this gives a pole at \(\epsilon=0\). Writing the remaining integrand as \(f(\mu_i)=f(0)+O(\mu_i)\), we find
\begin{equation}
	\int_0^\infty d\mu_i\,\mu_i^{\epsilon-1}f(\mu_i) =\frac{f(0)}{\epsilon}+O(1).
\end{equation}
Only the constant term contributes to the pole, so the residue is obtained by setting \(\mu_i=0\) in the remaining integrand. At \(\mu_i=0\), edge \(i\) contributes only a diagonal term at \(a=\mathsf a(i)\), giving
\begin{equation}
	A_g\big|_{\mu_i=0} = \det\!\left( I_V+L_{g\setminus i}^{T}L_{g\setminus i} +\lambda_i^2e_ae_a^T \right).
	\label{eq:graph-zero-mode-endpoint-polynomial}
\end{equation}
Put \(q=1+\lambda_i^2\) and write each remaining endpoint parameter at \(a\) as \(\sqrt q\) times a new parameter. Factoring \(\sqrt q\) from row \(a\) and column \(a\) then gives \(qA_{g\setminus i}\), with \(A_{g\setminus i}\) evaluated in the new parameters. Rescaling \(z\mapsto z/q\) then removes \(q\) from the exponential. Collecting the remaining factors gives
\begin{equation}
	\lim_{\rho_i\to t}(t-\rho_i)J_g(\boldsymbol{x}) = J_{g\setminus i}(\pi_i\boldsymbol{x}) \int_0^\infty\frac{d\lambda_i}{\lambda_i}\, \frac{\lambda_i^{2t}}{(1+\lambda_i^2)^{w_a+t}} = \frac{\Gamma(t)\Gamma(w_a)}{2\Gamma(w_a+t)} J_{g\setminus i}(\pi_i\boldsymbol{x}),
	\label{eq:graph-zero-mode-Euler-cut}
\end{equation}
where \(w_a\) is evaluated at \(\rho_i=t\), and the beta integral is understood by meromorphic continuation.

Restoring the normalization \eqref{eq:graph-final-normalization} cancels the vertex-weight gamma factors, leaving \(\Gamma(t)/(4\pi^{t+1})=1/(2t\Omega_{2t+1})\). Multiplying by \(\Omega_{2t+1}(t+\rho_i)\), and exchanging \(\lambda_i\) and \(\mu_i\) for the negative pole, gives
\begin{equation}
	\lim_{\rho_i\to\pm t} \Omega_{2t+1}(t^2-\rho_i^2)\, I_g(\boldsymbol{x}) = I_{g\setminus i}(\pi_i\boldsymbol{x}).
	\label{eq:graph-zero-mode-Schwinger-cut}
\end{equation}
For the general spectral cut \eqref{eq:graph-spectral-cut} of the same physical edge, the endpoint residue selects the coefficient of \(\mu_i^\ell\) at \(\rho_i=t+\ell\), or of \(\lambda_i^\ell\) at \(\rho_i=-(t+\ell)\), in the remaining Schwinger integrand.
\clearpage
\section{Analyticity and Growth}
\label{app:analyticity-and-growth}

\subsection{Petal}
\label{app:petal-analyticity-and-growth}

We establish the analyticity and growth properties used in Section~\ref{subsec:one-loop-petal} to determine \(\mathrm p(s;\rho)\).

\paragraph{Analyticity.}
The Schwinger-parameter form \eqref{eq:lattice-I-NJ-x}--\eqref{eq:lattice-normalization-x} of the petal integral at the physical point \(\bar{\boldsymbol x}\) \eqref{eq:lattice-physical-point} reduces, after the \(z\) integration, to
\begin{equation}
	I_{\texttt{V}^{\circ}}(s;\rho) = \int_0^\infty\frac{d\lambda}{\lambda} \int_0^\infty\frac{d\mu}{\mu}\, \frac{\lambda^{s+\rho}\mu^{s-\rho}} {[1+(\lambda-\mu)^2]^{s+1}}.
	\label{eq:one-loop-self-loop-after-z}
\end{equation}
Setting \(\lambda=\omega\mu\) and integrating over \(\mu\), we obtain
\begin{equation}
	I_{\texttt{V}^{\circ}}(s;\rho) =\frac1{2s}\int_0^\infty \omega^{s+\rho-1}|1-\omega|^{-2s}\,d\omega.
	\label{eq:one-loop-self-loop-ratio-integral}
\end{equation}
Splitting at \(\omega=1\) and inverting the second interval gives
\begin{equation}
	2s\,I_{\texttt{V}^{\circ}}(s;\rho) = \int_0^1 \left[ \omega^{s+\rho-1}+\omega^{s-\rho-1} \right](1-\omega)^{-2s}\,d\omega.
	\label{eq:app-petal-folded-integral}
\end{equation}
Convergence requires \(\operatorname{Re}(s\pm\rho)>0\) and \(\operatorname{Re}s<\tfrac12\). We work with real \(s\), so \(0<s<\tfrac12\) and \(|\operatorname{Re}\rho|<s\).

Subtracting the constant term of \((1-\omega)^{-2s}\) at \(\omega=0\) gives
\begin{equation}
	2s\,I_{\texttt{V}^{\circ}}(s;\rho) =\frac1{s+\rho}+\frac1{s-\rho} +\int_0^1 \left[ \omega^{s+\rho-1}+\omega^{s-\rho-1} \right] \left[(1-\omega)^{-2s}-1\right]\,d\omega.
	\label{eq:app-petal-subtracted-integral}
\end{equation}
Since \((1-\omega)^{-2s}-1=O(\omega)\) at zero, the remaining integral is holomorphic for \(|\operatorname{Re}\rho|<s+1\). This continues \(I_{\texttt{V}^{\circ}}\) to a neighbourhood of \(|\operatorname{Re}\rho|\leq\tfrac12\), with only simple poles at \(\rho=\pm s\). The factor \(\mathrsfs{G}(s;\rho)\) has the same simple poles and no zeros, so \(\mathrm p=I_{\texttt{V}^{\circ}}/\mathrsfs{G}\) is holomorphic throughout this strip. Antiperiodicity then extends \(\mathrm p\) to an entire function of \(\rho\).

\paragraph{Growth conditions.}
Set \(\rho=\sigma+\mathrm i t\), with \(|\sigma|\leq\tfrac12\) and \(|t|\geq1\). The terms in \eqref{eq:app-petal-subtracted-integral} obey
\begin{equation}
	\left|\frac1{s\pm\rho}\right| \leq\frac1{|t|}\leq1, \qquad \left|\omega^{s\pm\rho-1}\right| =\omega^{s\pm\sigma-1} \leq\omega^{s-\frac32}.
	\label{eq:app-petal-growth-factor-bounds}
\end{equation}
The last inequality follows from \(s\pm\sigma-1\geq s-\tfrac32\) and \(0<\omega<1\). Applying these bounds to the two rational terms and the two terms inside the integral gives
\begin{equation}
	\left|I_{\texttt{V}^{\circ}}(s;\sigma+\mathrm i t)\right| \leq \frac1{2s} \left[ 2+ 2\int_0^1 \omega^{s-\frac32} \left[(1-\omega)^{-2s}-1\right]\,d\omega \right] =:C_s.
	\label{eq:app-petal-integral-growth-bound}
\end{equation}
The integrand behaves as \(2s\,\omega^{s-\frac12}\) as \(\omega\to0\) and as \((1-\omega)^{-2s}\) as \(\omega\to1\). Both powers are integrable for \(0<s<\tfrac12\), so \(C_s\) is finite and independent of \(\sigma\) and \(t\).

Stirling's formula gives, uniformly for \(|\sigma|\leq\tfrac12\),
\begin{equation}
	\left|\mathrsfs{G}(s;\sigma+\mathrm i t)\right| = \frac{2\pi\Gamma(1-2s)} {\Gamma(1-s)\Gamma(s+1)} |t|^{2s-1}e^{-\pi|t|} \left[1+O(|t|^{-1})\right], \qquad t\to\pm\infty.
	\label{eq:app-petal-gamma-growth}
\end{equation}
Dividing \eqref{eq:app-petal-integral-growth-bound} by \eqref{eq:app-petal-gamma-growth} therefore yields
\begin{equation}
	\left|\mathrm p(s;\sigma+\mathrm i t)\right| = O\!\left(|t|^{1-2s}e^{\pi|t|}\right), \qquad t\to\pm\infty, \qquad |\sigma|\leq\tfrac12.
	\label{eq:app-petal-growth-bound}
\end{equation}

\subsection{Three-melon}
\label{app:three-melon-analyticity-and-growth}

We use the hypergeometric form of the propagators to study the analyticity and growth of the three-melon density. We first determine the locations and orders of its spectral and dimensional poles, then establish the growth bounds needed to fix the homogeneous ambiguity.

\paragraph{Analyticity.}

For \(D=2s+1\), the density representation \eqref{eq:three-melon-density-one-dimensional} gives
\begin{equation}
	\bar I_{\texttt{V\{3\}V}}(s;\boldsymbol{\rho}) = 2^{2s}\Omega_{2s} \int_0^1 d\mathrm w\,[\mathrm w(1-\mathrm w)]^{s-\frac12} \prod_{i=1}^3G_{\Delta_i^\pm}(\mathrm w),
	\label{eq:app-three-melon-angular}
\end{equation}
where \(\Delta_i^\pm=s\pm\rho_i\) and
\begin{equation}
	G_{\Delta_i^\pm}(\mathrm w) = \frac{\Gamma(s+\rho_i)\Gamma(s-\rho_i)} {(4\pi)^{s+\frac12}\Gamma(s+\frac12)} \,{}_2F_1\!\left( s+\rho_i,s-\rho_i;s+\frac12;\mathrm w \right).
	\label{eq:app-three-melon-propagator}
\end{equation}
We keep \(\boldsymbol{\rho}\) fixed at generic values. Here \(\mathrm w=(1+\cos\theta)/2\), where \(\theta\) is the angular separation of the vertices. The endpoints \(\mathrm w=0\) and \(\mathrm w=1\) correspond to antipodal and coincident vertices, respectively. Away from spectral poles, the integral converges directly in the strip \(-\frac12<\operatorname{Re}s<1\).

For \(0\leq\mathrm w<1\), the regularized hypergeometric function
\begin{equation}
	\frac{{}_2F_1(a,b;c;\mathrm w)}{\Gamma(c)} = \sum_{n=0}^{\infty} \frac{(a)_n(b)_n}{\Gamma(c+n)n!}\,\mathrm w^n
	\label{eq:app-three-melon-regularized-hypergeometric}
\end{equation}
is entire in \(a,b,c\). The possible spectral poles of the propagator \eqref{eq:app-three-melon-propagator} come from its numerator gamma functions,
\begin{equation}
	s=-n\pm\rho_i, \qquad n=0,1,2,\ldots.
	\label{eq:app-three-melon-spectral-poles}
\end{equation}
For fixed \(\boldsymbol{\rho}\), these lie to the left of a sufficiently far-right half-plane. The antipodal endpoint \(\mathrm w=0\) is integrable for \(\operatorname{Re}s>-\frac12\), so any additional poles in this half-plane must arise from the coincident points, i.e., \(\mathrm w=1\).

Set \(\mathrm x=1-\mathrm w\), \(a_i=\frac12+\rho_i\), and \(b_i=\frac12-\rho_i\). For \(s-\frac12\notin\mathbb Z\), the connection formula at \(\mathrm w=1\) separates the propagator into singular and regular branches,
\begin{equation}
	\begin{aligned}
		G_{\Delta_i^\pm}(1-\mathrm x) ={}& \frac{\Gamma(s-\frac12)}{(4\pi)^{s+\frac12}} \mathrm x^{\frac12-s} \,{}_2F_1\!\left(a_i,b_i;\frac32-s;\mathrm x\right) \\
		&+ \frac{\Gamma(\frac12-s)\Gamma(s+\rho_i)\Gamma(s-\rho_i)} {(4\pi)^{s+\frac12}\Gamma(a_i)\Gamma(b_i)} \,{}_2F_1\!\left(s+\rho_i,s-\rho_i;s+\frac12;\mathrm x\right).
	\end{aligned}
	\label{eq:app-three-melon-connection}
\end{equation}
Both hypergeometric factors are analytic at \(\mathrm x=0\). At half-integer \(s\), the formula is understood by taking the limit of the complete expression; the apparent singularities of the separate branches cancel at separated points.

We substitute the connection formula \eqref{eq:app-three-melon-connection} into \eqref{eq:app-three-melon-angular} and select \(k\) singular branches from the three propagators. Including the measure and expanding the remaining analytic factors gives powers
\begin{equation}
	\mathrm x^{(1-k)(s-\frac12)+n}, \qquad k=0,1,2,3, \qquad n=0,1,2,\ldots.
	\label{eq:app-three-melon-endpoint-powers}
\end{equation}
Their endpoint integrals continue meromorphically as
\begin{equation}
	\int_0^\delta\mathrm x^{(1-k)(s-\frac12)+n}\,d\mathrm x = \frac{\delta^{(1-k)(s-\frac12)+n+1}} {(1-k)(s-\frac12)+n+1}, \qquad 0<\delta<1.
	\label{eq:app-three-melon-endpoint-integrals}
\end{equation}
The denominator does not vanish for \(k=0,1\) in \(\operatorname{Re}s>-\frac12\). For two or three singular branches, it gives the possible dimensional poles
\begin{equation}
	\begin{aligned}
		k=2:\quad& s=\frac32+n, &&D=4+2n, \\
		k=3:\quad& s=1+\frac n2, &&D=3+n,
	\end{aligned} \qquad n=0,1,2,\ldots.
	\label{eq:app-three-melon-dimensional-poles}
\end{equation}

At positive integer \(s\), corresponding to odd dimensions \(D=3,5,\ldots\), the branch coefficients in \eqref{eq:app-three-melon-connection} are regular in \(s\), and the two coincidence exponents differ by a noninteger, so the local expansion contains no logarithms. Equation~\eqref{eq:app-three-melon-endpoint-integrals} therefore gives at most simple dimensional poles.

At \(s=N+\frac12\), with \(N\geq1\), the exponents differ by an integer. Taking the limit of \eqref{eq:app-three-melon-connection} gives a finite singular part together with analytic and logarithmic terms,
\begin{equation}
	G_{\Delta_i^\pm}(1-\mathrm x) = \sum_{j=0}^{N-1}\mathrsfs{G}_{ij}\,\mathrm x^{j-N} +\mathrsfs{H}_i(\mathrm x)\log\mathrm x+\mathrsfs{J}_i(\mathrm x),
	\label{eq:app-three-melon-logarithmic-expansion}
\end{equation}
where \(\mathrsfs{H}_i\) and \(\mathrsfs{J}_i\) are analytic at \(\mathrm x=0\). The explicit forms of \(\mathrsfs{G}_{ij}\), \(\mathrsfs{H}_i(\mathrm x)\), and \(\mathrsfs{J}_i(\mathrm x)\) are not needed. This is the logarithmic limiting form of the hypergeometric connection formula.

A product containing two logarithmic factors contains at most one singular factor, whose strongest divergence is \(\mathrm x^{-N}\). Multiplication by the measure \(\mathrm x^N(1-\mathrm x)^N\,d\mathrm x\) makes these terms integrable. Terms with three logarithmic factors are also integrable. Consequently, only terms with at most one logarithm can contribute dimensional poles. Since
\begin{equation}
	\int_0^\delta\mathrm x^{\upsilon-1}\log\mathrm x\,d\mathrm x = \delta^\upsilon \left( \frac{\log\delta}{\upsilon} -\frac{1}{\upsilon^2} \right),
	\label{eq:app-three-melon-logarithmic-poles}
\end{equation}
the poles at \(s=N+\frac12\), corresponding to even dimensions \(D=4,6,\ldots\), are at most double.

\subparagraph{Special cases.}

At \(s=\frac12\) (\(D=2\)), each propagator has only a logarithmic coincidence singularity. The integrand behaves as \(O(|\log\mathrm x|^3)\), which is integrable, so there is no dimensional pole.

At \(s=1\), corresponding to \(D=3\), only the product of the three leading singular terms in \eqref{eq:app-three-melon-connection} is nonintegrable at coincidence. Only the \(n=0\) term in \eqref{eq:app-three-melon-endpoint-integrals} contributes to the pole at \(s=1\), giving
\begin{equation}
	\int_0^\delta \mathrm x^{1-2s}\,d\mathrm x = \frac{\delta^{2-2s}}{2(1-s)} = -\frac{1}{2(s-1)}+O(1), \qquad s\to1.
	\label{eq:app-three-melon-D3-endpoint-integral}
\end{equation}
The coefficient of \(\mathrm x^{-1/2}\) in each propagator at \(s=1\) is \(\Gamma(\tfrac12)/(4\pi)^{3/2}=1/(8\pi)\), independent of \(\rho_i\). Using \(\Omega_2=4\pi\), we obtain the universal residue
\begin{equation}
	\operatorname*{Res}_{s=1} \bar I_{\texttt{V\{3\}V}}(s;\boldsymbol{\rho}) = -\frac{2^2\Omega_2}{2(8\pi)^3} = -\frac1{64\pi^2}.
	\label{eq:app-three-melon-D3-residue}
\end{equation}
Equivalently, setting \(s=1-\epsilon/2\) gives
\begin{equation}
	\bar I_{\texttt{V\{3\}V}} (D=3-\epsilon;\boldsymbol{\rho}) = \frac{1}{32\pi^2\epsilon}+O(1).
	\label{eq:app-three-melon-D3-density-pole}
\end{equation}

\paragraph{Growth conditions.}

The solution of the dimensional recurrence in Section~\ref{sec:three-melon-analytical-solution} contains an undetermined \(2\)-periodic coefficient \(\mathrm p(s;\boldsymbol{\rho})\). To resolve this ambiguity, we study the asymptotic behaviour of the corresponding holomorphic periodic function \(\mathrm b(s;\boldsymbol{\rho})=\sin(\pi s)\cos^2(\pi s)\,\mathrm p(s;\boldsymbol{\rho})\) as the imaginary part of \(s\) tends to \(\pm\infty\).

We keep \(\boldsymbol{\rho}\) fixed and write \(s=\sigma+\mathrm i t\), with \(\tfrac12<\sigma<1\). Specifically, we will show that
\begin{equation}
	\frac{\mathrm b(s;\boldsymbol{\rho})}{\cos^2(\pi s)} = \sin(\pi s)\,\mathrm p(s;\boldsymbol{\rho}) \longrightarrow-1, \qquad t\to\pm\infty.
	\label{eq:app-three-melon-growth-condition}
\end{equation}
Writing the recurrence solution as \(\bar I_{\texttt{V\{3\}V}}=\mathrm p\,\mathrsfs{H}+\bar I_{\texttt{V\{3\}V}}^{\,\mathrm{part}}\) gives
\begin{equation}
	\frac{\mathrm b(s;\boldsymbol{\rho})}{\cos^2(\pi s)} = \sin(\pi s)\, \frac{\bar I_{\texttt{V\{3\}V}}(s;\boldsymbol{\rho})} {\mathrsfs{H}(s;\boldsymbol{\rho})} - \sin(\pi s)\, \frac{\bar I_{\texttt{V\{3\}V}}^{\,\mathrm{part}}(s;\boldsymbol{\rho})} {\mathrsfs{H}(s;\boldsymbol{\rho})}.
	\label{eq:app-three-melon-growth-ratio-decomposition}
\end{equation}
We show that the first term tends to \(-1\) and the second to zero as \(t\to\pm\infty\).

We first evaluate the density ratio. Euler's transformation gives
\begin{equation}
	{}_2F_1\!\left(s+\rho_i,s-\rho_i;s+\tfrac12;\mathrm w\right) = (1-\mathrm w)^{\frac12-s} {}_2F_1\!\left(a_i,b_i;s+\tfrac12;\mathrm w\right),
	\label{eq:app-three-melon-growth-euler-transform}
\end{equation}
with \(a_i=\tfrac12+\rho_i\) and \(b_i=\tfrac12-\rho_i\), independent of \(s\). Substituting into \eqref{eq:app-three-melon-angular}, we expand the three hypergeometric functions and integrate each monomial using the beta integral. Dividing by \(\mathrsfs{H}\) and multiplying by \(\sin(\pi s)\) gives, with \(n=n_1+n_2+n_3\),
\begin{equation}
	\sin(\pi s)\, \frac{\bar I_{\texttt{V\{3\}V}}(s;\boldsymbol{\rho})} {\mathrsfs{H}(s;\boldsymbol{\rho})} = \frac{ 2^{6-4s}\pi^2 \displaystyle\prod_{i=1}^{3}\Gamma(s+\rho_i)\Gamma(s-\rho_i) }{ (\tfrac32-s)\, \Gamma(s+\tfrac12)^2 \displaystyle\prod_{\boldsymbol\sigma\in\{\pm1\}^3} \Gamma\!\left(\frac{s+\rho_{\boldsymbol\sigma}}2\right) } \sum_{n_1,n_2,n_3\ge0} \frac{(s+\tfrac12)_n}{(\tfrac52-s)_n} \prod_{i=1}^{3} \frac{(a_i)_{n_i}(b_i)_{n_i}} {(s+\tfrac12)_{n_i}\,n_i!}.
	\label{eq:app-three-melon-growth-integrated-series}
\end{equation}

Next, we obtain the limit of the sum in \eqref{eq:app-three-melon-growth-integrated-series} as \(t\to\pm\infty\). In order to interchange the limit and sum, we need a convergent bound independent of \(t\). We use
\begin{equation}
	\left| \frac{(s+\tfrac12)_n}{(\tfrac52-s)_n} \right|\le1, \qquad \left|(s+\tfrac12)_{n_i}\right| \ge(\sigma+\tfrac12)_{n_i}.
	\label{eq:app-three-melon-growth-pochhammer-bounds}
\end{equation}
The first inequality is obtained simply by comparing the numerator and the denominator: their imaginary parts have equal magnitude, while \(\sigma+\tfrac12<\tfrac52-\sigma\). The second follows because each factor's modulus is at least its positive real part. Consequently, the absolute sum is bounded by
\begin{equation}
	\prod_{i=1}^{3} \sum_{n_i=0}^{\infty} \frac{|(a_i)_{n_i}(b_i)_{n_i}|} {(\sigma+\tfrac12)_{n_i}\,n_i!}.
	\label{eq:app-three-melon-growth-majorant}
\end{equation}
For large \(n_i\), Stirling's formula gives us
\begin{equation}
	\frac{|(a_i)_{n_i}(b_i)_{n_i}|} {(\sigma+\tfrac12)_{n_i}\,n_i!} = O\!\left(n_i^{-\sigma-\frac12}\right),
	\label{eq:app-three-melon-growth-majorant-tail}
\end{equation}
so each series converges for \(\sigma>\tfrac12\). For fixed \(n_1,n_2,n_3\), every term in the triple sum in \eqref{eq:app-three-melon-growth-integrated-series} with \(n>0\) decays as \(O(|t|^{-n})\) as \(|t|\to\infty\), while the term with \(n=0\) equals one. The convergent bound \eqref{eq:app-three-melon-growth-majorant} is independent of \(t\), so dominated convergence permits interchanging the limit and sum. Therefore, the triple sum tends to \(1\) as \(t\to\pm\infty\).

It remains to evaluate the prefactor in \eqref{eq:app-three-melon-growth-integrated-series}. For fixed \(\boldsymbol{\rho}\), using Stirling's formula, we get
\begin{equation}
	\frac{ \displaystyle\prod_{i=1}^{3}\Gamma(s+\rho_i)\Gamma(s-\rho_i) }{ \displaystyle\Gamma(s+\tfrac12)^2 \prod_{\boldsymbol\sigma\in\{\pm1\}^3} \Gamma\!\left(\frac{s+\rho_{\boldsymbol\sigma}}2\right) } = \frac{2^{4s-6}s}{\pi^2} [1+O(s^{-1})], \qquad t\to\pm\infty.
	\label{eq:app-three-melon-growth-gamma-quotient}
\end{equation}
Now, using \eqref{eq:app-three-melon-growth-gamma-quotient} and the limit of the triple sum in \eqref{eq:app-three-melon-growth-integrated-series}, we obtain
\begin{equation}
	\sin(\pi s)\, \frac{\bar I_{\texttt{V\{3\}V}}(s;\boldsymbol{\rho})} {\mathrsfs{H}(s;\boldsymbol{\rho})} \longrightarrow-1, \qquad t\to\pm\infty.
	\label{eq:app-three-melon-growth-full-integral}
\end{equation}

We next treat the particular-solution term in \eqref{eq:app-three-melon-growth-ratio-decomposition}. For its Barnes representation, initially assume \(\max_i|\Re\rho_i|<\sigma\), and \(\tfrac12<\sigma<1\) as above. Dividing \eqref{eq:three-melon-analytic-particular-sum} by \(\mathrsfs{H}\) gives 
\begin{equation}
	\mathrsfs{P}(s) := \frac{\bar I_{\texttt{V\{3\}V}}^{\,\mathrm{part}}(s;\boldsymbol\rho)} {\mathrsfs{H}(s;\boldsymbol\rho)} = -\sum_{\mathrm{cyc}(k,i,j)} \sum_{n=0}^{\infty} \frac{\mathsf C_{s,k\vert ij}(s+2n;\boldsymbol\rho)} {\mathrsfs{H}(s+2n+2;\boldsymbol\rho)}.
	\label{eq:app-three-melon-particular-ratio}
\end{equation}
The sum in \eqref{eq:app-three-melon-particular-ratio} then becomes
\begin{equation}
	\mathrsfs{P}(s) = -\sec^2(\pi s)\sum_{n=0}^{\infty}\mathsf S(s+2n),
	\label{eq:app-three-melon-particular-tail}
\end{equation}
where \(\mathsf S(u):=\sum_{\mathrm{cyc}(k,i,j)}\mathsf S_{k\vert ij}(u)\), with the source \eqref{eq:three-melon-analytic-source-sum} and \(\mathrsfs{H}\) \eqref{eq:three-melon-analytic-propagator} giving
\begin{equation}
	\mathsf S_{k\vert ij}(u) = \frac{ \pi^{3/2}2^{-2u-4}\Gamma(u) \cos(\pi\rho_i)\cos(\pi\rho_j) \Gamma(u+\rho_i)\Gamma(u-\rho_i) \Gamma(u+\rho_j)\Gamma(u-\rho_j) }{ \Gamma(u+\tfrac32) \displaystyle\prod_{\boldsymbol\sigma\in\{\pm1\}^3} \Gamma\!\left(\frac{u+2+\rho_{\boldsymbol\sigma}}2\right) } \left.\mathsf M(\mathsf z_k)\right|_{s=u}.
	\label{eq:app-three-melon-particular-R-explicit}
\end{equation}
The factor \(\sec^2(\pi s)\) in \eqref{eq:app-three-melon-particular-tail} originates from the two petal densities \eqref{eq:three-melon-recurrence-density}. The poles can come from gamma functions in the numerator, but \(\mathsf S(u)\) is holomorphic for \(\Re u>\max_i|\Re\rho_i|\). 

Next, we find an analytic continuation of the sum \eqref{eq:app-three-melon-particular-ratio} through a Barnes representation and use that representation to study the growth of the particular solution. Alternatively, we could use \eqref{eq:three-melon-connection-formula}, but the presence of several hypergeometric functions makes that approach quite cumbersome.

To obtain the Barnes representation of the sum, we can restore the generating variable \(\zeta\) of \eqref{eq:three-melon-tail-generating-function} by replacing the factor \(16^n\) in \eqref{eq:three-melon-tail-summand-factorization} by \(\zeta^n\), i.e.,
\begin{equation}
	\mathrsfs{P}(s;\zeta) := -\sec^2(\pi s) \sum_{n=0}^{\infty} \mathsf S(s+2n)\left(\frac{\zeta}{16}\right)^n, \qquad |\zeta|<1.
	\label{eq:app-three-melon-particular-generating-function}
\end{equation}
%The required value \(\mathrsfs{P}(s)\) is the symmetric boundary value of this function at \(\zeta=16\).
The generating function \eqref{eq:app-three-melon-particular-generating-function} has the Barnes representation
\begin{equation}
	\mathrsfs{P}(s;\zeta) = -\frac{\sec^2(\pi s)}{2\pi\mathrm i} \int_L \Gamma(-v)\Gamma(1+v)\mathsf S(s+2v) \left(-\frac{\zeta}{16}\right)^v\,dv, \qquad |\arg(-\zeta)|<\pi,
	\label{eq:app-three-melon-barnes-generating-function}
\end{equation}
where \(L\) runs upwards along \(\Re v=\eta\), with
\begin{equation}
	-1<\eta<0, \qquad \sigma+2\eta>\max_i|\Re\rho_i|.
	\label{eq:app-three-melon-barnes-contour}
\end{equation}
The poles of \(\Gamma(-v)\), at \(v=0,1,2,\ldots\), lie to the right of \(L\); and those of \(\Gamma(1+v)\), at \(v=-1,-2,\ldots\), lie to its left. The second condition given in \eqref{eq:app-three-melon-barnes-contour} ensures that \(\mathsf S(s+2v)\) is holomorphic on and to the right of the contour \(L\).

To prove \eqref{eq:app-three-melon-barnes-generating-function}, we begin by taking \(-1<\zeta<0\) and closing the contour \(L\) to the right through a rectangle whose right edge is at \(\Re v=N+\tfrac12\). The contour then encloses the poles at \(v=0,\ldots,N\). The top and bottom edges vanish exponentially due to the factor \(\Gamma(-v)\Gamma(1+v)=-\pi/\sin(\pi v)\) as the height of the rectangle tends to infinity, while Stirling's formula applied to \eqref{eq:app-three-melon-particular-R-explicit} bounds the integral on the right edge by \(O(|\zeta|^N N^{-1/2})\), which also vanishes when \(N\to\infty\).

At each enclosed pole, we obtain the residue 
\begin{equation}
	\begin{aligned}
		\operatorname*{Res}_{v=n} \left[ \Gamma(-v)\Gamma(1+v)\mathsf S(s+2v) \left(-\frac{\zeta}{16}\right)^v \right] &= \frac{(-1)^{n+1}}{n!}\,n!\, \mathsf S(s+2n)\left(-\frac{\zeta}{16}\right)^n \\
		&= -\mathsf S(s+2n)\left(\frac{\zeta}{16}\right)^n.
	\end{aligned}
	\label{eq:app-three-melon-barnes-residues}
\end{equation}
The contour runs clockwise, so the residue theorem supplies an additional minus sign multiplying the contribution in \eqref{eq:app-three-melon-barnes-residues}. Summing the residues and using the prefactor in \eqref{eq:app-three-melon-barnes-generating-function} reproduces \eqref{eq:app-three-melon-particular-generating-function}.

At \(\zeta=16\pm\mathrm i0\), the factor \((-\zeta/16)^v\) becomes \(e^{\mp\mathrm i\pi v}\). Taking the symmetric average given in Section~\ref{sec:three-melon-analytical-solution} and using \(\Gamma(-v)\Gamma(1+v)=-\pi/\sin(\pi v)\) in \eqref{eq:app-three-melon-barnes-generating-function}, we obtain
\begin{equation}
	\mathrsfs{P}(s) = \frac{\sec^2(\pi s)}{2\mathrm i} \int_L \cot(\pi v)\,\mathsf S(s+2v)\,dv.
	\label{eq:app-three-melon-barnes-symmetric-kernel}
\end{equation}
We now write \(s=\sigma+\mathrm i t\) and estimate \(\mathrsfs{P}(s)\) as \(t\to\pm\infty\). First, parametrize \(L\) by \(v=\eta+\tfrac{\mathrm i}{2}(w-t)\), with \(w\in\mathbb R\), and set \(\gamma=\sigma+2\eta\), such that the argument \(s+2v=\gamma+\mathrm i w\) of \(\mathsf S\) is independent of \(t\). Then, we use \(dv=\mathrm i\,dw/2\) and obtain
\begin{equation}
	\mathrsfs{P}(s) = \frac{\sec^2(\pi s)}4 \int_{\mathbb R}K_t(w)\mathsf S(\gamma+\mathrm iw)\,dw, \qquad K_t(w) := \cot\!\left[ \frac{\pi}{2}\bigl(\gamma-\sigma+\mathrm i(w-t)\bigr) \right].
	\label{eq:app-three-melon-particular-barnes}
\end{equation}
Here, we keep \(\sigma\) and \(\gamma\) fixed, with \(\max_i|\Re\rho_i|<\gamma<\sigma\) and \(0<\sigma-\gamma<2\).

We now use \eqref{eq:app-three-melon-particular-barnes} to determine the large-\(|t|\) behaviour of \(\mathrsfs{P}(\sigma+\mathrm it)\). For fixed \(\sigma\), the prefactor \(\sec^2(\pi s)\) decays as \(e^{-2\pi|t|}\). To establish the decay of \(\mathrsfs{P}\), we must also determine how the integral depends on \(t\). We will show that this integral is \(O(1)\) as \(t\to\pm\infty\), and therefore \(\mathrsfs{P}\) decays similarly. 

Stirling's formula applied to \eqref{eq:app-three-melon-particular-R-explicit} gives
\begin{equation}
	\mathsf S_{k\vert ij}(u) = -\frac{15}{2\sqrt\pi} \cos(\pi\rho_i)\cos(\pi\rho_j)\, 4^u u^{-1/2}[1+O(u^{-1})].
	\label{eq:app-three-melon-particular-R-asymptotic}
\end{equation}
On \(u=\gamma+\mathrm iw\), we separate the oscillating factor by writing
\begin{equation}
	\mathsf S(\gamma+\mathrm iw)=e^{\mathrm iw\log4}a(w), \qquad a(w)=O(|w|^{-1/2}), \qquad |w|\to\infty.
	\label{eq:app-three-melon-particular-oscillatory-source}
\end{equation}
Since the contour contains no poles, \(a\) is smooth for every real \(w\) and tends to zero at infinity, so it is bounded. This bound is independent of \(t\).

The kernel \(K_t(w)\) in \eqref{eq:app-three-melon-particular-barnes} is also bounded uniformly in \(w\) and \(t\). Indeed, its cotangent argument has fixed real part \(\frac{\pi}{2}(\gamma-\sigma)\in(-\pi,0)\), so it encounters no poles, and \(K_t(w)\to\mp\mathrm i\) as \(w-t\to\pm\infty\).

However, these bounds alone do not establish convergence of the integral: the \(|w|^{-1/2}\) decay in \eqref{eq:app-three-melon-particular-oscillatory-source} is not integrable at infinity. Integration by parts gives
\begin{equation}
	\int_{\mathbb R}K_t(w)\mathsf S(\gamma+\mathrm iw)\,dw = -\frac1{\mathrm i\log4} \int_{\mathbb R}e^{\mathrm iw\log4} \bigl[a'(w)K_t(w)+a(w)K_t'(w)\bigr]\,dw.
	\label{eq:app-three-melon-barnes-integral-bound}
\end{equation}
The boundary term \(\left[e^{\mathrm iw\log4}a(w)K_t(w)/(\mathrm i\log4)\right]_{-\infty}^{\infty}\) vanishes because \(a(w)\to0\) and \(K_t(w)\) is bounded. We can now bound the two terms in \eqref{eq:app-three-melon-barnes-integral-bound}. Differentiating the expansion \eqref{eq:app-three-melon-particular-R-asymptotic} gives \(a'(w)=O(|w|^{-3/2})\). Since \(a'\) is smooth at finite \(w\), this decay makes it absolutely integrable. Multiplication by the uniformly bounded kernel \(K_t\) therefore gives a bound on the first integral that is independent of \(t\). For the second term, differentiating the kernel gives
\begin{equation}
	|K_t'(w)| = \frac{\pi/2}{ \sin^2\!\left[\frac{\pi}{2}(\gamma-\sigma)\right] +\sinh^2\!\left[\frac{\pi}{2}(w-t)\right] }.
	\label{eq:app-three-melon-barnes-kernel-bounds}
\end{equation}
The sine term is fixed and positive, so \(K_t'\) is finite at every \(w\). The exponential decay as \(|w-t|\to\infty\) makes its absolute integral finite, and changing \(t\) merely shifts the integration variable, leaving that integral unchanged. Since \(a\) is bounded independently of \(t\), the second integral is also bounded independently of \(t\). Thus the integral in \eqref{eq:app-three-melon-particular-barnes} is \(O(1)\) as \(t\to\pm\infty\).

Substituting this bound into \eqref{eq:app-three-melon-particular-barnes} and using \(|\sin(\pi s)|=O(e^{\pi|t|})\), we obtain
\begin{equation}
	\mathrsfs{P}(\sigma+\mathrm it)=O(e^{-2\pi|t|}), \qquad \sin(\pi s)\, \frac{\bar I_{\texttt{V\{3\}V}}^{\,\mathrm{part}}(s;\boldsymbol\rho)} {\mathrsfs{H}(s;\boldsymbol\rho)} = O(e^{-\pi|t|}) \longrightarrow0.
	\label{eq:app-three-melon-growth-particular-solution}
\end{equation}
Combining this with \eqref{eq:app-three-melon-growth-full-integral} in the decomposition \eqref{eq:app-three-melon-growth-ratio-decomposition} establishes that
\begin{equation}
	\frac{\mathrm b(s;\boldsymbol\rho)}{\cos^2(\pi s)} = \sin(\pi s)\,\mathrm p(s;\boldsymbol\rho) \longrightarrow-1, \qquad t\to\pm\infty.
	\label{eq:app-three-melon-growth-condition-established}
\end{equation}

\bibliographystyle{unsrtnat}
\bibliography{VacuumGraphsOnSphere}
\end{document}